\documentclass[a4paper,11pt]{article}
\pdfoutput=1 
\usepackage{jheppub} % for details on the use of the package, please
\usepackage{graphicx,dcolumn,bm,epsfig,cancel,hyperref}
\usepackage{gensymb}
\usepackage{multirow}
\usepackage{hyperref}
\usepackage{aas_macros}
\usepackage{amsmath}
\usepackage{amssymb, times}
\usepackage[utf8]{inputenc}
\newcommand{\comment}[1]{{}}
\usepackage{color}

\newcommand{\be}{\begin{equation}}
\newcommand{\ee}{\end{equation}}
\newcommand{\bea}{\begin{eqnarray}}
\newcommand{\eea}{\end{eqnarray}}

\newcommand {\xmm} {\textsl{XMM-Newton}}
\newcommand {\chandra} {\textsl{Chandra}}
\newcommand {\swift} {\textsl{Swift}}
\newcommand {\rxte} {\textsl{RXTE}}

\newcommand {\nustar} {\textsl{NuSTAR}}
\newcommand {\suzaku} {\textsl{Suzaku}}
\newcommand {\integral} {\textsl{INTEGRAL}}
\newcommand{\fdg}{\mbox{\ensuremath{.\!\!^\circ}}}% 
\newcommand {\commoncaption}[1]{
The theoretical curve (black dashed) displayed corresponds to ALP mass  $m_a = 10^{-6} \text{eV}$ and
 the value of $G_{an} \times g_{a \gamma\gamma} = #1 \text{GeV}^{-2}$, corresponding to the star. \textbf{Right Panel}: The $95\%$ CL upper limits on $G_{an} \times g_{a \gamma\gamma}$ versus $m_a$ resulting from our analysis of this magnetar without (solid) and with (dashed) incorporating superfluidity. The four sets of curves, distinguished by their colors, correspond to core temperatures $T_c=10^8$ K, $5\times 10^8$ K, $10^9$ K and $5\times 10^9$ K, from top to bottom, respectively.
}

\def \sgr {SGR$\,$1806$-$20}
\def\aa {1E$\,$1547.0$-$5408}
\def\flux {\mbox{erg~cm$^{-2}$~s$^{-1}$}}

\def\sgrnew{SGR$\,$0501+4516}
\def\fu{4U$\,$0142+61}
\def \sgraa {SGR$\,$1900+14}
\def \e {1E$\,$1841$-$045}
\def \enew {1E$\,$2259+586}
\def \rxs {1RXS$\,$J1708$-$4009}
\DeclareMathOperator{\sign}{sign}

\begin{document}
\preprint{INT-PUB-21-001}
\title{Magnetars and Axion-like Particles: Probes with the Hard X-ray Spectrum
}
\author[a]{Jean-Fran\c{c}ois Fortin,}
\author[b]{Huai-Ke Guo,}
\author[c,d]{Steven P. Harris,}
\author[e]{Elijah Sheridan,}
\author[b]{Kuver Sinha,}
\affiliation[a]{D\'epartement de Physique, de G\'enie Physique et d'Optique,\\Universit\'e Laval, Qu\'ebec, QC G1V 0A6, Canada}
\affiliation[b]{Department of Physics and Astronomy, University of Oklahoma, Norman, OK 73019, USA}
\affiliation[c]{Physics Department, Washington University, St. Louis, Missouri 63130, USA}
\affiliation[d]{Institute for Nuclear Theory, University of Washington, Seattle, WA 98195, USA}
\affiliation[e]{Department of Physics and Astronomy, Vanderbilt University, Nashville, TN 37212, USA}
\emailAdd{jean-francois.fortin@phy.ulaval.ca}
\emailAdd{ghk@ou.edu}
\emailAdd{harrissp@uw.edu}
\emailAdd{elijah.sheridan@vanderbilt.edu}
\emailAdd{kuver.sinha@ou.edu}

%\pacs{11.30.Er, 11.30.Fs, 11.30.Hv, 12.60.Fr, 31.30.jp}

\abstract
{Quiescent hard X-ray and soft gamma-ray emission from neutron stars constitute a promising frontier to explore axion-like-particles (ALPs). ALP production in the core peaks at energies of a few keV to a few hundreds of keV; subsequently, the ALPs escape and  convert to photons in the magnetosphere. The emissivity goes as $\sim T^6$ while the conversion probability is enhanced for large magnetic fields, making magnetars, with their high core temperatures and strong magnetic fields, ideal targets for probing ALPs. We  compute the  energy spectrum of photons resulting from conversion of ALPs in the magnetosphere and then compare it   against hard X-ray data from \nustar, \integral, and \xmm\,for a set of eight magnetars for which such data exists. Upper limits are placed on the product of the ALP-nucleon and ALP-photon couplings. For the  production in the core, we perform a calculation of the ALP emissivity in degenerate nuclear matter modeled by a relativistic mean field theory.  The reduction of the emissivity due to improvements to the one-pion exchange approximation is incorporated, as is the suppression of the emissivity due to proton superfluidity in the neutron star core. A range of core temperatures is considered, corresponding to different models of the steady heat transfer from the core to the stellar surface.  For the subsequent conversion, we solve the coupled differential equations mixing ALPs and photons in the magnetosphere. The conversion occurs due to a competition between the dipolar magnetic field and the photon refractive index induced by the external magnetic field. Semi-analytic expressions are provided alongside the full numerical results.  We also present an analysis of the uncertainty on the axion limits we derive due to the uncertainties in the magnetar masses, nuclear matter equation of state, and the proton superfluid critical temperature.}

\maketitle

\section{Introduction}
A major focus of the search for physics beyond the Standard Model is the QCD axion \cite{Weinberg:1977ma, Wilczek:1977pj, Peccei:1977hh} and more general pseudo-scalar axion-like particles (ALPs), which are ubiquitous in string theory \cite{Arvanitaki:2009fg, Cicoli:2012sz}. The conversion of axions or ALPs in magnetic fields has been a long-standing  method to search for these particles in astroparticle physics. Much of this effort has centered on the conversion of ALPs in large-scale magnetic fields (we refer to \cite{Mirizzi:2006zy, Csaki:2001yk} for some representative papers in this vast literature). 

A perhaps less-explored but nevertheless very interesting alternative search method  that has witnessed a resurgence lately is the conversion of ALPs to photons near \textit{localized} sources like neutron stars.  ALPs ($a$) produced in the core by nucleon ($N$) bremsstrahlung (from the term $\mathcal{L} \supset G_{an} (\partial_\mu a)\bar{N}\gamma^\mu \gamma_5 N$) convert to photons in the magnetosphere (from the term $\mathcal{L} \supset \frac{1}{4} g_{a \gamma \gamma} a F_{\mu \nu} \tilde{F}^{\mu \nu}$).\footnote{We will assume that the ALP-proton and ALP-neutron couplings are equal, and denote both by $G_{an}$.} To establish notation for the rest of the paper, we collect the parts of the ALP Lagrangian that are relevant for us:
\be
\mathcal{L} \supset  G_{an} (\partial_\mu a)\bar{N}\gamma^\mu \gamma_5 N - \frac{1}{4} g_{a \gamma \gamma} a F_{\mu \nu} \tilde{F}^{\mu \nu} \,\,.
\ee
Such ALP-produced photons provide an exotic source of emission from neutron stars which cannot exceed the actual observed emission in any energy bin; this results in constraints on the product of ALP couplings $G_{an} \times g_{a \gamma \gamma}$. One should immediately note that searches based on \textit{conversion} are distinct from searches based purely on \textit{cooling} limits, which are sensitive  to the ALP-nucleon coupling $G_{an}$ (which controls bremsstrahlung emission from the core) or ALP-electron coupling $g_{aee}$ (emission from the crust) but not the ALP-photon coupling $g_{a\gamma\gamma}$ \cite{Sedrakian:2018kdm}. One should also note that the goal here is \textit{exclusion} of regions of ALP parameter space, and not the modeling of the actual observed spectrum from neutron stars using ALP-induced emission; claiming an excess over astrophysical background would require one to   carefully model such backgrounds. That is not the goal of our paper.

Since a stronger magnetic field favors the conversion probability $P_{a\rightarrow \gamma}$, neutron stars with extremely strong magnetic fields -- magnetars -- are the natural targets for such investigations. Moreover, since the production of ALPs in neutron stars is proportional to $T^6$ where $T$ is the core temperature, magnetars -- with their internal temperatures reaching $\sim 10^9$ K -- serve as a natural site where ALPs, if they exist, would be copiously produced. The large production rate in conjunction with an enhanced conversion probability make magnetars an ideal site to probe ALP-induced emission. Imaging, spectroscopy, timing, and polarimetry of X-ray emission from magnetars are major targets of several future experiments, making this a particularly opportune moment to add fundamental physics as a component of these missions.

A major challenge in the program of constraining ALPs with galactic-scale magnetic fields is the modeling of the  field in the diverse  environments that the ALP-photon system must traverse to reach Earth. Simulations with many  assumptions are typically employed to model the field in the host galaxy of the source, the intergalactic medium, and the Milky Way. For \textit{localized} conversions near compact objects, on the other hand,  the magnetic field -- sometimes reaching critical strengths of $B \sim \mathcal{O}(10^{14})$ G -- is precisely measured by data and  well-approximated by a dipole. Moreover, the ALP only traverses a small distance before it converts, typically $\sim \mathcal{O}(1000 r_0)$, where $r_0$ is the neutron star radius. The confluence of these factors endows such scenarios with the possibility of performing precision calculations of ALP-induced modulations to the photon spectrum, and hence precision constraints on ALP parameter space.

The dipole nature of the magnetic field makes solving the propagation equations analytically and obtaining the correct  behavior of the conversion probability non-trivial, although the general behavior is intuitive. The ALP-photon mixing matrix involves an off-diagonal term that is proportional to the magnetic field and whose parametric dependence on the radial distance $r$ is $\sim g_{a\gamma\gamma}B_0/r^3$, where $B_0$ is the magnetic field at the surface. The diagonal terms include the ALP mass $m_a$ and the photon mass, which can be derived from the refractive indices and whose parametric dependence on $r$ is $\sim 1/r^6$. At the surface $r = r_0$ the photon mass term dominates and the mixing angle (see Appendix \ref{appconv}) is small; far away $r \gg r_0$ the ALP mass term $m_a$ dominates and the mixing angle is again small; in the middle, around the ``radius of conversion" $r_{a \rightarrow \gamma} \sim \mathcal{O}(1000 r_0)$, the off-diagonal mixing term becomes the same order as the diagonal photon mass term and the mixing angle becomes appreciable. It is here that ALP-photon conversion occurs.

To our knowledge, the first attempt at obtaining $P_{a\rightarrow \gamma}$ in the dipolar magnetic field near neutron stars was performed by Morris \cite{Morris:1984iz}, who vastly overestimated the probability of conversion by failing to take into account the non-zero refractive indices of the photon in the strong magnetic field. The classic paper by Raffelt and Stodolsky \cite{{Raffelt:1987im}} properly accounted for the refractive indices, but only calculated $P_{a\rightarrow \gamma}$ near the neutron star surface, and concluded that it was too small to be observable. This is true for typical values of the magnetic field and the values of the ALP-photon coupling of interest $g_{a \gamma \gamma} \sim 10^{-10}$ GeV$^{-1}$. However, as we have pointed out, the probability of conversion actually does become substantial near the radius of conversion $r_{a \rightarrow \gamma}$. The numerical solution to the full propagation equations and semi-analytic expressions for $P_{a \rightarrow \gamma}$ were presented by a subset of the current authors in \cite{Fortin:2018ehg, Fortin:2018aom}, where the applications to hard X-ray emission from neutron stars were emphasized.\footnote{Some related work appeared subsequently in \cite{Buschmann:2019pfp,Dessert:2019dos}. Soft X-rays have been explored in \cite{Perna:2012wn, Lai:2006af}.} In a subsequent paper, the importance of probing ALPs in soft gamma-ray emission from neutron stars was emphasized by some of the authors in \cite{Lloyd:2020vzs, sinhawip}.

The purpose of this paper is to apply the methods developed in \cite{Fortin:2018ehg, Fortin:2018aom, Lloyd:2020vzs} to a systematic study of existing hard X-ray data from magnetars. There are two main features of the current study:

$(i)$  We perform a calculation of the production of ALPs due to nucleon bremsstrahlung processes in the neutron star core. These processes, occurring in the degenerate nuclear matter core of the magnetar, are calculated using a modern nuclear equation of state (EoS)  coming from a relativistic mean field theory.  The calculation is performed using a one-pion exchange interaction between nucleons; we also add a correction factor to the production rate that incorporates improvements to the nuclear interaction, decreasing the axion emissivity by approximately a factor of four.  In addition, we include the effects of proton superfluidity in the magnetar core, which introduces a gap in the proton energy spectrum, strongly suppressing the axion production rate from any process involving protons.  The precise core temperature of observed magnetars is unknown (see Appendix \ref{a1}), so we examine a range of core temperatures which correspond to different models of heat transfer between the magnetar core and crust.

$(ii)$ The spectrum of hard X-ray photons produced from ALP to photon conversion in the magnetosphere is calculated for the set of eight magnetars  listed in Tab.~\ref{tab:catalog}.  The theoretical spectrum is plotted on the $\nu\,F_{\nu}$ plane and subsequently compared against data from \suzaku, \integral, \xmm, \nustar\, and \rxte. Limits are then placed on the product of couplings $G_{an} \times g_{a\gamma\gamma}$ for each magnetar, for several benchmark core temperatures, with and without accounting for superfluidity.

This paper is structured as follows. After a short introduction to magnetars, we first calculate the production of ALPs from the core for non-superfluid (or, ``ungapped") and superfluid nuclear matter in Sections \ref{magnintro}, \ref{ungap}, \ref{superflu}, respectively. We then incorporate the conversion of the produced ALPs into photons in Section \ref{conv1}. In Section \ref{magnresults}, we present our results for the magnetars listed in Tab.~\ref{tab:catalog}. We end with our Conclusions. A series of appendices contain the details of our calculations: a discussion of the range of core temperatures considered in this work appears in Appendix \ref{a1}; the details of ALP emissivity in degenerate nuclear matter can be found in Appendix \ref{sec:np_emissivity};  a discussion of improvements to the one-pion exchange approximation appears in Appendix \ref{a3}; calculations for ALP emissivity with superfluid protons are presented in Appendix \ref{a4}; details of the ALP-photon mixing matrix, propagation equations, and conversion probability are gathered in Appendix \ref{appconv}; and a discussion of the uncertainties in our derived axion coupling constraints due to possible variations in the magnetar masses, nuclear EoS, and proton critical temperature is given in Appendix \ref{sec:uncertainties}.

%%%%%%%%%%%%%%%%%%%%%%%%%%%%%%%%%%%%%%%%%%%%%%%%%%%%%%%%%%%%%%%%%%%%%%%%%%%

\section{Magnetars: ALP Production, Conversion, and Spectrum}

In this section, we calculate the emissivity and spectrum of ALPs produced in the core of magnetars.  We then incorporate the ALP-photon conversion probability to obtain the spectrum of photons produced from the conversion process. We begin with a short introduction to this class of neutron stars. 

\subsection{Magnetars} \label{magnintro}

Magnetars are a group of  neutron stars whose measured spin periods ($P\sim
2$--$12$~s) and spin-down rates ($\dot{P}\sim10^{-15}$--$10^
{-10}$~s~s$^{-1}$) provide evidence for dipole magnetic fields with strengths up to $10^
{14}$--$10^{15}$~G (we refer to \cite{Turolla:2015mwa, Kaspi:2017fwg, Enoto:2019vcg} for reviews). Magnetars emit short X-ray bursts ($L_{\rm X}\sim 
10^{36}$--$10^{43}$~erg~s$^{-1}$ with duration $\Delta t\sim 
0.01$--$50$~s) and giant 
flares ($L_{\rm X}\sim10^{44}$--$10^{47}$~erg~s$^{-1}$ and 
$\Delta t\sim100$--$1000$~s). For the purposes of our work, we are interested in the persistent emission from magnetars 
 in the $\sim
0.5$--$200$~keV band, with $L_{\rm X}\sim10^{31}$--$10^{36}$~erg~s$^{-1}$.  The persistent emission exhibits two distinct components: soft
quasi-thermal emission up to around 10~keV, and  very flat hard X-ray tails extending to beyond 200~keV. The soft X-ray emission can be fit with an absorbed blackbody with temperature $kT \sim 0.5$~keV in conjunction with a power-law component  $dN/dE \propto E^{-
\Gamma}$ at suprathermal energies. The power-law component of the soft thermal emission is quite steep, with index $\Gamma
\sim 1.5-4 $. The hard X-ray component, on the other hand, is much flatter and has index $\Gamma \sim 0.7 - 1.5$. Data from  \integral~\cite{Papitto:2020tgi, Kuiper:2004ya, Kuiper:2006is,Hartog:2008tq,Hartog:2008tp, Mereghetti:2004sx, Molkov:2004sy, Gotz:2006cx}, \suzaku~\cite{Morii:2010vi,Enoto:2011fg, Enoto:2010ix, Enoto:2017dox}, \rxte ~\cite{Levine:1996du}, \swift ~\cite{Kuiper_2012}, \xmm~\cite{Rea:2009nh}, ASCA and \nustar~\cite{An:2013xui, Vogel:2014xfa, Younes:2017sme} have revealed the hard X-ray component for around nine magnetars. Observations by the COMPTEL instrument on the Compton Gamma-Ray Observatory put upper limits on the hard spectral component, indicating a spectral turnover in the range 200-500 keV. \footnote{The McGill Online Magnetar Catalog website, \href{http://www.physics.mcgill.ca/~pulsar/magnetar/TabO4.html}{http://www.physics.mcgill.ca/~pulsar/magnetar/TabO4.html}, lists  pulsed and total hard X-ray fluxes in the 20–150 keV range for ten magnetars, with HEASARC's WebPIMMS being used to estimate the fluxes in the absence of direct data.  Ref.~\cite{Olausen:2013bpa} describes the quantities in the McGill table in more detail.}

The soft X-ray component is generally ascribed to thermal emission from the surface, modified by the magnetosphere. The origin of the hard X-ray component is a difficult problem and one of ongoing research, with resonant inverse Compton scattering of thermal photons by ultra-relativistic charges being advanced as the most efficient mechanism leading to such emission. We refer to \cite{Beloborodov:2012ug, Baring:2017wvb, Wadiasingh:2017rcq} and references therein for major work in the astrophysics literature in this direction. Since we will be interested in deriving upper limits on ALP couplings in our work and not claiming excess over astrophysical ``background", we will not recapitulate the detailed astrophysical  models introduced in these papers.

We will use published hard X-ray spectra for the magnetars listed in Tab.~\ref{tab:catalog} gleaned from a variety of sources. The experimental data will be compared to the theoretical spectrum derived in this Section. The derivation of the theoretical spectrum consists of two steps:

$(i)$ Calculating the emissivity of ALPs produced in the core: this will be performed in Sections  \ref{ungap} and \ref{superflu}, with more details provided in Appendices \ref{a1}, \ref{sec:np_emissivity}, \ref{a3} and \ref{a4}.

$(ii)$ Calculating the probability of conversion of ALPs to photons in the magnetosphere and the resulting photon spectrum: this will be performed in Section \ref{conv1}, with details provided in Appendix~\ref{appconv}.

We now perform these steps.

\subsection{ALP Production: Preliminaries}

In our analysis of axion (we will use the terms ``axions" and ``ALPs" interchangably in the remainder of the paper) production in magnetars, we assume that the magnetar is a $1.4M_{\odot}$ neutron star, composed of charge neutral, beta equilibrated nuclear matter.  This matter is uniform in the neutron star core, but decreases in density as the surface is approached, eventually transitioning into a solid crust.  We will focus on axion production from the neutron star core, as it is expected to dominate over axion emission from the crust \cite{Sedrakian:2018kdm}.  In our calculations of the axion production, we can ignore axion reabsorption because the mean free path of axions in cold nuclear matter is large compared to the size of a magnetar \cite{Harris:2020qim,Burrows:1990pk}.  We use the IUF equation of state \cite{Fattoyev:2010mx} to model the uniform nuclear matter in the core, which consists of neutrons, protons, electrons, and muons.  We do not consider the possibility of exotic phases of matter in the core \cite{Annala:2019puf,Dexheimer:2019pay,Han:2019bub}.  The outer crust is described by the EoS of Ruester \textit{et al.} (2006) \cite{Ruester:2005fm} and the inner crust is described by the EoS of Douchin and Haensel (2001) \cite{Douchin:2001sv}.  A $1.4M_{\odot}$ star with this equation of state has a radius of 12.6 km.  The inner 11.3 km of the star is a uniform nuclear matter core, and the outer 1.3 km is the crust.  
%%%%%%%%%%%%%%%%%%%%%%%%%%%%%%%%%%%%%%%%%%%%%%%%%%%%%%%%%%%%%
\subsection{Ungapped nuclear matter} \label{ungap}
In dense matter modeled by a relativistic mean field theory (like the IUF EoS we use here) \cite{glendenning2000compact}, neutron and proton quasiparticles behave like free fermions with (equal) Dirac effective masses $m_*$ and effective chemical potentials $\mu_n^*$ and $\mu_p^*$.  The chemical potentials include the rest mass of the particle.  The nucleons have energy dispersion relations
\begin{equation}
    E_i = \sqrt{p^2+m_*^2}+U_i \equiv E_i^*+U_i,\label{eq:E_ungapped}
\end{equation}
where $U_i$ is the nuclear mean field experienced by the neutron or proton \cite{Roberts:2016mwj}.  The electrons and muons are treated as free Fermi gases.  In magnetars, where the temperature is much less than an MeV, the neutrino mean free path is large compared to the size of the magnetar.  Consequently, no Fermi sea of neutrinos builds up in the star \cite{Yakovlev:2000jp}.

In the uniform nuclear matter core of a magnetar, since the process $N\rightarrow N+a$ (where $N$ is a neutron or proton) is kinematically forbidden, a spectator nucleon $N'$ is required to conserve energy and momentum and thus axions are produced in the nucleon bremsstrahlung reaction $N+N'\rightarrow N+N'+a$.  In ungapped nuclear matter, this is the primary mechanism of axion production in the magnetar core.  Below, we review the calculations of the axion emissivity due to $n+n\rightarrow n+n+a$ and $p+p\rightarrow p+p+a$ in the limit of strongly degenerate nuclear matter.  In addition, we improve upon the calculation of the axion emissivity from $n+p\rightarrow n+p+a$ that is presented in \cite{Iwamoto:1992jp}.  We defer the discussion of axion production in superfluid nuclear matter to Section. \ref{superflu}.

Following Friman \& Maxwell \cite{1979ApJ...232..541F}, we model the interaction between the nucleons in this bremsstrahlung process by one-pion exchange \cite{Machleidt:2017vls}.  The matrix element, determined by summing the eight tree-level Feynman diagrams which describe this process, was calculated by Brinkmann \& Turner \cite{PhysRevD.38.2338}.  Assuming that the neutron and proton couple with equal strength to the axion, the matrix element for $n+n\rightarrow n+n+a$ and $p+p\rightarrow p+p+a$ is
\begin{equation}
    S\sum_{\text{spins}}\vert\mathcal{M}\vert^2 = \frac{256}{3}(m_n/m_{\pi})^4 f^4 G_{an}^2\left[\frac{\mathbf{k}^4}{(\mathbf{k}^2+m_{\pi}^2)^2}+\frac{\mathbf{l}^4}{(\mathbf{l}^2+m_{\pi}^2)^2}+\frac{\mathbf{k}^2\mathbf{l}^2-3(\mathbf{k}\cdot \mathbf{l})^2}{(\mathbf{k}^2+m_{\pi}^2)(\mathbf{l}^2+m_{\pi}^2)}\right],
\end{equation}
where $f\sim 1$ is the pion-nucleon coupling constant \cite{1979ApJ...232..541F} and $\mathbf{k}=\mathbf{p_2}-\mathbf{p_4}$ and $\mathbf{l}=\mathbf{p_2}-\mathbf{p_3}$ are 3-momentum transfers between the nucleons.  The factor of $m_n$ comes from the definition of the pion-nucleon coupling, and is thus the vacuum mass of the neutron, not the effective mass.  The symmetry factor $S=1/4$ accounts for the presence of identical particles in both the initial and final states. The matrix element for $n+p\rightarrow n+p+a$
\begin{equation}
    S\sum_{\text{spins}}\vert\mathcal{M}\vert^2 = \frac{1024}{3}(m_n/m_{\pi})^4 f^4 G_{an}^2\left[\frac{\mathbf{k}^4}{(\mathbf{k}^2+m_{\pi}^2)^2}+\frac{4\mathbf{l}^4}{(\mathbf{l}^2+m_{\pi}^2)^2}-2\frac{\mathbf{k}^2\mathbf{l}^2-3(\mathbf{k}\cdot \mathbf{l})^2}{(\mathbf{k}^2+m_{\pi}^2)(\mathbf{l}^2+m_{\pi}^2)}\right] ,\label{eq:npmatrixelement} 
\end{equation}
was calculated originally in \cite{PhysRevD.38.2338}, but a minus sign error was corrected by \cite{Carenza:2019pxu}.  In this process, $S=1$ because neither the initial nor the final state contains identical particles.

The nucleons in the core of magnetars are strongly degenerate, as the core temperature is much smaller than the nucleon Fermi energy.  In degenerate nuclear matter, the dominant contribution to the bremsstrahlung rate comes from nucleons near their Fermi surface, and thus the calculation of the axion emissivity can be performed using the Fermi surface approximation, which is discussed in Appendix \ref{sec:np_emissivity} and in Ref.~\cite{Harris:2020qim}.  In this approximation, the axion emissivity from the process $n+n\rightarrow n+n+a$ is
\begin{equation}
Q^0_{nn} = \frac{31}{2835\pi}C_{\pi}(m_n/m_{\pi})^4f^4G_{an}^2p_{Fn}F(c)T^6,  \label{eq:Q0nn}  
\end{equation}
where $c=m_{\pi}/(2p_{Fn})$, $p_{Fn}$ is the neutron Fermi momentum, and 
\begin{equation}
    F(y) = 4-\frac{1}{1+y^2}-5y\arctan{\left(\frac{1}{y}\right)}+\frac{2y^2}{\sqrt{1+2y^2}}\arctan{\left(\frac{1}{\sqrt{1+2y^2}}\right)}.
\end{equation}
The factor of $C_{\pi}$ is discussed at the end of this section and in Appendix \ref{a3}.  The expression for the axion emissivity $Q^0_{pp}$ from the process $p+p\rightarrow p+p+a$ is the same, but with the neutron Fermi momentum $p_{Fn}$ replaced with the proton Fermi momentum $p_{Fp}$ and $c=m_{\pi}/(2p_{Fn})$ replaced with $d=m_{\pi}/(2p_{Fp})$.  These expressions for $Q^0_{nn}$ and $Q^0_{pp}$ are standard in the literature, having been derived in \cite{Iwamoto:1984ir} (see also \cite{Iwamoto:1992jp,Stoica:2009zh,Harris:2020qim}). The emissivity of axions from the $n+p\rightarrow n+p+a$ process in strongly degenerate nuclear matter is\footnote{\label{footnote:raffelt}This emissivity was calculated in \cite{Iwamoto:1992jp}, but as Raffelt notes in \cite{Raffelt:1996wa}, the interference between Feynman diagrams was improperly treated.  In addition, the emissivity of this process in the degenerate limit was analytically calculated in \cite{PhysRevD.38.2338}, but under the assumption of a momentum-independent matrix element.}
\begin{equation}
    Q^0_{np} = \frac{124}{2835\pi}C_{\pi}(m_n/m_{\pi})^4f^4G_{an}^2p_{Fp}G(c,d)T^6.
    \label{eq:Q0_np}
\end{equation}
The derivation of this quantity and the expression for $G(c,d)$ are provided in Appendix \ref{sec:np_emissivity}.

In these expressions for $Q^0_{nn}$, $Q^0_{pp}$, and $Q^0_{np}$, we have introduced a multiplicative factor $C_{\pi}$ to encapsulate the overestimate of the strong interaction rate by the one-pion exchange approximation \cite{Beznogov:2018fda}.  In the OPE approximation, $C_{\pi}=1$, but in this paper, we choose $C_{\pi}=1/4$ based on calculations from \cite{Hanhart:2000ae}, which we describe in Appendix \ref{a3}.

In this analysis of hard X-ray emission from magnetars, we need the differential emissivity, as we are only concerned with the emitted axions that are able to convert to hard X-rays.   Below, we present the differential emissivity $\mathop{dQ}/\mathop{dx}$, where $x=\omega/T$ and $\omega$ is the energy of the emitted axion.  The differential emissivity of axions due to $n+n\rightarrow n+n+a$ is
\begin{equation}
    \frac{\mathop{dQ^0_{nn}}}{\mathop{dx}} = \frac{1}{36\pi^7}C_{\pi} (m_n/m_{\pi})^4 f^4 G_{an}^2p_{Fn} F(c)T^6\frac{x^3(x^2+4\pi^2)}{e^x-1} .
    \label{eq:axion_nn_emissivity}
\end{equation}
For the process $p+p\rightarrow p+p+a$, the differential emissivity $\mathop{dQ^0_{pp}}/\mathop{dx}$ is the same except that $p_{Fn}$ is replaced with $p_{Fp}$ and $c=m_{\pi}/(2p_{Fn})$ is replaced with $d=m_{\pi}/(2p_{Fp})$.  The emissivity due to the process $n+p\rightarrow n+p+a$ is given by
\begin{equation}
    \frac{\mathop{dQ^0_{np}}}{\mathop{dx}} = \frac{1}{9\pi^7}C_{\pi}(m_n/m_{\pi})^4 f^4 G_{an}^2p_{Fp} G(c,d)T^6\frac{x^3(x^2+4\pi^2)}{e^{x}-1}.
    \label{eq:axion_np_emissivity}
\end{equation}
The total differential emissivity of axions from ungapped nuclear matter is given by the sum 
\begin{equation}
    \frac{\mathop{dQ^0_{\text{total}}}}{\mathop{dx}} = \frac{\mathop{dQ^0_{nn}}}{\mathop{dx}} + \frac{\mathop{dQ^0_{pp}}}{\mathop{dx}}+\frac{\mathop{dQ^0_{np}}}{\mathop{dx}}.
    \label{eq:total_normal_emissivity}
\end{equation}

To obtain the differential luminosity $\mathop{dL^0_{\text{total}}}/\mathop{dx}$ of axions from a $1.4M_{\odot}$ magnetar, we integrate the differential emissivity over the magnetar core 
\begin{equation}
    \frac{\mathop{dL^0_{\text{total}}}}{\mathop{dx}} = \int \mathop{d^3r}\frac{\mathop{dQ^0_{\text{total}}}}{\mathop{dx}} = 4\pi \int_0^{R_{\text{crust}}}\mathop{dr}r^2\frac{\mathop{dQ^0_{\text{total}}}}{\mathop{dx}}.\label{eq:dL0}
\end{equation}
%%%%%%%%%%%%%%%%%%%%%%%%%%%%%%%%%%%%%%%%%%%%%%%%%%%%%%%%%%%%
\subsection{Superfluid nuclear matter} \label{superflu}
When the temperature is less than the critical temperature, which depends on the baryon density, nucleons in high density nuclear matter undergo Cooper pairing \cite{PhysRev.108.1175}, forming a superfluid.  Cooper pairing of nucleons - which is typically considered in the singlet or triplet channels - is possible because of the long-range attractive nature of the nuclear force.  The formation of Cooper pairs, either $\{nn\}$ or $\{pp\}$, opens a gap near the Fermi surface in the energy spectrum of the paired species.  Thus, superfluidity does not change the equation of state significantly, but has a large impact on properties such as the specific heat and the rate of neutrino production inside the neutron star.  Overviews of superfluidity in nuclear matter and its impact on transport are given in \cite{Yakovlev:1999sk,Sedrakian:2018ydt,Page:2013hxa}.

Beznogov \textit{et al.} \cite{Beznogov:2018fda} ran simulations of neutron star cooling (due to neutrino emission only) and compared to data from the cooling supernova remnant HESS J1731-347 in order to put constraints on the critical temperatures of neutron and proton superfluidity in the neutron star core.  Combining the results of their Markov-Chain Monte Carlo analysis with the theoretical expectation that the proton $^1S_0$ critical temperature is higher than the neutron $^3P_2$ critical temperature \cite{Page:2013hxa,Sedrakian:2018ydt}, they found that the critical temperature for proton singlet pairing must be above $4\times 10^9$ K in most of the core, while the critical temperature for neutron triplet pairing must be lower than $3\times 10^8$ K throughout the entire core.  Similar conclusions were reached by \cite{Beloin:2016zop}.  

\begin{figure}
    \centering
    \includegraphics[width=0.45\textwidth]{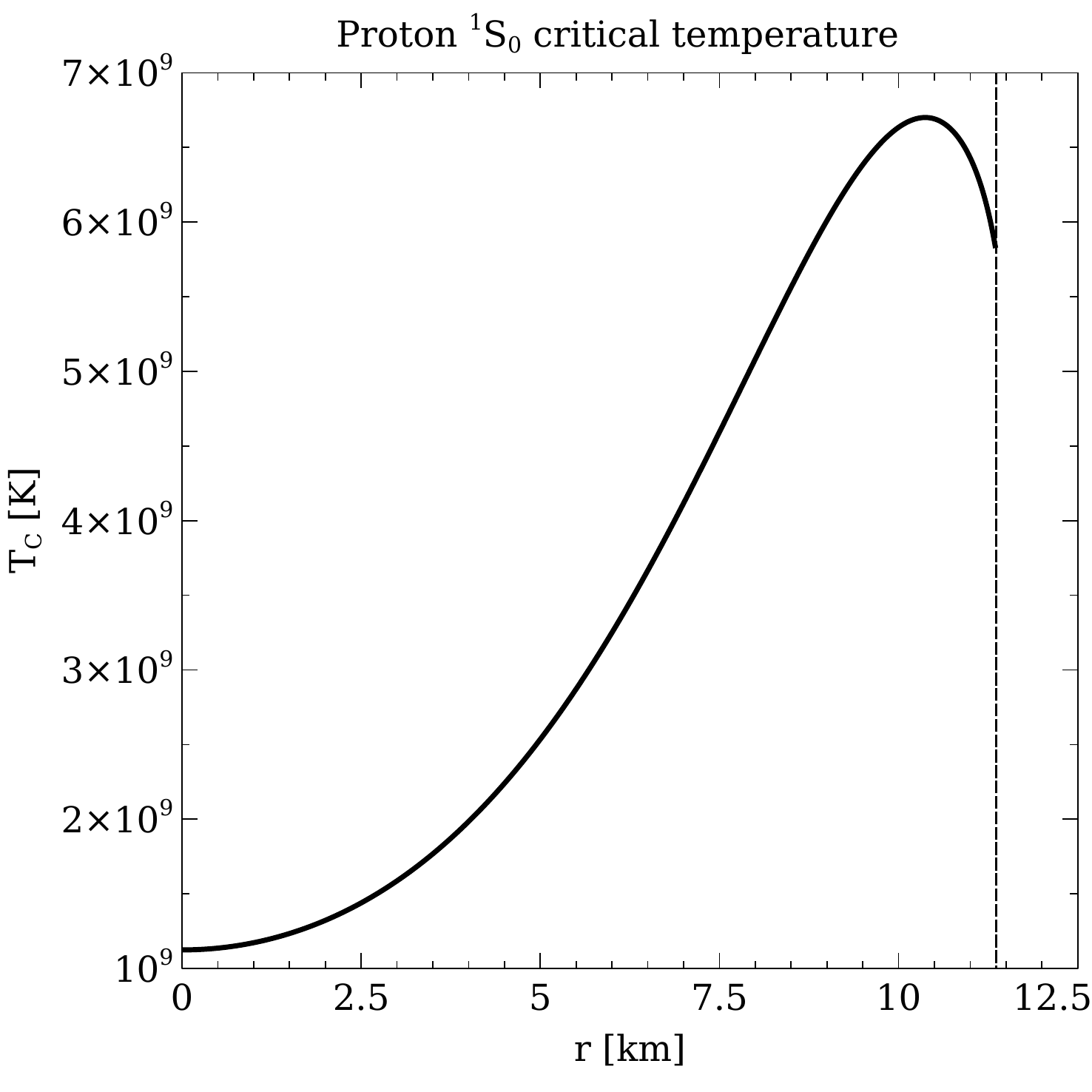}
    \includegraphics[width=0.45\textwidth]{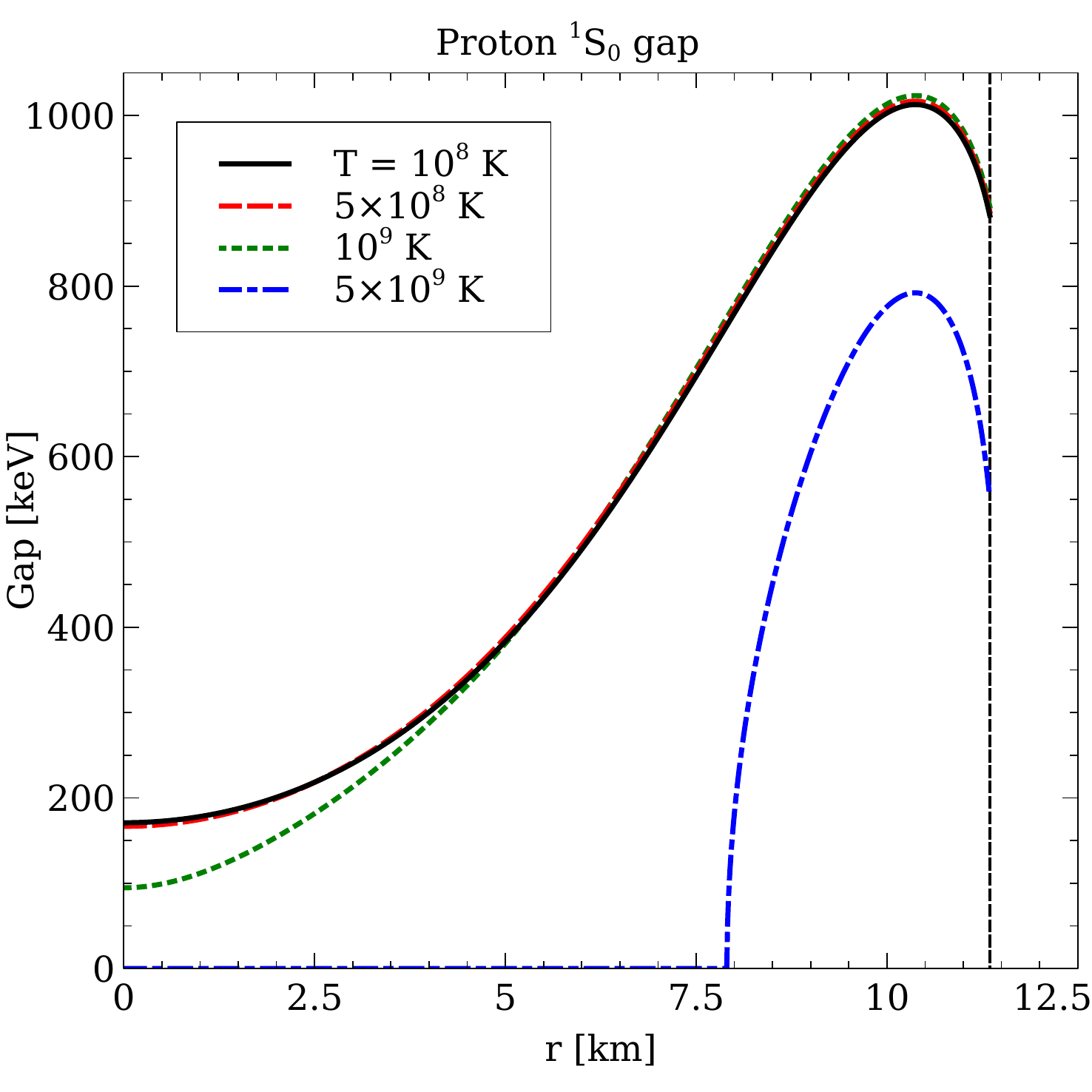}
    \caption{\textbf{Left panel:} Profile of the critical temperature for $^1S_0$ proton superfluidity in a $1.4M_{\odot}$ magnetar, where the proton gap is calculated using the Ho \textit{et al.} \cite{Ho:2014pta} parametrization of the CCDK model \cite{Chen:1993bam}.  The dashed vertical line indicates the transition from the uniform nuclear matter core to the crust.  \textbf{Right panel:} Profile of the gap $\Delta(n_B,T)$ in the proton energy spectrum, for four different magnetar core temperatures.}
    \label{fig:criticaltemp}
\end{figure}

In this work, we assume the core temperature of the magnetars in Tab.~\ref{tab:catalog} is larger than $10^8 \text{ K}$, allowing us to neglect neutron triplet pairing\footnote{If the magnetar core temperature is below $10^8 \text{ K}$ (see, for example, \cite{An:2018nze}), we would likely have to consider neutron triplet pairing, causing the axion production rate from nucleon bremsstrahlung to drop dramatically.  However, the presence of superfluid neutrons introduces a new channel of axion production from the breaking and formation of neutron Cooper pairs.} and consider proton singlet pairing that has a critical temperature $T_c(n_B)$ that is consistent with the constraints from \cite{Beznogov:2018fda,Beloin:2016zop}.  To describe the $^1S_0$ proton pairing, we choose the CCDK model \cite{Chen:1993bam} which predicts the size of the zero-temperature proton gap as a function of density.  The functional form of $\Delta(T=0,n_B)$ is parametrized by Ho \textit{et al.} \cite{Ho:2014pta} (see Eq.~2 and Tab.~2 in their paper).  For singlet pairing, the zero-temperature gap $\Delta(T=0,n_B)$ and the critical temperature $T_c(n_B)$ are related by \cite{Yakovlev:1999sk}
\begin{equation}
    T_c(n_B)=0.5669 \Delta(T=0,n_B).
\end{equation}
We plot the resultant critical temperature profile in a $1.4M_{\odot}$ magnetar in the left panel of Fig.~\ref{fig:criticaltemp}.  

The size of the gap at finite temperature is obtained by solving the gap equation -- see for example, Eq.~7 in \cite{Yakovlev:1999sk}.  The results have been fit using the functional form \cite{Yakovlev:1999sk,1994ARep...38..247L}
\begin{equation}
    \frac{\Delta(T,n_B)}{T} \sim \sqrt{1-\tau}\left(1.456-\frac{0.157}{\sqrt{\tau}}+\frac{1.764}{\tau}\right), \label{eq:vA}
\end{equation}
where $\tau = T/T_c(n_B)<1$.  For $\tau\geq 1$, $\Delta(T,n_B)=0$ and the matter is no longer in the superfluid phase.  The proton gap throughout the $1.4M_{\odot}$ magnetar is plotted in the right panel of Fig.~\ref{fig:criticaltemp} for several values of the core temperature.  As the temperature approaches the critical temperature from below, the size of the gap decreases and when the critical temperature is exceeded, the gap is zero.

Near the Fermi surface, the proton energy spectrum becomes
\begin{equation}
    E_p-\mu_p =  \begin{cases} 
      -\sqrt{\Delta^2+v_{Fp}^2(p-p_{Fp})^2} & p<p_{Fp} \\
      +\sqrt{\Delta^2+v_{Fp}^2(p-p_{Fp})^2} & p\geq p_{Fp}, \\
   \end{cases}\label{eq:superfluid_energy}
\end{equation}
where $v_{Fp}=\partial E_p/\partial p \vert_{p=p_{Fp}}$ is the proton Fermi velocity.  

In nuclear matter where the protons are superfluid, there are now four mechanisms of axion emission.  The three nucleon bremsstrahlung processes discussed in Sec. \ref{ungap} are still active, although those involving protons are suppressed by a function of $\exp{(-\Delta/T})$ when the protons are in the superfluid phase, due to the gap in the proton energy spectrum \cite{Page:2013hxa,Alford:2016cee}.  There is also an additional mechanism of axion production, originating from the temperature-induced breaking and formation of proton Cooper pairs \cite{Keller:2012yr,Buschmann:2019pfp,1976ApJ...205..541F}.  When a proton Cooper pair is formed, energy is liberated that can be taken away by an axion.  This process happens slowly when the temperature is much less than the critical temperature, but the rate increases significantly as the critical temperature is approached from below.  Above the critical temperature, the Cooper pair breaking and formation processes cannot occur.

The emissivity of the bremsstrahlung processes involving superfluid protons was calculated in \cite{Yakovlev:1999sk,1995A&A...297..717Y}, which we review below.  Like the calculation in ungapped nuclear matter (Section \ref{ungap}), the axion emissivity from $p+p\rightarrow p+p+a$ with superfluid protons is calculated using the Fermi surface approximation, but with the gapped proton dispersion relations Eq.~\ref{eq:superfluid_energy}.  The details of the calculation are given in Appendix \ref{a4}.  The axion emissivity with superfluid protons can be written as
\begin{equation}
    Q^S_{pp} = Q^0_{pp}R_{pp}(n_B,T),\label{eq:QSppwithR}
\end{equation}
where $R_{pp}(n_B,T)$, given in Eq.~\ref{eq:ppsuperfluidintegral}, is a factor that is less than one, representing the reduction of the normal-matter rate due to the gap $\Delta(n_B,T)$ in the proton's energy spectrum.  The emissivity due to $n+p\rightarrow n+p+a$ can similarly be written
\begin{equation}
    Q^S_{np} = Q^0_{np}R_{np}(n_B,T),\label{eq:QSnpwithR}
\end{equation}
where $R_{np}$ is given in Eq.~\ref{eq:npsuperfluidintegral}.  The emissivity due to Cooper pair formation $p+p\rightarrow \{pp\}+a$ is \cite{Keller:2012yr}
\begin{equation}
    Q_{CP} = \frac{G_{an}^2}{3\pi}\nu(0)v_{Fp}^2\Delta^2T^3\int_0^{\infty}\mathop{dx} \frac{x^3}{(1+e^{x/2})^2}\frac{\theta(x-2\Delta/T)}{\sqrt{x^2-4\Delta^2/T^2}}.\label{eq:Q_CP}
\end{equation}
Here, $\nu(0)= m_*^Lp_{Fp}/(\pi^2)$ is the density of states at the proton Fermi surface \cite{coleman2015introduction} and $m_*^L=\sqrt{p_{Fp}^2+m_*^2}$ is the Landau effective mass of the protons \cite{Maslov:2015wba,Li:2018lpy}.  The proton Fermi velocity is given by $v_{Fp}=p_{Fp}/\sqrt{p_{Fp}^2+m_*^2}$.

\begin{figure}[t]
  \centering
  \includegraphics[width=0.48\textwidth]{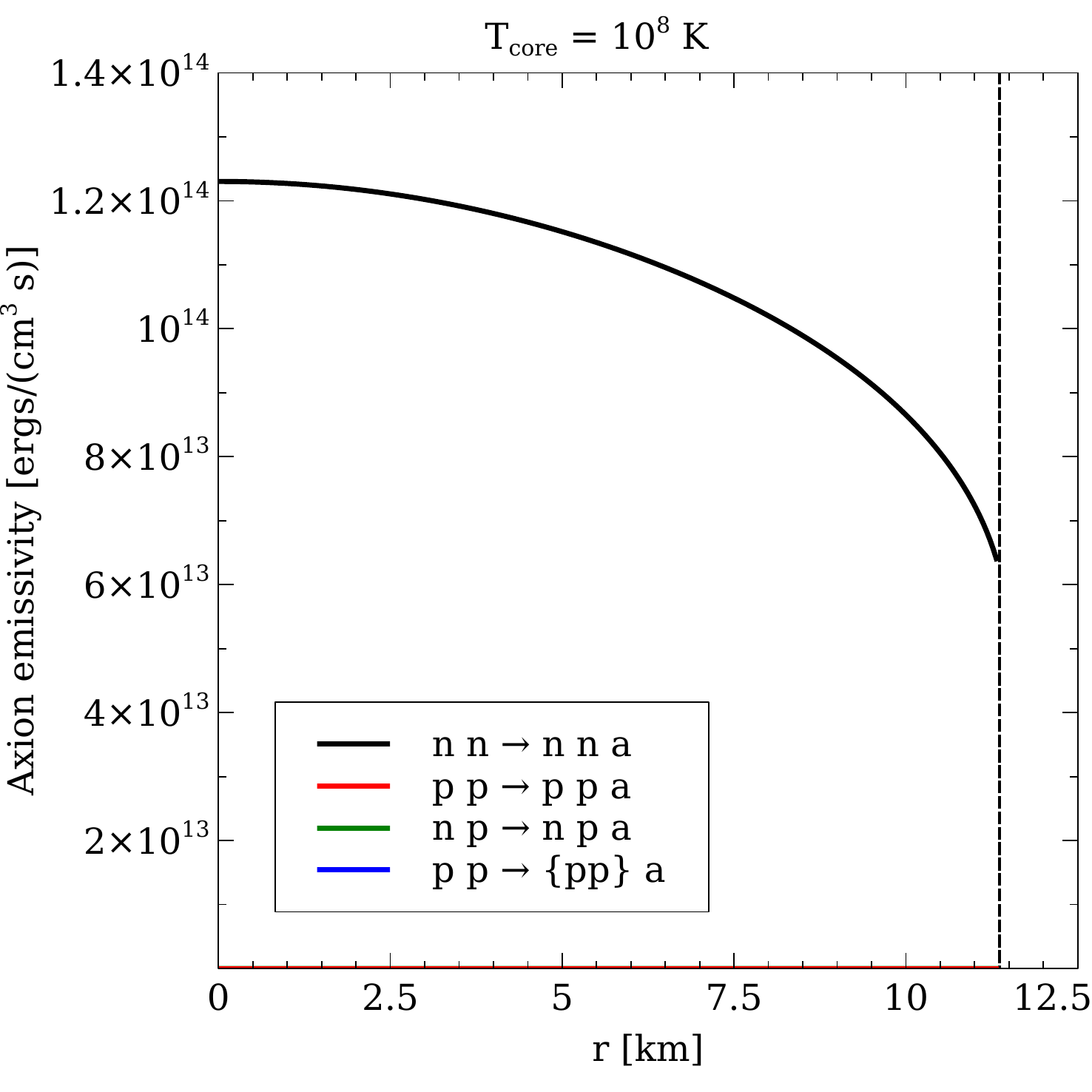}
  \includegraphics[width=0.48\textwidth]{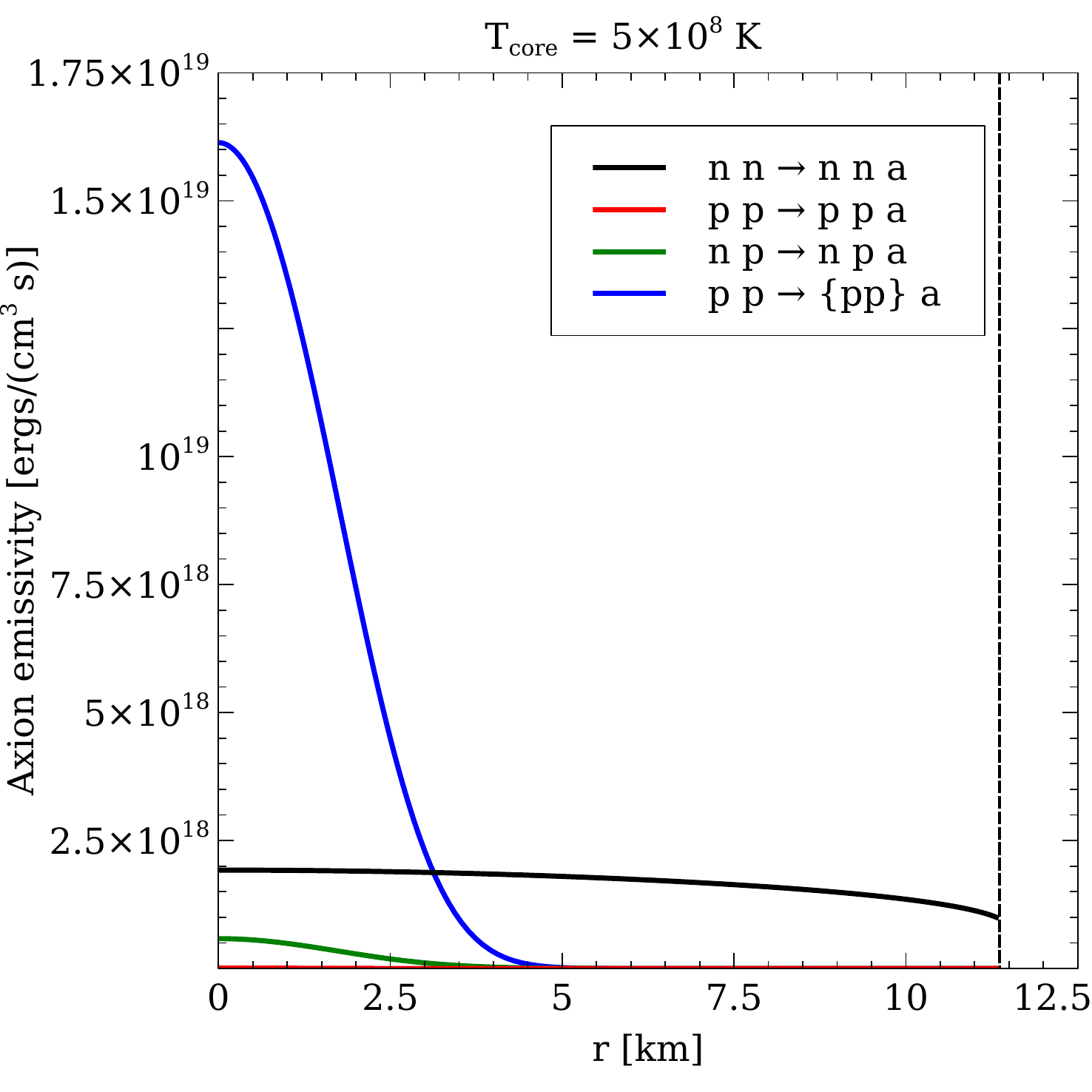} \\
  \includegraphics[width=0.48\textwidth]{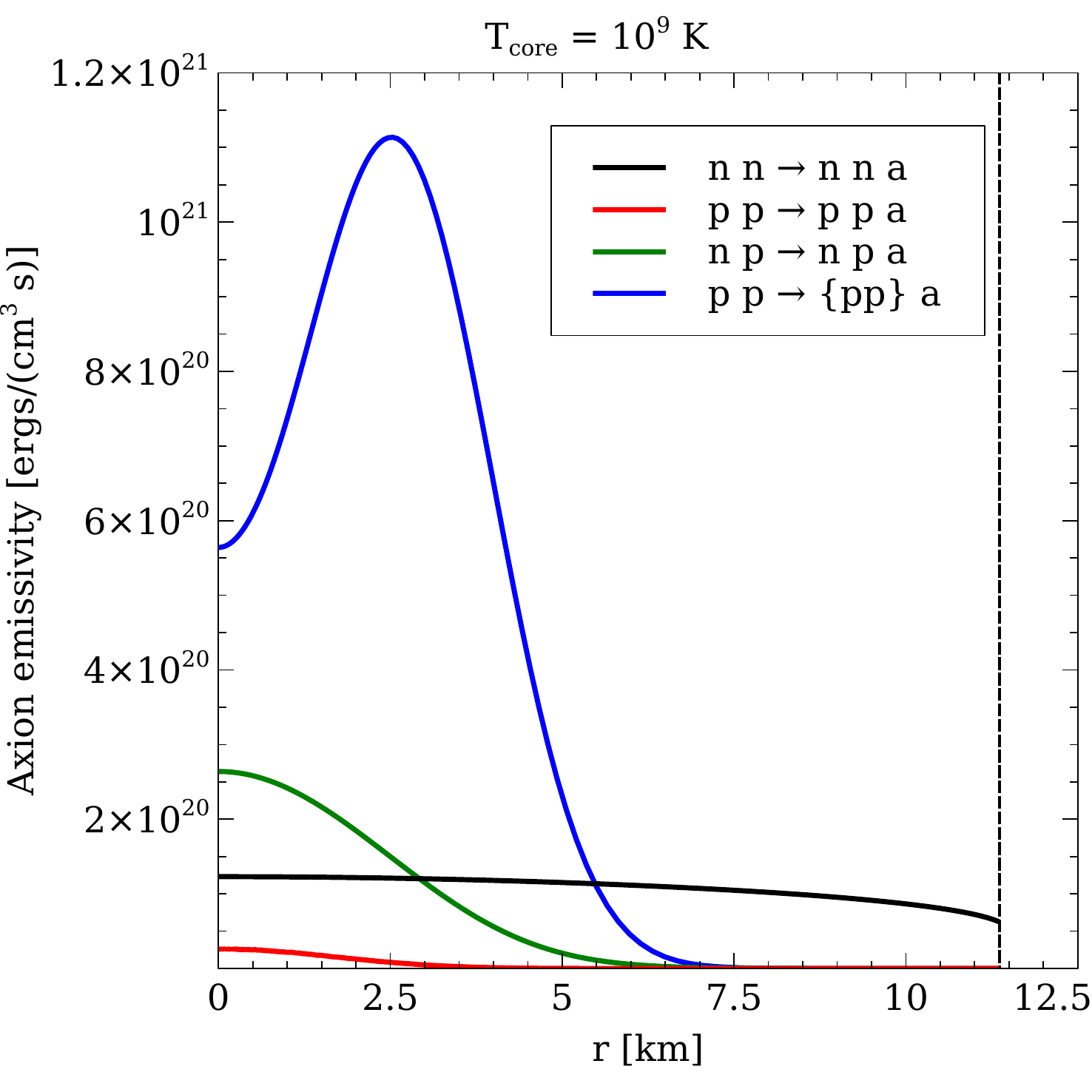}
  \includegraphics[width=0.48\textwidth]{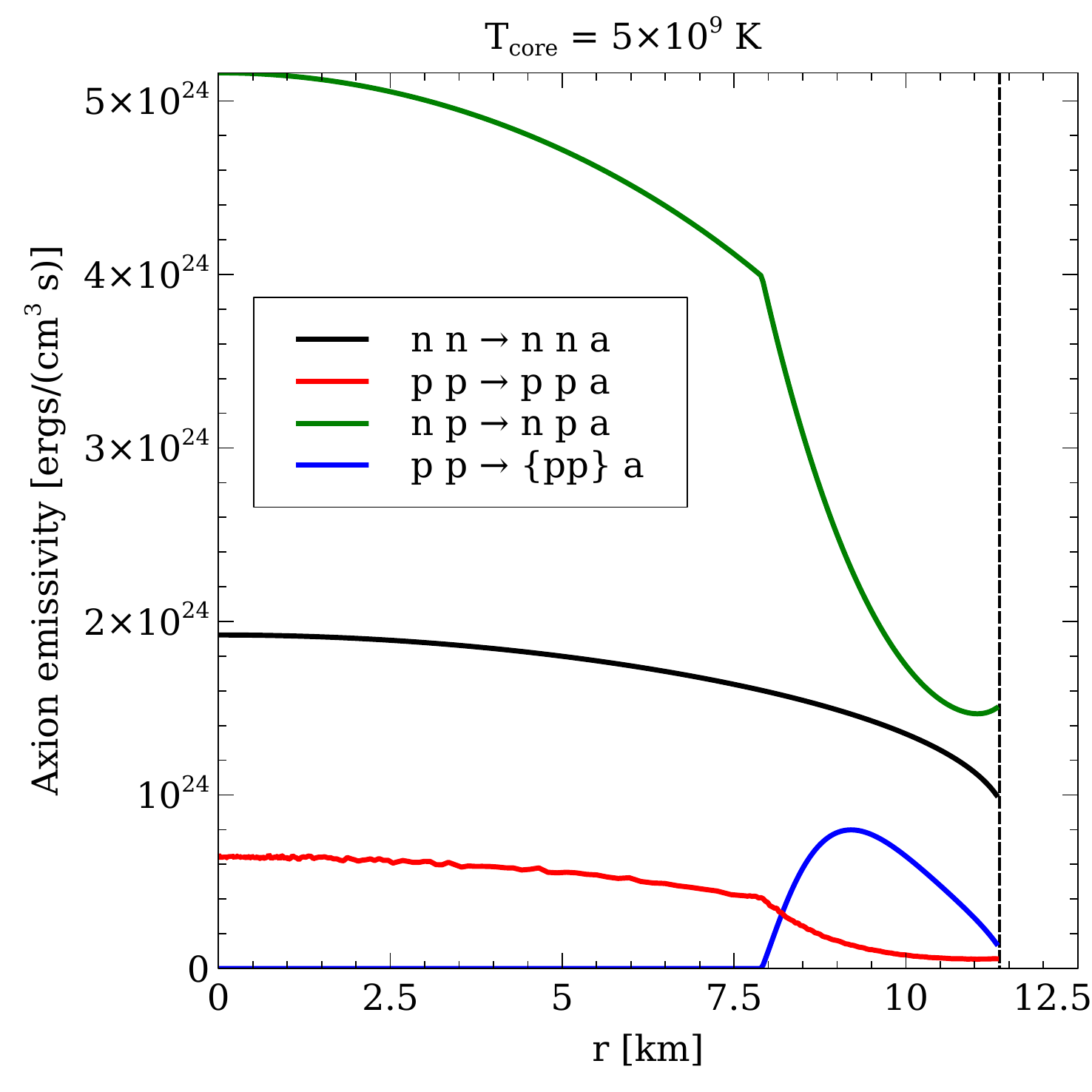} \\
  \caption{
Axion emissivity due to different processes occurring in the core of the magnetar, as a function of distance $r$ from the center of the star.  We assume the neutron star core has a uniform temperature, different for each of the four panels.  At the lowest temperature $T=10^8 \text{ K}$, the proton spectral gap is large and thus the neutron bremsstrahlung process dominates.  As temperature increases in the subsequent panels, the superfluid gap closes and the Boltzmann suppression (some function of $\exp{(-\Delta/T)}$) of the processes involving protons lessens dramatically.  The Cooper pair breaking process is dominant in sections of the neutron star where the temperature is just below the critical temperature.  The emissivity expressions used in these plots are Eqns.~\ref{eq:Q0nn}, \ref{eq:QSppwithR}, \ref{eq:QSnpwithR}, and \ref{eq:Q_CP}.  The axion-nucleon coupling constant has been chosen to be at the upper limit set by SN1987a $G_{an} = 7.9\times 10^{-10} \text{ GeV}^{-1}$ \cite{Graham:2015ouw}.}
\label{fig:bremsstrahlung_profiles}
\end{figure}

In Fig.~\ref{fig:bremsstrahlung_profiles}, we plot the emissivity due to these bremsstrahlung and Cooper pair breaking processes in a $1.4M_{\odot}$ magnetar, where the proton $^1S_0$ superfluid gap is described by the CCDK model.  Each panel corresponds to a different value of the magnetar core temperature.  We assume the core has a uniform temperature, because the timescale for thermal equilibration of dense nuclear matter is short compared to the magnetar lifetime \cite{Pons:2019zyc}. 

If the core temperature is $10^8 \text{ K}$ (top left panel of Fig.~\ref{fig:bremsstrahlung_profiles}), the proton gap is 170 keV in the center of the magnetar, rising to a maximum of 1 MeV near the core-crust boundary.  Thus, axion production processes involving protons are exponentially suppressed by the size of the gap over the temperature.  In all parts of the magnetar core, the $n+n\rightarrow n+n+a$ process dominates.  The Cooper pair breaking process is also suppressed at this temperature, as the core temperature is well below the critical temperature in all parts of the magnetar core.  

At a temperature of $T=5\times 10^8 \text{ K}$ (top right panel), the proton gap is almost the same size throughout the magnetar core as when the core temperature was $T=10^8 \text{ K}$.  However, the Boltzmann suppression of processes involving protons is significantly less, as the temperature is a factor of five larger.  Even still, $\Delta/T$ is still large enough that the dominant bremsstrahlung process is again $n+n\rightarrow n+n+a$.  However, the rate of the Cooper pair formation process is much higher at this temperature, which is much closer to the critical temperature in the center of the magnetar.  Axion production from Cooper pair formation dominates in the center of the magnetar, and neutron bremsstrahlung dominates the axion emissivity from the mantle region where the critical temperature is higher.

If the magnetar core temperature is $10^9 \text{ K}$ (lower left panel), then the center of the magnetar is right below the critical temperature and the Cooper pair breaking mechanism of axion emission is dominant.  Since the gap in the proton spectrum is small (about 95 keV) in the center of the magnetar, and $\Delta/T$ is no longer large, the $n+p\rightarrow n+p+a$ process is no longer significantly Boltzmann suppressed, and provides a greater contribution to the axion production than $n+n\rightarrow n+n+a$ in the very center of the star.  However, in the mantle, the proton gap still reaches 1 MeV and the neutron bremsstrahlung process dictates the net axion emission from this region.

Finally, we show in the bottom right panel of Fig.~\ref{fig:bremsstrahlung_profiles} the axion emissivity in the case of a core temperature of $5\times 10^9\text{ K}$.  Such a high core temperature is likely not sustainable for the lifetime of the magnetars we examine here, due to the enhanced neutrino emission at these high temperatures.  Nevertheless, if the magnetar did have such a high core temperature, the protons would only be superfluid in the magnetar mantle, where, starting at about 8 km from the center of the magnetar, the gap rapidly increases from 0 to a maximum of almost 800 keV when the crust is reached.  In the center of the magnetar, the protons are not superfluid and thus $n+p\rightarrow n+p+a$ produces the most axions, as is expected in ungapped nuclear matter.  As the crust is approached, the superfluid gap in the proton spectrum appears and widens dramatically within a couple kilometers, but due to the high core temperature the rate of $n+p\rightarrow n+p+a$ is only modestly suppressed ($\Delta/T$ is not too large) and remains the largest contributor to axion production.

In this paper, we only consider axions that can convert to hard X-rays with energies from 20-150 keV.  Therefore, we can restrict our analysis to just the axions produced with energies below a couple hundred keV.  As the Cooper pair formation process only produces axions with energies greater than twice the gap and thus typically in the range of several hundreds of keV to 2 MeV, we will ignore this production mechanism in our analysis\footnote{In magnetars where the protons are superfluid in only part of the magnetar - for example, when the core temperature is $5\times10^9 \text{ K}$ - the gap can be arbitrarily small but nonzero in part of the star.  This region corresponds to a thin shell in the magnetar, and likely does not contribute much to the total axion emsission (see the right panel of Fig.~\ref{fig:criticaltemp}, for example).}.  We write down the spectrum of axions $\mathop{dQ}^S/\mathop{dx}$ produced by the three nucleon bremsstrahlung process, where proton superfluidity is treated in the CCDK model.  Since neutrons are not paired, the rate of $n+n\rightarrow n+n+a$ is unchanged from the ungapped case (\ref{eq:Q0nn}).  The process $p+p\rightarrow p+p+a$ is highly suppressed because all four nucleons in the process have gaps in their energy spectra.  Thus, in parts of the neutron star where the proton gap is finite, we ignore this process entirely.  We do consider it in parts of the star where $T>T_c(n_B)$, where the emissivity is the ungapped result $Q^0_{pp}$.  The last process, $n+p\rightarrow n+p+a$, is suppressed when the protons are paired, but is not as strongly suppressed as $p+p\rightarrow p+p+a$ because the neutrons remain unpaired.  Therefore, we use the superfluid expression for the differential emissivity from $n+p\rightarrow n+p+a$ 
\begin{equation}
    \frac{\mathop{dQ^S_{np}}}{\mathop{dx}} = \frac{2}{3\pi^7}C_{\pi}(m_n/m_{\pi})^4f^4G_{an}^2p_{Fp}G(c,d)T^6x^2I^S_{np}(x,n_B,T),
    \label{eq:dQSnpdx}
\end{equation}
where
\begin{equation}
    I^S_{np}(x,n_B,T)=\int_{-\infty}^{\infty}\mathop{dx_2}\mathop{dx_4}\frac{z_4+x-z_2}{(1+e^{z_2})(1+e^{-z_4})(e^{z_4+x-z_2}-1)},
\end{equation}
where $z_i$ is as defined in Eq.~\ref{eq:z}.  When the gap goes to zero and the protons are no longer superfluid, 
\begin{equation}
    I^S_{np}(x,n_B,T)\vert_{\Delta=0} = \frac{1}{6}\frac{x(x^2+4\pi^2)}{e^x-1},
\end{equation}
which reduces Eq.~\ref{eq:dQSnpdx} to the expression in nonsuperfluid matter Eq.~\ref{eq:axion_np_emissivity}.  Therefore, $I^S_{np}$ smoothly transitions from a superfluid expression to an ungapped expression as the temperature rises above the critical temperature for superfluidity. The total differential emissivity of axions from nuclear matter, properly accounting for proton superfluidity in regions of the magnetar where $T<T_c(n_B)$ is given by the sum 
\begin{equation}
    \frac{\mathop{dQ^S_{\text{total}}}}{\mathop{dx}} = \frac{\mathop{dQ^0_{nn}}}{\mathop{dx}} + \frac{\mathop{dQ^S_{pp}}}{\mathop{dx}}+\frac{\mathop{dQ^S_{np}}}{\mathop{dx}}+\frac{\mathop{dQ^S_{CP}}}{\mathop{dx}},
    \label{eq:total_superfluid_emissivity}
\end{equation} 
using the expressions from Eqs.~\ref{eq:axion_nn_emissivity}, \ref{eq:Q_CP}, and \ref{eq:dQSnpdx}.

The differential luminosity $dL^S_{\text{total}}/dx$ is obtained by integrating $dQ^S_{\text{total}}/dx$ over the magnetar core, as in Eq. \ref{eq:dL0}.  As discussed above, in our analysis we will neglect the Cooper pair breaking process because it produces axions with energies too large to produce hard X-rays, and we will only consider $p+p\rightarrow p+p+a$ in parts of the magnetar where $T>T_c(n_B)$, as the process is highly suppressed otherwise.
%%%%%%%%%%%%%%%%%%%%%%%%%%%%%%%%%%%%%%%%%%%%%

\subsection{ALP-Photon Conversion} \label{conv1}

The ALPs produced in the core escape to the magnetosphere, where they are subsequently converted to photons. ALP conversion to hard X-ray in the magnetosphere was treated in full detail in previous papers by a subset of the authors \cite{Fortin:2018ehg,Fortin:2018aom}.  For completeness, in this Section and in Appendix \ref{appconv}, we review and collect the most important results for a self-contained analysis.  

We first assume a dipolar magnetic field defined by 
\begin{equation}
B=B_0\left(\frac{r_0}{r}\right)^3,
\label{dipolarB}
\end{equation}
where $B_0$ is magnetic field at the surface of the magnetar and $r_0$ is the magnetar radius.  As shown below, since the conversion radius is a few thousand radii $r_0$ away from the star, the dipolar approximation is justified\footnote{The recent NICER observation of the pulsar J0030+0451 hints that the magnetic field outside of that pulsar may be non-dipolar, perhaps suggesting the need to reexamine standard assumptions about pulsar magnetic fields \cite{Bilous:2019knh}.}.

The evolution equations for ALPs and photons propagating radially outwards, in terms of the dimensionless distance from the magnetar defined by $x=r/r_0$,\footnote{In this Section and in Appendix \ref{appconv}, $x=r/r_0$ is the dimensionless distance from the magnetar surface, which must not be confused with $x=\omega/T$ used in other sections.} were derived in \cite{Raffelt:1987im} and are given by
\begin{equation}
i\frac{d}{dx}\left(\begin{array}{c}a\\E_\parallel\\E_\perp\end{array}\right)=\left(\begin{array}{ccc}\omega r_0+\Delta_ar_0&\Delta_Mr_0&0\\\Delta_Mr_0&\omega r_0+\Delta_\parallel r_0&0\\0&0&\omega r_0+\Delta_\perp r_0\end{array}\right)\left(\begin{array}{c}a\\E_\parallel\\E_\perp\end{array}\right),
\label{EqnDiffMat}
\end{equation}
where
\begin{equation}
\Delta_a=-\frac{m_a^2}{2\omega},\qquad\Delta_\parallel=(n_\parallel - 1) \omega,\qquad\Delta_\perp=(n_\perp - 1) \omega,\qquad\Delta_M=\frac{1}{2} g_{a \gamma \gamma} B\sin\theta.
\label{propequa2}
\end{equation}
In Eq.~\ref{EqnDiffMat} and \ref{propequa2}, the ALP field is represented by $a(x)$ while the parallel and perpendicular electric fields are denoted by $E_\parallel(x)$ and $E_\perp(x)$, respectively.  Moreover, $\omega$ is the energy of the particles, $m_a$ is the ALP mass, $g_{a\gamma\gamma}$ is the ALP-photon coupling, and $\theta$ is the angle between the direction of propagation and the magnetic field.

Finally, the photon refractive indices $n_\parallel$ and $n_\perp$, given explicitly by
\begin{equation}
n_\parallel=1+\frac{1}{2}q_\parallel\sin^2\theta,\qquad n_\perp=1+\frac{1}{2}q_\perp\sin^2\theta,
\end{equation}
originate from the photon polarization tensor.  They can be computed at one-loop level, and in the limit of interest here with $\omega\lesssim2m_e$ and $B\ll B_c$, the refraction indices are
\begin{equation}
q_\parallel=\frac{7\alpha}{45\pi}b^2,\qquad\qquad q_\perp=\frac{4\alpha}{45\pi}b^2,
\label{EqnEH}
\end{equation}
where $b=B/B_c$ is the ratio of the magnetic field to the quantum critical magnetic field $B_c=m_e^2/e=4.413\times10^{13}\,\text{G}$.  Here $e=\sqrt{4\pi\alpha}$ is the charge given in terms of the fine structure constant $\alpha\sim1/137$. We note that these results for the photon refractive indices need to be modified in regimes of energies $\sim 500$ keV - 1 MeV, where the Euler-Heisenberg approximation breaks down. 

We now discuss some order of magnitude estimates to understand the general behavior of the ALP-photon system. From the mixing matrix in  Eq.~\ref{EqnDiffMat}, it is clear that a larger value of $\Delta_M \propto g_{a\gamma\gamma}B$ favors stronger mixing of the ALP-photon system, while larger values of the diagonal ALP mass $\Delta_a \propto m^2_a$ or photon mass term $\Delta_{\parallel} \propto n_\parallel$ suppress the mixing. The conversion clearly becomes negligible for large ALP masses (for typical values of other parameters, this turns out to be $m_a \sim \mathcal{O}(10^{-4})$ eV, as will be seen in our results). At the surface, using a typical value  of $B_0 \sim 10^{14}$ G = $1.95\times 10^{-6}$ GeV$^{2}$, $\omega \sim 100$ keV, $g_{a\gamma\gamma} \sim 10^{-10}$ GeV$^{-1}$, $m_a \sim 10^{-5}$ eV {and $\theta=\pi/2$}, one obtains $\Delta_M =9.8 \times 10^{-8}\text{eV}$, $\Delta_a = -5.0 \times 10^{-16}\text{eV}$, and $\Delta_{\parallel} = 50.1\text{eV}$. Clearly, at the surface, the photon mass term dominates, leading to a suppressed conversion probability. Near the radius of conversion $r_{a\rightarrow \gamma} \sim \mathcal{O}(1000 r_0)$ where $\Delta_M \sim \Delta_{\parallel}$, the conversion probability becomes appreciable. Beyond the radius of conversion, $\Delta_a$ dominates over $\Delta_M$ and $\Delta_{\parallel}$ and the conversion probability becomes negligible again. 
The behavior of the conversion probability $P_{a\rightarrow\gamma}$, the mixing angle $\varphi_\text{mix}$, and the conversion radius $r_{a\rightarrow\gamma}$ are discussed in Appendix \ref{appconv}. 

\begin{figure}[t]
    \centering
    \includegraphics[width=0.6\textwidth]{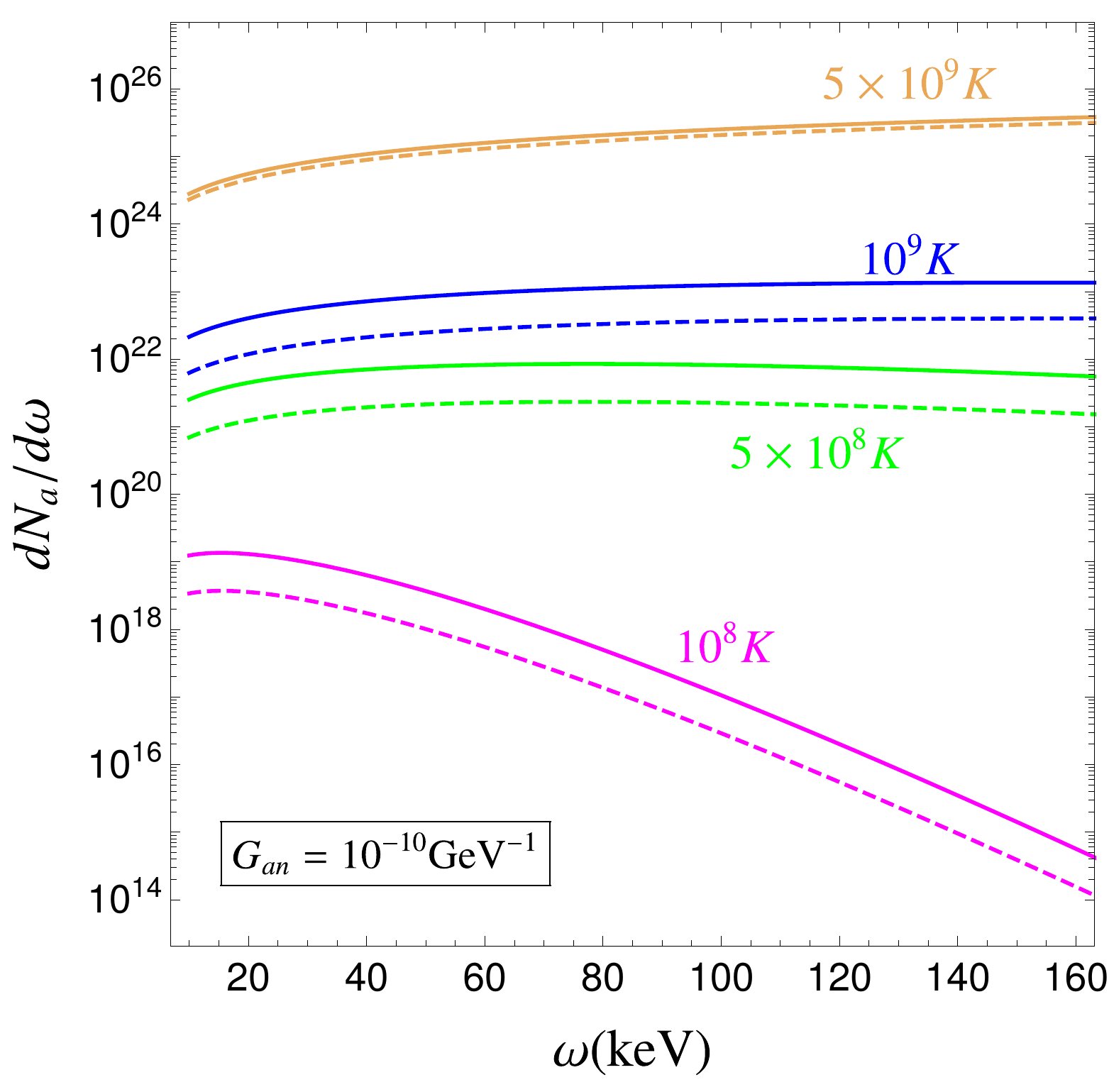}
    \caption{The dimensionless (in natural units) ALP emission spectrum $d N_a/d\omega$ defined in Eq.~\ref{eq:dNadw} as a function of energy at fixed $G_{an}=10^{-10} \text{GeV}^{-1}$. Four core temperatures are shown and correspond to the four different colors. The solid lines are
    obtained without accounting for superfluidity while the dashed ones include superfluidity in parts of the magnetar where the temperature is below the critical temperature.  The results in units of  $s^{-1}\text{keV}^{-1}$ can be obtained by multiplying by a factor of 
    $1.5 \times 10^{18}$.
    }
    \label{fig:dNadw}
\end{figure}

\section{Results} \label{magnresults}

In this Section, we collect all our previous derivations to obtain the theoretically predicted curve on the $\nu F_{\nu}$ plane for magnetars. We then compare the theory curve against observational data.

\subsection{Master Equation}
The differential luminosity $\mathop{dL_a}/\mathop{dx}$ of axions from the magnetar core is found by integrating the differential emissivity Eq.~\ref{eq:total_normal_emissivity} as indicated in Eq.~\ref{eq:dL0} ignoring superfluidity, and similarly integrating Eq.~\ref{eq:total_superfluid_emissivity} when including the effects of proton superfluidity.  One can also find the axion emission spectrum, i.e., number per time per energy, 
\begin{equation}
\frac{d N_a}{d \omega} =  \frac{1}{\omega}\frac{d L_a}{d \omega} = \frac{1}{T^2} \frac{1}{x} \frac{d L_a}{d x} .
\label{eq:dNadw}
\end{equation}
In natural units, this is dimensionless and is shown in Fig.~\ref{eq:dNadw}.

The total energy of photon emission from axion conversion in the magnetic field of the magnetar is obtained by multiplying Eq.~\ref{eq:dNadw} by the axion to photon conversion probability~\cite{Fortin:2018ehg}:
\begin{eqnarray} \label{master1}
L_{a \rightarrow \gamma} = \int_0^{\infty} d \omega
\frac{1}{2 \pi} \int_0^{2 \pi} d \theta   \cdot \omega \cdot \frac{d N_a}{d \omega} \cdot P_{a \rightarrow \gamma} (\omega,\theta) .
\end{eqnarray}
The photon conversion probability $P_{a \rightarrow \gamma}$ is worked out in Appendix \ref{appconv}.

The photon spectrum per area observed at Earth is obtained by dividing the above photon spectrum $d L_{a \rightarrow \gamma}/d \omega$ by  $4 \pi D^2$, with $D$ being the
distance between the magnetar and Earth. Our theoretical spectrum will be presented on the $\nu F_{\nu}$ plane, which is related to the above definition by
the following relation:
\begin{equation}\label{nuFnu}
\boxed{
\nu F_{\nu}(\omega) = \omega^2 \frac{1}{4 \pi D^2}
\frac{1}{\omega}\frac{d L_{a \rightarrow \gamma}}{d \omega} .}
\end{equation}
The experimental spectral data is usually expressed by various collaborations on the same plane; thus, this representation is suited for a comparison between theory and observation.
The $\nu F_{\nu}$ spectral representation in  \eqref{nuFnu} will be referred to as ``the spectrum" in the following sections, where we will perform our  analysis.

%%%%%%%%%%%%%%%%%%%%%%%%%%%%%%%%%%%%%%%%%%%%%%%%%%%%%%%%%%%%%%%%%%%%%%%%%%%%%%%%%%%%%%%%%%%%%%%%%%%%%%%%%%%%%%%%%%%%%
\begin{table*}[t!]
\centering
%\resizebox{\textwidth}{!}{
\begin{tabular}{c|c|c}
\hline
\hline
Name                  & $B$ ($10^{14} G$)  & $D$ (kpc) \\
SGR 1806-20           & {7.7}            & 8.7   \\
1E 1547.0-5408        & {6.4}         & 4.5   \\
4U 0142+61            & 1.3             & 3.6   \\ 
SGR 0501+4516         & 1.9             & {3.3}     \\
1RXS J170849.0-400910 & 4.7            & 3.8   \\
1E 1841-045           & 7             & 8.5  \\
SGR 1900+14           & 7              & 12.5 \\
1E 2259+586           & 0.59           & 3.2  \\
\hline
\hline
\end{tabular}
%}
\caption{
\label{tab:catalog} 
The list of magnetars used in this work. For details, we refer to the text. }
\end{table*}

\subsection{Comparison with Observation: Preliminaries}

The magnetars we will study are shown in Tab.~\ref{tab:catalog}. The spectral data is collected from several studies and is based on observations by \suzaku, \integral, \xmm,  \nustar, and \rxte. For each magnetar, we first discuss the observational data presented on the $\nu F_\nu$ plane. Since the hard X-ray spectrum of a given magnetar has sometimes been studied by different satellites, or different observation runs of the same satellite, and analyzed by various authors, our strategy is to take representative (and often the latest) data available in the literature. The data available is sometimes in the form of upper limits when no emission is observed; for some magnetars, we analyze both actually observed emission as well as available upper limits.

To compare the observational data with the spectrum derived from ALP-photon conversion from Eq.~\ref{nuFnu}, we adopt the following steps. Firstly, for a fixed $m_a$, the theoretical photon spectrum is calculated as a function of the photon energy for a given magnetar. This spectrum is 
proportional to $(G_{an}\times g_{a\gamma\gamma})^2$. 

When actual experimental data is available for an observed emission, we  construct a $\chi^2$:
\begin{eqnarray}
\chi^2(N) = \sum_i^N \frac{(\nu F_{\nu, \text{obs}} - \nu F_{\nu, \text{ALP}})^2}{\sigma_i^2} ,
\end{eqnarray}
where $i=1,2,\cdots, N$ labels the index of the energy bins which contain experimental data points, with $\nu F_{\nu, \text{ALP}}$ being the theoretically predicted spectral value;  $\nu F_{\nu, \text{obs}}$ being the average measured luminosity within a single bin; $\sigma_i$ being the experimental $1\sigma$ error bar; and the degrees of freedom of the $\chi^2$ being $N$. This follows from the assumption that
the noises in different bins, when the measured $\nu F_{\nu}$ spectrum is subtracted by the putative emission from axion conversion, are statistically independent. 
Using this $\chi^2$ directly to derive the $95\%$ confidence level (CL) upper limits on the product of couplings
$G_{an} \times g_{a\gamma\gamma}$ is problematic, since some theory curves do not fit the experimental data well, with
a value of $\chi^2$ larger than what would be excluded for the hypothesis that the
spectrum consists only of contribution from axion photon conversions. This is of course not unexpected; there is certainly a substantial contribution to the emission from astrophysical processes completely unrelated to the putative ALP-photon conversion. A leading hypothesis of the hard X-ray emission is that coming from resonant inverse Compton scattering of thermal photons by ultra-relativistic charges \cite{Beloborodov:2012ug, Baring:2017wvb, Wadiasingh:2017rcq}, which depends on many parameters and is model-dependent. Since our goal is to derive upper limits
on the product of couplings $G_{an} \times g_{a\gamma\gamma}$, rather than advancing ALP-photon conversion as a precise fit to the experimental data, we choose to
firstly minimize the $\chi^2$ (with value $\chi^2_{\text{min}}$) by varying the coupling $G_{an} \times g_{a\gamma\gamma}$. We then
determine the $95\%$ CL upper limits on $\chi^2_{\text{min}}$ using  $\Delta \chi^2 = \chi^2 -\chi^2_{\text{min}}$. For each mass $m_a$, two values of $G_{an} \times g_{a\gamma\gamma}$ are obtained on either side of the $\chi_{\text{min}}$
and the values above the larger of these two are excluded as they yield stronger
signals above the experimental data set.
The ALP mass $m_a$ is then varied and the  analysis is repeated, to finally yield a contour in the plane of $G_{an} \times g_{a\gamma\gamma}$ vs. $m_a$.

For the case where only upper limits for the relevant energy bins are available, we compare in each energy bin the upper limit with 
the averaged luminosity within that bin and find, among all the energy bins, the largest value of $G_{an} \times g_{a\gamma\gamma}$  that yields a spectrum
not exceeding any of the upper limits. For magnetars that  only have observational data points and no observational upper limits, 
the $95\%$ CL upper limits on $G_{an} \times g_{a\gamma\gamma}$ obtained from the $\Delta \chi^2$ analysis is the final result that we report. For  magnetars that 
have both observational data points and observational upper limits, we keep the more stringent one for each $m_a$. This process is iterated over a list of axion masses
and the couplings corresponding to the $95\%$ CL exclusion are found for these masses. We note that
in principle the experimental upper limits can be
combined with the $\chi^2$ constructed from the measured
fluxes, should a knowledge of the likelihood used in
deriving these upper limits be available. In addition,
the joint likelihood constructed this way for each
magnetar can be combined for all magnetars. 
These should give slightly improved results but would require a much more involved 
analysis of the raw experimental data points and is beyond the scope of
this paper.

In the following sections, we discuss in detail the observational status of
each magnetar used in our study, and present the corresponding upper limits on $G_{an} \times g_{a\gamma\gamma}$ based on the procedure explained above.

\subsection{\texorpdfstring{SGR 1806$-$20}{SGR 1806-20}}

\begin{figure}
    \centering
    \includegraphics[width=0.48\textwidth]{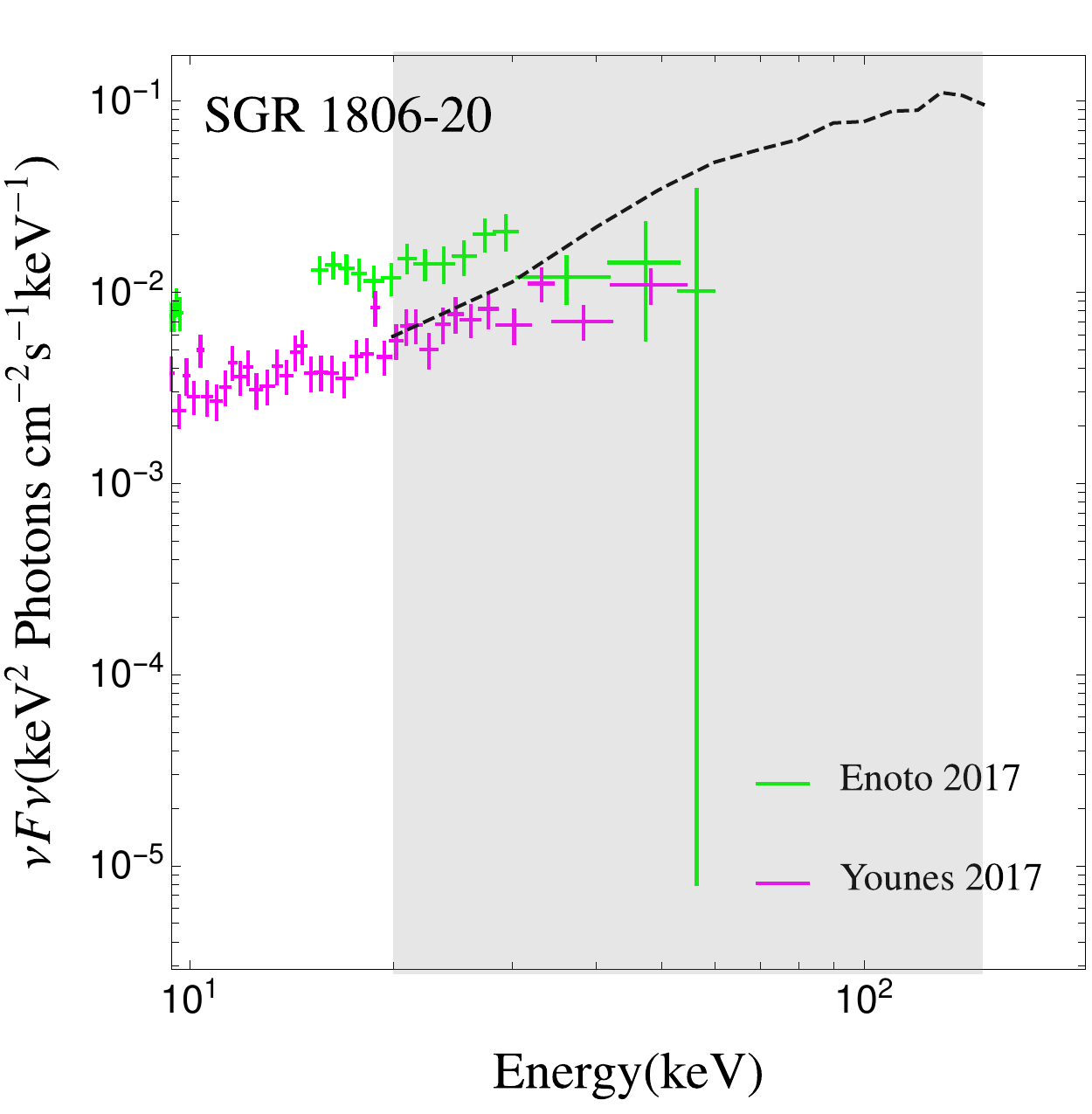}
    \quad
     \includegraphics[width=0.485\textwidth]{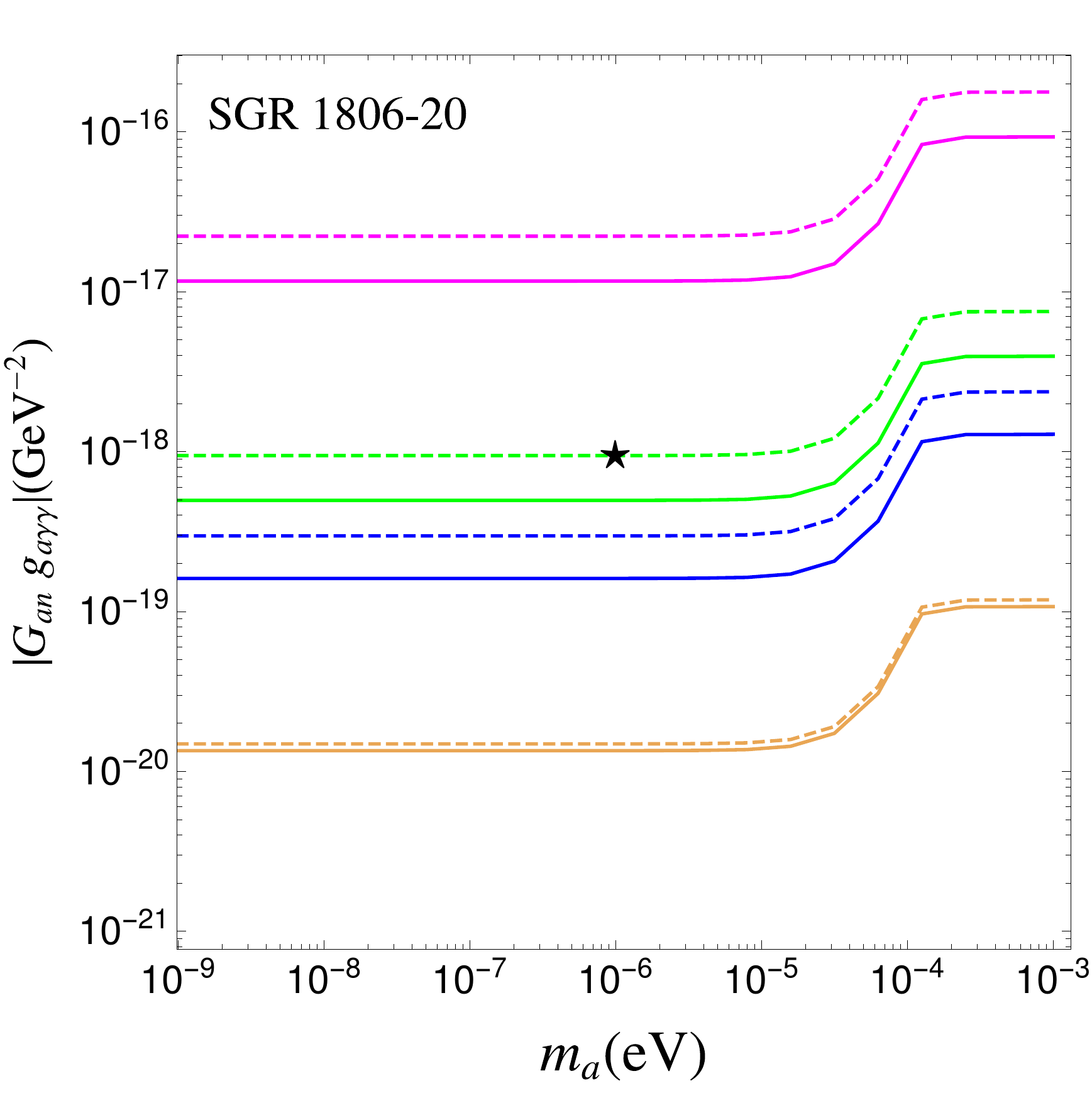}
      \caption{\label{fig:SGR1806-20}
\textbf{Left panel}: The theoretical spectrum (black dashed line) and observational data on the $\nu F_{\nu}$ plane for \sgr\ . The observational data is obtained from \nustar\  \cite{Younes:2017sme} ({in magenta for one Observation ID 30102038009, and four more observations are used in the analysis, see text}) and \suzaku\  \cite{Enoto:2017dox}  (green). \commoncaption{9.4\times 10^{-19}}
 }

\end{figure}

Our main reference for \sgr\ is the  analysis performed by~\cite{Younes:2017sme}. 
The authors analyzed five \nustar\ observations dating from April 2015 - April 2016, which was more than a decade after its major bursting episode. The data were processed using \texttt{nustardas} version v1.5.1 and analyzed using \texttt{nuproducts} task and HEASFT version 6.19. The spectral analysis was performed using XSPEC version 12.9.0k. The spectra were binned to have a minimum of five counts per bin and the Cash statistic in XSPEC was used for error calculation. All quoted errors are at $1\sigma$ level. The authors fit all the \nustar\ spectra with an absorbed power-law and blackbody model, obtaining a  power-law photon index $\Gamma=1.33\pm0.03$ and a blackbody temperature
$kT=0.62\pm0.06$~keV. The 3-79~keV flux was determined to be 
$(3.07\pm0.04)\times10^{-11}$~erg~s$^{-1}$~cm$^{-2}$. If the \nustar\ model is extended down to 0.5~keV, the  flux between 0.5-79~keV is obtained to be 
$(3.68\pm0.05)\times10^{-11}$~erg~s$^{-1}$~cm$^{-2}$ and the luminosity is obtained as 
$(3.33\pm0.06)\times10^{35}$~erg~s$^{-1}$, for a distance of
8.7~kpc.  For details of the data analysis and fitting, we refer to the original paper. The spectrum corresponding to Observation ID 30102038009 from~\cite{Younes:2017sme} is depicted in the left panel of Fig.~\ref{fig:SGR1806-20} in magenta where the bars denote
$1\sigma$ uncertainties. In the analysis, we use data from all 5 Observation IDs, including also 30102038002, 30102038004, 30102038006 and 30102038007.

The authors of \cite{Enoto:2017dox} studied sixteen \suzaku\ observations of nine persistently bright sources, including three observations of \sgr\ (September 2006, as well as March and October 2007).  Of these three observations, we utilized the one from September 2006 (Observation ID 401092010)  for our work. The authors reprocessed data from the  X-ray Imaging Spectrometer (XIS) in the range 0.2-12 keV and the Hard X-ray Detector (HXD) in the range of energies 10-600 keV using HEADAS-6.14. In the soft range, the full window mode of XIS was analyzed, while for the hard X-rays, only the HXD-PIN data was utilized. For details of the subtraction of non X-ray and cosmic X-ray backgrounds from the HXD-PIN spectrum, we refer to the original paper. The final data after background subtraction is shown in Fig.~\ref{fig:SGR1806-20} in green, with $1\sigma$ statistical and systematic uncertainties. The systematic uncertainty in the HXD-PIN data includes the contributions from the non X-ray (1\% systematic uncertainty) and cosmic X-ray (10\% systematic uncertainty) backgrounds. The HXD-PIN spectrum was binned to have $> 5\sigma$ significance or $>30$ counts in each bin.  The phase-averaged spectrum was fit using a Comptonization-like power-law model, obtaining a photon index $\Gamma = 1.62$ and flux $F_{15-60}$\,$=$\,$33.7$\,$\times$\,$10^{-12}$ \flux (Table 5 of \cite{Enoto:2017dox}).

Hard X-ray emission from \sgr\ has also been studied by several other groups. Relevant results for the photon index include $\Gamma = 1.5 \pm 0.3$ obtained by analyzing 2004 \integral-IBIS observations  \cite{Mereghetti:2005jq};  $\Gamma = 2.0 \pm 0.2$
from 2006 \suzaku\ HXD-PIN data in the range 10-40 keV \cite{Esposito:2007wx}; and $\Gamma = 1.7 \pm 0.1$ obtained with 2007 \suzaku\ data \cite{Enoto:2010ix}.

The results depicted in Fig.~\ref{fig:SGR1806-20} should thus be regarded as representative of results emanating from X-ray studies of \sgr\, and not necessarily the only exclusive picture. Clearly, adopting a different dataset or combination of datasets would change the resulting constraints on the ALP coupling and mass, although the change would be relatively small, given the general concurrence of the data across the references listed above. 

Finally, we make a few comments about the benchmark values of the magnetic field and the  distance for \sgr\ that are adopted in our work. Both  \cite{Younes:2017sme} and \cite{Enoto:2017dox}  agree on the distance, which is taken to be $d = 8.7$ kpc. However, based on \nustar\ data from 2015-2016,  \cite{Younes:2017sme} obtain the spin derivative $\dot \nu = (-1.25 \pm 0.03) \times 10^{-12}$ Hz s$^{-1}$, which agrees with the historical minimum $\dot \nu = (-1.22 \pm 0.17) \times 10^{-12}$ Hz s$^{-1}$ derived from observations in 1996. Since \cite{Younes:2017sme} perform their analysis based on data a long time after major burst activities, the level of torque derived by them can be considered to be the quiescent state magnetic configuration and corresponds to a value of $B = 7.7 \times 10^{14}$ G. On the other hand, the magnetic field derived by \cite{Enoto:2017dox} based on 2006 \suzaku\ data is higher by a factor of $\sim 2.5$.

{
The $95\%$ CL upper limits on the product coupling $G_{an} \times g_{a\gamma\gamma}$ is shown on the right panel of Fig.~\ref{fig:SGR1806-20}, for 4 different core temperatures
$10^8$ K, $5\times 10^8$ K, $10^9$ K and $5\times 10^9$ K, depicted with 
different colors. Because a higher core temperature leads to more
abundant axion production, the limits are stronger, corresponding to lower $G_{an} \times g_{a\gamma\gamma}$, as the core temperature increases.
We can see that with a core temperature of $10^8$ K, the upper limit on the product of  couplings is the weakest,
of the order of $\sim 10^{-17}\text{GeV}^{-2}$. For a more realistic
value of the core temperature $5\times 10^8$ K, the results are improved by approximately an order of magnitude. The solid and dashed curves are obtained without and with considering superfluidity. With superfluidity, the constraints become weaker. While the difference is substantial for lower core temperatures,  for a core temperature of $5 \times 10^9$, the reduction is very minor. This is consistent with the behavior shown in Fig.~\ref{fig:dNadw}.
We also show on the left panel a representative theory spectrum as a
black-dashed curve for illustration, corresponding to the black star on the green-dashed limit curve in the right panel which has $m_a = 10^{-6}$ eV and $G_{an} \times g_{a\gamma\gamma} = 9.4\times 10^{-19}\text{GeV}^{-2}$. 

}

\subsection{\texorpdfstring{1E$\,$1547.0$-$5408}{1E 1547.0-5408}}

\begin{figure}
    \centering
    \includegraphics[width=0.48\textwidth]{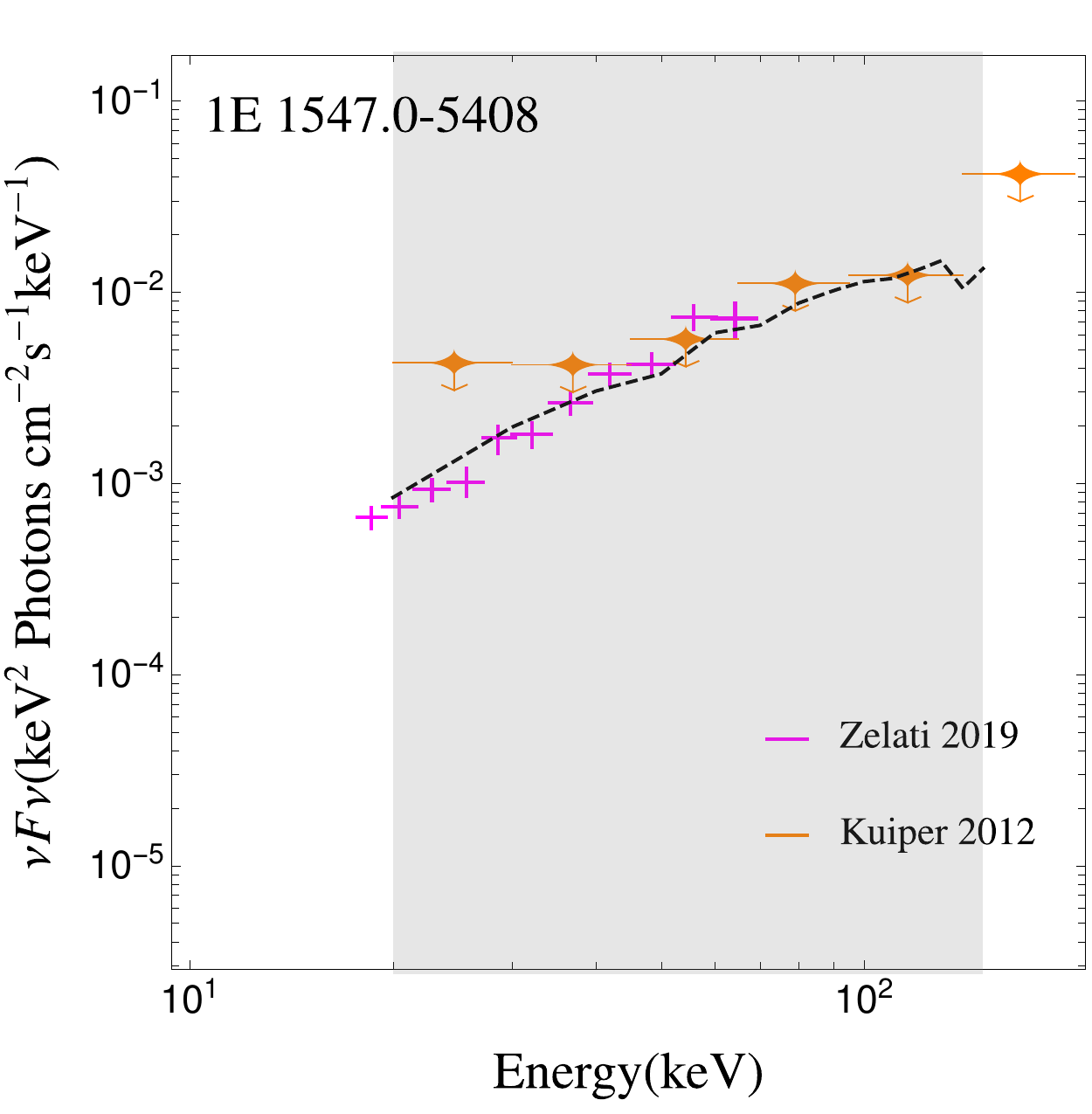}
    \quad
     \includegraphics[width=0.485\textwidth]{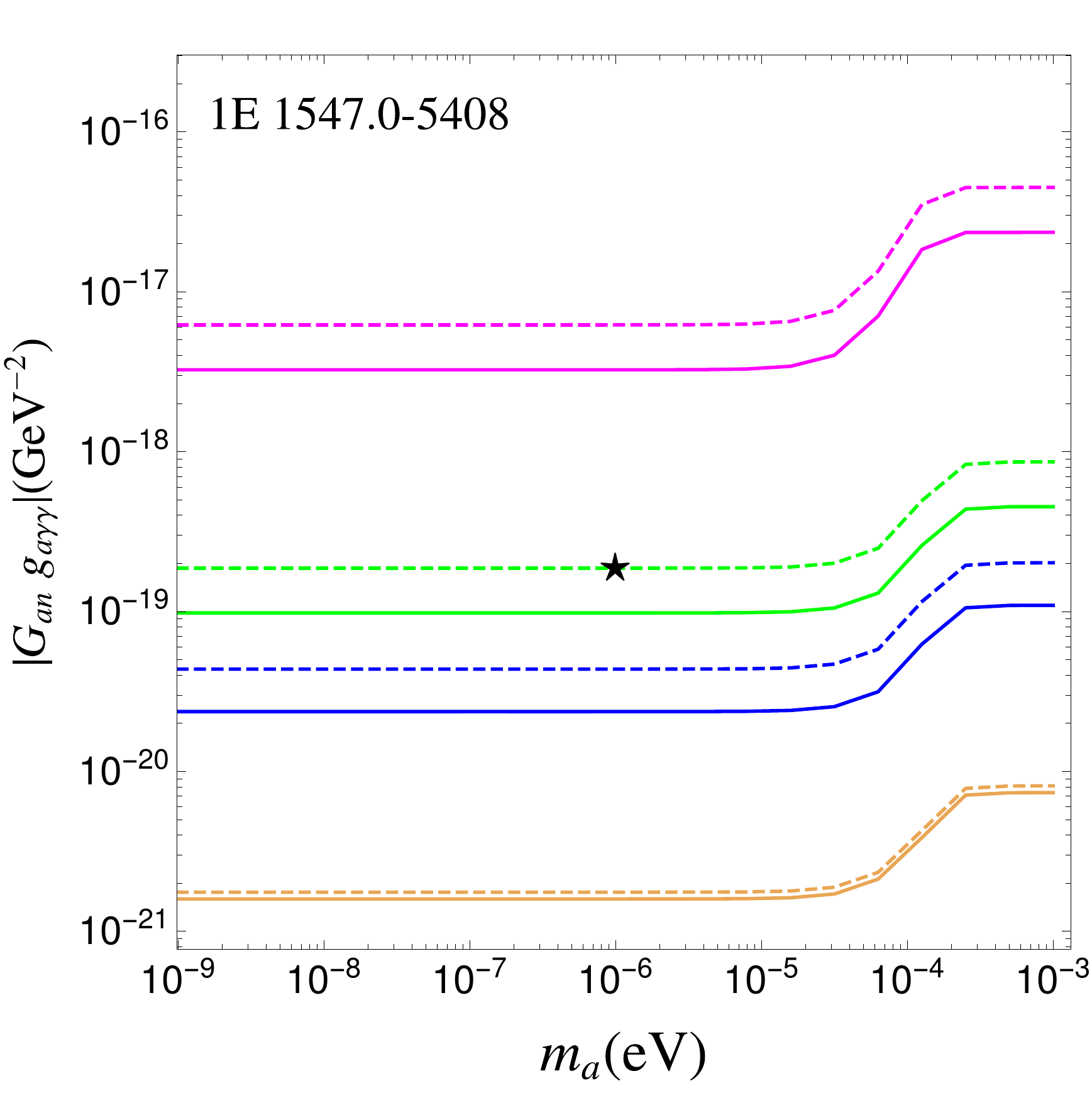}
        \caption{\label{fig:1E1547.0-5408}\textbf{Left panel}: The theoretical spectrum (black dashed line) and observational data on the $\nu F_{\nu}$ plane for \aa\ . The observational data is obtained from \nustar\  \cite{Zelati:2019dpc} (magenta) and \integral\  \cite{2008ATel.1774....1K, Kuiper_2012}  (orange). \commoncaption{1.9\times 10^{-19}}
    }
\end{figure}

Soft and hard X-ray emission from the  magnetar \aa\ has been measured by several satellites: $(i)$ an observed flux of $\sim$\,2$\times$10$^{-12}$\,\flux\ in the energy range 0.3--10\,keV by the \textit{Einstein} satellite and $(ii)$ flux of $\sim$\,(3-4)\,$\times$\,10$^{-13}$\,\flux\ in the range $0.3-10$ keV by \xmm\ and \chandra\ in 2004 and 2006. Following bursts in 2007-2009, the observed flux in the same soft X-ray range increased to $\sim$\,$8 \times 10^{-11}$\,\flux\, while emission was also detected in the hard X-ray band up to 200 keV, with the flux in the hard X-ray range being larger than that in the soft X-ray range by a factor of $\sim 5$.

Our main references for \aa\ will be the work performed by the authors of \cite{Zelati:2019dpc}, who analyzed  observations made by \swift\ since 2012 and \nustar\ in 2016 and 2019. \nustar\ data corresponding to Observation ID 30101035002 (April 23-24 2016) and Observation ID 30401008002 (February 15-17 2019) constitute our source for the hard X-ray regime. The authors processed the data using \texttt{nustardas} version 1.9.3 and extracted the background-subtracted spectra using \texttt{nuproducts}. Photons outside the 3-79 keV were vetoed, the datasets corresponding to the \nustar\ observations were merged, and the  final background-subtracted spectrum was  grouped to contain at least 20 photons per energy bin.  
The spectrum was fitted with several multi-component models: a double blackbody plus power-law model, a blackbody plus broken power-law model, and a resonant Compton scattering plus power-law model. Using the double blackbody plus power-law model,  the observed flux in  the 1--10\,keV range was obtained as $F_{1-10}$\,$=$\,$7.5^{+0.2}_{-1.0}$\,$\times$\,$10^{-12}$ \flux\  while for the  15--60\,keV energy range the observed flux was $F_{15-60}$\,$=$\,$4.6^{+0.1}_{-0.6}$\,$\times$\,$10^{-12}$ \flux. 

The value of the flux hardness ratio was $\eta$\,$=$\,$F_{15-60}/F_{1-10}$\,$\sim$\,0.6. The authors also performed a pulse phase-resolved spectral analysis of the two \nustar\ observations using combined datasets from  focal plane modules A and B. However, we do not use the phase-resolved spectrum in our work and refer to \cite{Zelati:2019dpc} for more details.

The hard X-ray emission has also been studied by several other groups over the last decade. These include \cite{Enoto:2017dox} who used a Comptonization-like power-law model to fit \suzaku\ and \nustar\ data and obtained $F_{15-60}$\,$=$\,$13.3$\,$\times$\,$10^{-12}$ \flux with $\Gamma = 1.15 \pm 0.12$ (Table 6 of \cite{Enoto:2017dox}). An earlier study \cite{Kuiper_2012} used a blackbody and power-law model to fit \integral\ and \swift\ data from 2009-2010 to obtain a range of fluxes and photon indices extending from $F_{20-150}$\,$=$\,$80$\,$\times$\,$10^{-12}$ \flux and $\Gamma = 0.87 \pm 0.07$ to $F_{20-150}$\,$=$\,$252$\,$\times$\,$10^{-12}$ \flux and $\Gamma = 1.41 \pm 0.06$ (see Tab.~8 of \cite{Kuiper_2012}). The temporal evolution of the hard X-ray flux and hardness after the outbursts in 2007-2009 has been studied in detail by  \cite{Zelati:2019dpc}. 

An even earlier study \cite{2008ATel.1774....1K} prior to the outbursts used \integral\ observations from 2003 - 2006 (performed between Revs. 46 and 411) to put $2\sigma$ upper limits on the hard X-ray component. The data is displayed in Figure 13 of \cite{Kuiper_2012} and corresponds to upper limits on the flux from \aa\ derived from a deep 4 Ms \integral\ mosaic  targeting PSR$\,$J1617$-$5055, which is located  within 5\fdg 5 from \aa. These upper limits are at least ten times lower than the flux levels reached during \integral\ observations from 2009-2010.

For the purposes of our work, we take both  \cite{Zelati:2019dpc}, as well as the $2\sigma$ upper limits from \cite{2008ATel.1774....1K, Kuiper_2012}, shown in the left panel of Fig.~\ref{fig:1E1547.0-5408}. The justification is that both these studies are faithful to the hard X-ray flux during quiescent periods: the first since it is several years after the bursts, and the second because it is prior to the bursts. We take a value of 4.5\,kpc as the distance, and assuming a spin-down rate $\dot{P}$\,$\sim$\,$4.77\times10^{-11}$\,s\,s$^{-1}$ the dipolar component of the magnetic field at the polar caps  is obtained to be $B_{\rm{p}}$\,$\sim$\,$6.4 \times 10^{19} (P\dot{P})^{1/2}$\,G $\sim$\,$6.4\times10^{14}$\,G \cite{Zelati:2019dpc}.

With these sets of experimental data, the $95\%$ CL upper limits on $G_{an} \times g_{a\gamma\gamma}$ are shown in the right panel of Fig.~\ref{fig:1E1547.0-5408}. Here since both experimental data points and upper limits are available, the better is chosen
for each $m_a$. For the lowest temperature of $10^8 K$, the limits based on the data points are slightly better than that based on the experimental upper limits for all $m_a$. 
This changes when increasing temperature and for the other three
temperatures, the experimental upper limits give better results.
The theory spectrum on the left panel, which corresponds to $5\times 10^8 K$ with superfluidity, just saturates the upper limit in the 
right-most energy bin in the gray band and is responsible for this
upper limit of $G_{an}\times g_{a\gamma\gamma} \sim 1.9\times 10^{-19}  \text{GeV}^{-2}$.

\subsection{\texorpdfstring{SGR$\,$0501+4516}{SGR 0501+4516}}

\begin{figure}
    \centering
    \includegraphics[width=0.48\textwidth]{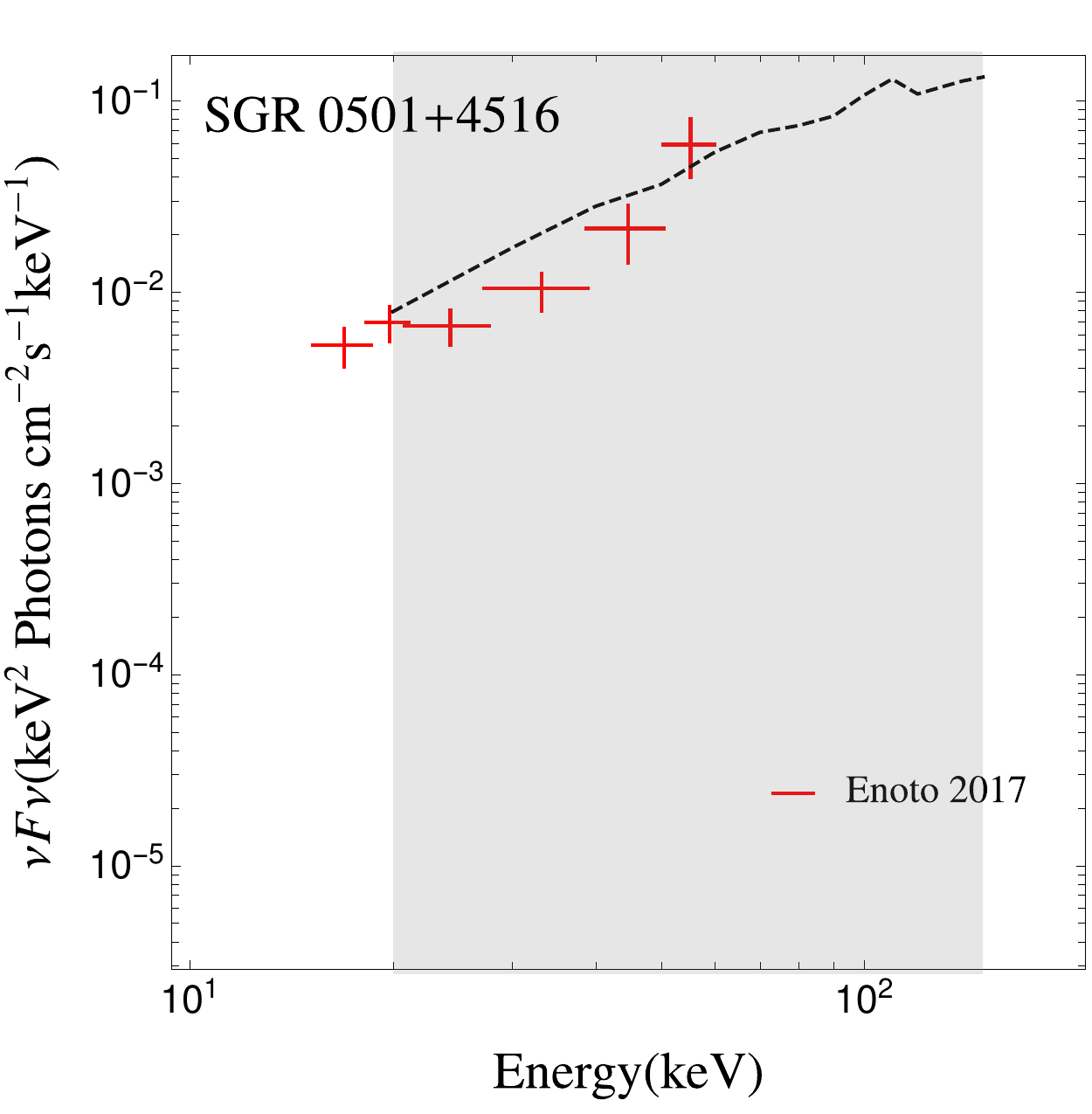}\quad
     \includegraphics[width=0.485\textwidth]{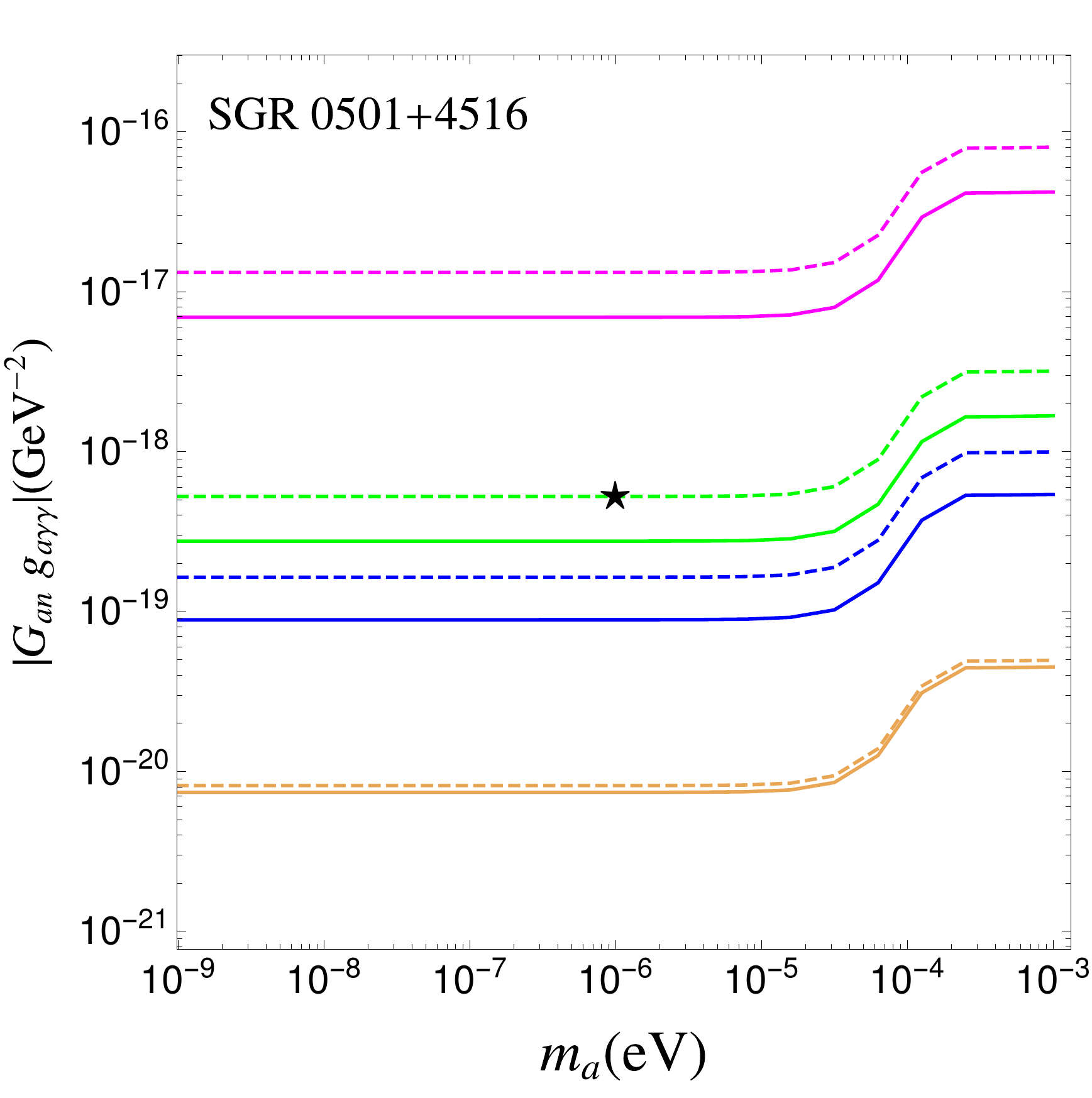}
      \caption{\label{fig:SGR0501+4516}\textbf{Left panel}: The theoretical spectrum (black dashed line) and observational data on the $\nu F_{\nu}$ plane for \sgrnew\ . The observational data is obtained from \suzaku\  \cite{Enoto:2017dox} (red). \commoncaption{5.2\times 10^{-19}}
    }
\end{figure}

\sgrnew\ was discovered with the Burst Alert Telescope on board \swift\ in 2008 and has been subsequently observed   with \chandra, \xmm, \rxte, \suzaku, and \nustar.   A wealth of information about the persistent X-ray emission \cite{Enoto:2017dox, 2010ApJ...722..899G} as well as its outburst behavior \cite{Lin_2011, Qu_2015} has been obtained. The soft X-ray spectrum of \sgrnew\ has been studied in quiescence  by several groups \cite{Camero_2014, Benli_2015, Mong:2018jyi}, and the burst-induced changes of its persistent soft X-ray emission has also been extensively investigated  \cite{2010ApJ...722..899G}. 

 The persistent hard X-ray spectrum following outbursts  has been investigated by several groups \cite{Enoto:2017dox, Guo:2014sia, 2011ASSP...21..323N}. Our main source for the analysis of \sgrnew\ will be \cite{Enoto:2017dox}, which analyzed the hard X-ray component using  \suzaku\ data from 2006 to 2013. The authors provided the temporal decay of the soft X-ray component as a function of time following the burst activity. Depending on the decay model, the unabsorbed flux in the 2-10 keV range can be attenuated by a factor of one order of magnitude (plateau model of decay) to two orders of magnitude (exponential model of decay) within 100 days after the burst, as shown in Fig.~18 of \cite{Enoto:2017dox}. An equivalent modeling of the attenuation of the hard X-ray component is not available. We will thus use the hard X-ray spectrum reported during the burst; taking into account the attenuation would strengthen our results.

The authors of \cite{Enoto:2017dox} studied ten target of opportunity observations of six transient objects, including \sgrnew\ . The data used by them for \sgrnew\ corresponds to \suzaku\ data from August 26, 2008 (Observation ID 903002010). The analysis of the HXD-PIN data from \suzaku\ and the spectral modeling has been described in our discussion of \sgr\ and a similar method was used for \sgrnew\ . We refer to  \cite{Enoto:2017dox}  for more details.

Assuming the location of \sgrnew\ in the Perseus arm of the galaxy, the distance is taken as 3.3 kpc in  \cite{Enoto:2017dox}, although distances of 2-5 kpc have also been used \cite{Mong:2018jyi}. The magnetic field is taken to be 1.9 $\times$ $10^{14}$ G \cite{Enoto:2017dox}.
The data and the $95\%$ CL upper limits on $G_{an} \times g_{a\gamma\gamma}$ are shown in the left and right panels of 
Fig.~\ref{fig:SGR0501+4516}, respectively.

\subsection{\texorpdfstring{4U$\,$0142+61}{4U 0142+61}}

\begin{figure}
    \centering
    \includegraphics[width=0.48\textwidth]{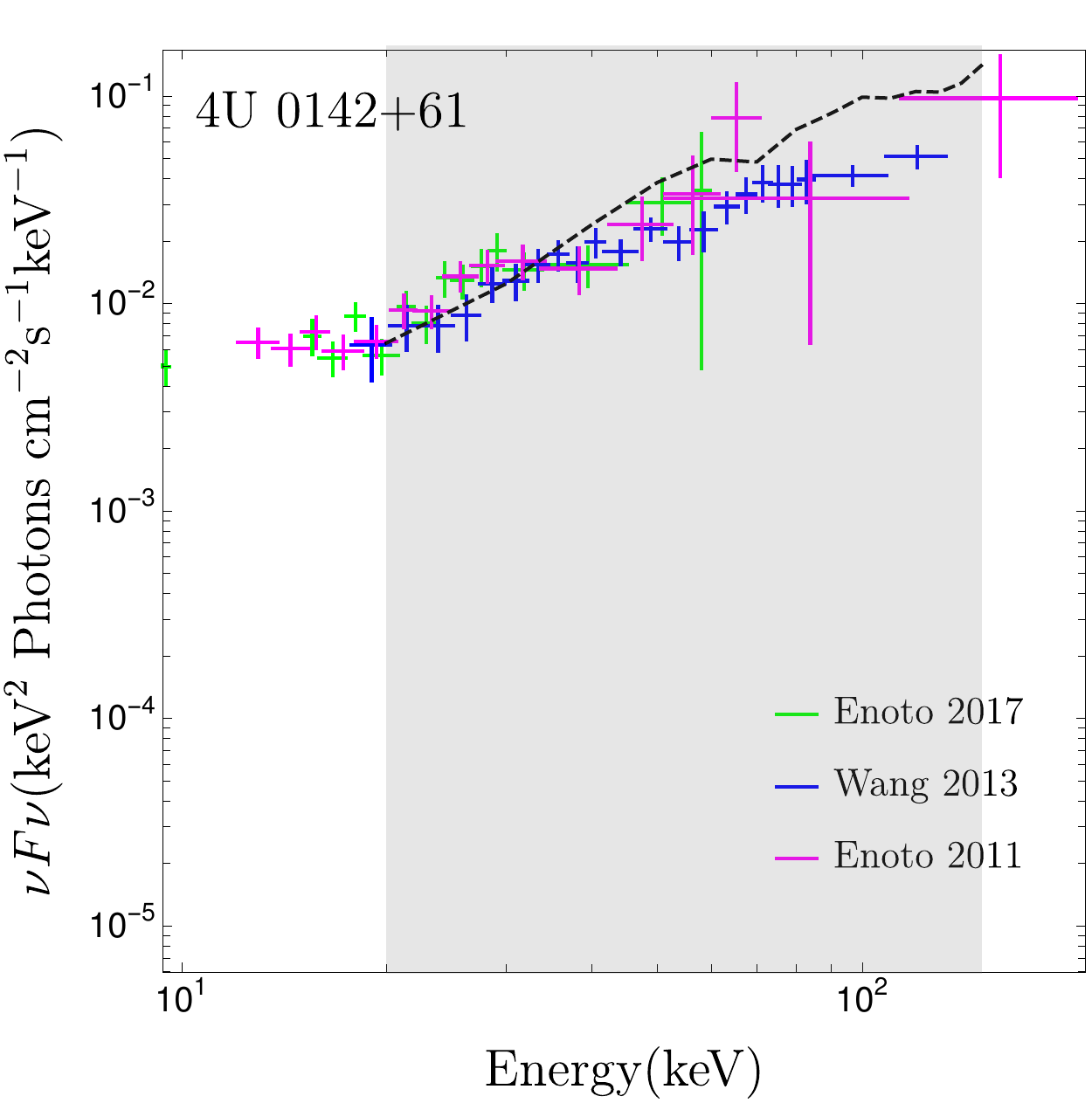} \quad
     \includegraphics[width=0.485\textwidth]{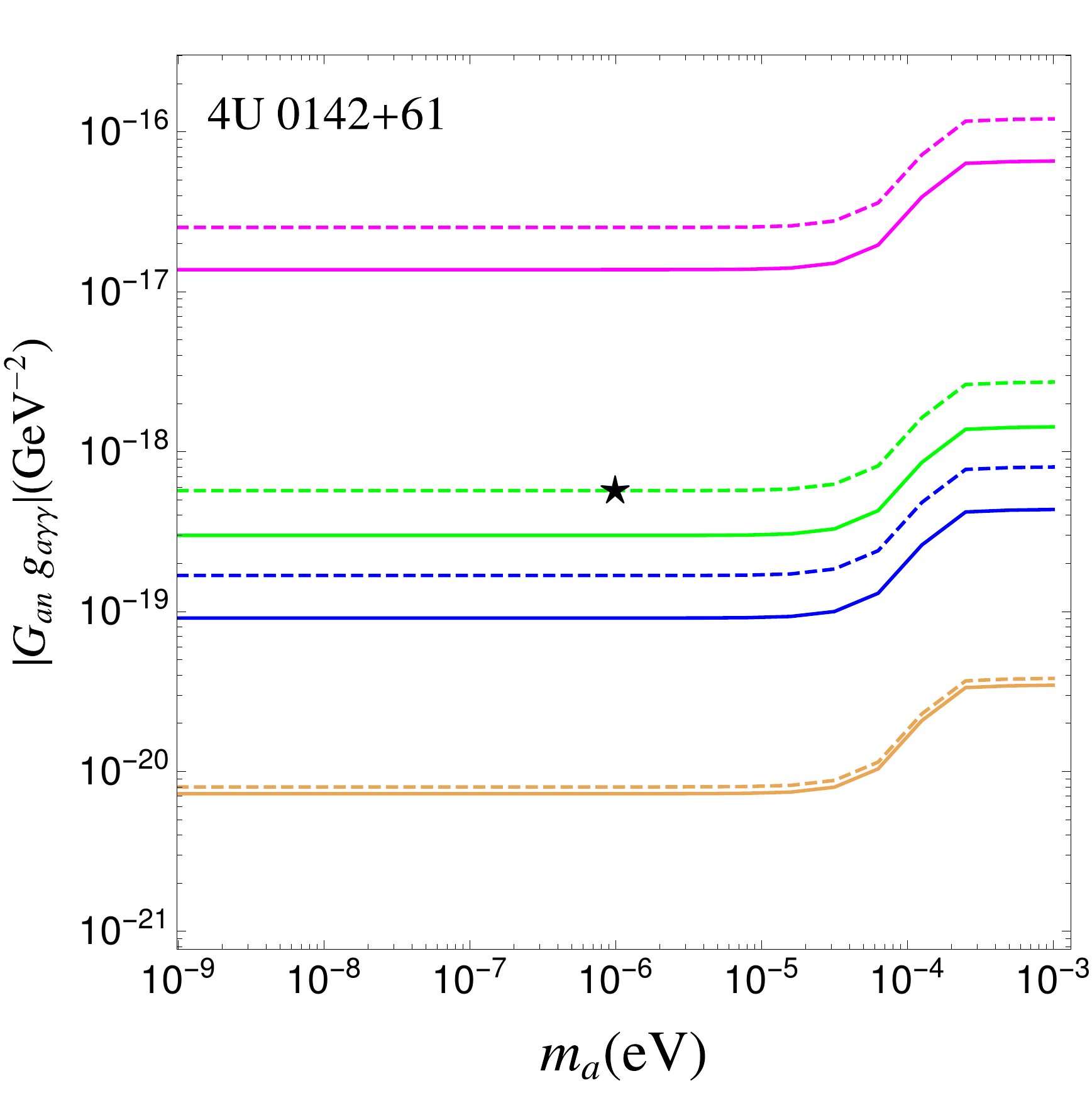}
    
    \caption{\label{fig:4U-0142+61}\textbf{Left panel}: The theoretical spectrum (black dashed line) and observational data on the $\nu F_{\nu}$ plane for \fu\ . The observational data is obtained from \integral\ (blue) \cite{Wang:2013pka}, and \suzaku\ \cite{Enoto:2011fg} (magenta) and ~\cite{Enoto:2017dox} (green). \commoncaption{5.7\times 10^{-19}} }
\end{figure}

The anomalous X-ray pulsar \fu\ was observed with \suzaku\ in 2007. There have been many studies of its hard X-ray spectrum: data from \swift\ and \rxte\ was used to model the spectrum above 20 keV with a photon index of $\sim 0.9-1$ \cite{Kuiper:2006is}; all contemporaneous data from \integral\ , \xmm\ , and \rxte\ was used by the authors of  \cite{Hartog:2008tq} to find a mean photon index of $\sim 0.9$ in the range 20-150 keV; and \suzaku\ data was used to fit the spectrum up to 70 keV with a power-law of $\Gamma \sim 0.9$ and a possible cutoff energy $\sim$ 150 keV \cite{Enoto:2011fg}. 

A detailed study of the hard X-ray spectrum using IBIS observations from \integral\ in the period 2003-2011 was performed by \cite{Wang:2013pka}. Archival data from the \integral\ Science Data Center was used where \fu\ was within $\sim 12$ degrees of the pointing direction of IBIS observations, with a total corrected on-source time of about 3.8 Ms. The data was analyzed using \integral\ off-line scientific analysis version 10, and the individual pointings were patched to create images in given energy ranges. The spectrum was then analyzed using XSPEC 12.6.0q and three models were used to fit the average spectrum: a power-law, bremsstrahlung, and a power-law with a cutoff to account for the fact that \fu\ was not detected by COMPTEL. It was argued that a cutoff power-law fit is the most appropriate one to model the hard X-ray spectrum and maintain compatibility with COMPTEL upper limits. This fit yielded a photon index $\Gamma \sim 0.51 \pm 0.11$, a cutoff energy of $128.6 \pm 17.2$, and a flux in the 18-200 keV range of $1.27 \pm 0.12 \times \,10^{-10}$ \flux\ (from Tab.~1 of \cite{Wang:2013pka}).

The authors of \cite{Enoto:2011fg} performed an analysis of the spectrum of \fu\ using \suzaku\ data from August 13-15, 2007. After screening, a net exposure of 99.7 ks with the XIS and 94.7 ks with the HXD were archived. The non X-ray and cosmic X-ray backgrounds for HXD-PIN spectrum were removed and the hard component was fitted by a power-law with $\Gamma = 0.89 \pm 0.11$, with the soft spectrum being modeled by a resonant cyclotron scattering model. The high-energy cutoff was obtained to be $> 180$ keV. The flux in the 10-70 keV range was obtained to be $4.4 \times \,10^{-11}$ \flux\ and the unabsorbed 1-10 keV flux was obtained to be $1.8  \times \,10^{-10}$ \flux\ . For an extension of the hard component to 200 keV, the flux in the 10-200 keV range is $1.5  \times \,10^{-10}$ \flux\  (see Fig.~7 of  \cite{Enoto:2011fg}).

An even more recent study of \fu\ by the authors of \cite{Enoto:2017dox} used data from \nustar\ (2013-2014) and \suzaku\ (2006-2013). We will use data corresponding to \suzaku\ observations from August 13, 2007 (Observation ID 402013010). The spin period and period derivative were 8.689 s and $0.20 \times 10^{-11}$ s s$^{-1}$ respectively, and the magnetic field is $1.3 \times 10^{14}$ G (see Tab.~6 of \cite{Enoto:2017dox}), which are the values we take. For details of the data processing and  background subtraction, we refer to our discussion under \sgr\ and the original paper. The  soft and hard X-ray spectrum were fit using a quasi-thermal Comptonization-like power-law tail and a hard power-law model. The photon index was obtained as $\Gamma = 0.24$ and the flux in the 15-60 keV range was obtained to be $35.1  \times \,10^{-12}$ \flux\ (see Tab.~5 of \cite{Enoto:2017dox}). The same reference also discusses \nustar\ observations of \fu\ from March 2014 (Observation ID 30001023002). For the \nustar\ data, the photon index was obtained as $\Gamma = 0.61$ and the flux $21.7  \times \,10^{-12}$ \flux\ (again from Tab.~5 of \cite{Enoto:2017dox}). %The spectrum corresponding to \suzaku\ is displayed in Figure 8 of \cite{Enoto:2017dox}.

In our work, we take into account all three studies referred to above: the \integral\ study from \cite{Wang:2013pka}, and the \suzaku\ studies from \cite{Enoto:2011fg} and \cite{Enoto:2017dox}.
The data and the $95\%$ CL upper limits on $G_{an} \times g_{a\gamma\gamma}$ are shown in the left and right panels of 
Fig.~\ref{fig:4U-0142+61}, respectively.

\subsection{\texorpdfstring{SGR$\,$1900+14}{SGR 1900+14}}

\begin{figure}
    \centering
    \includegraphics[width=0.48\textwidth]{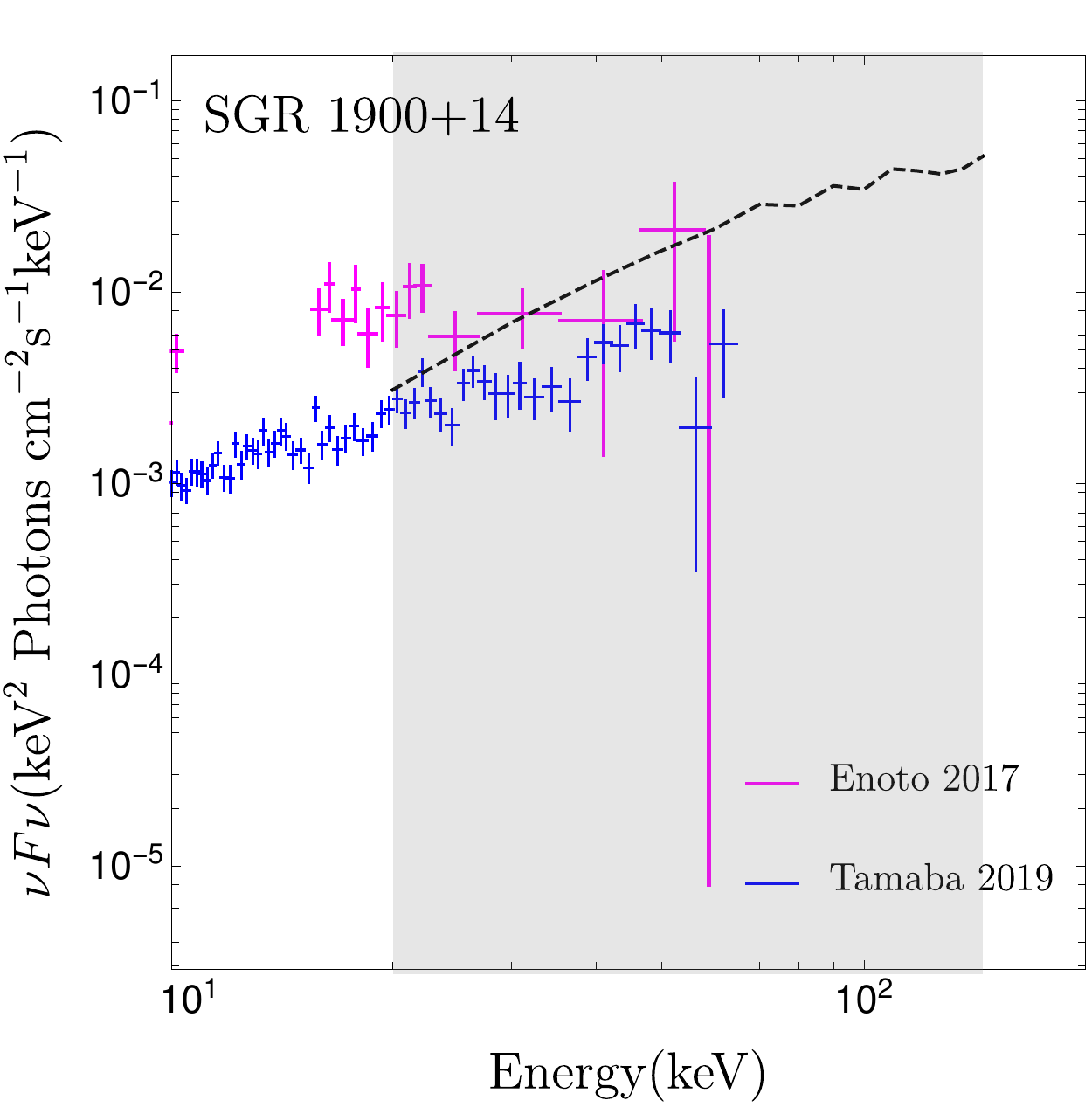} \quad
     \includegraphics[width=0.485\textwidth]{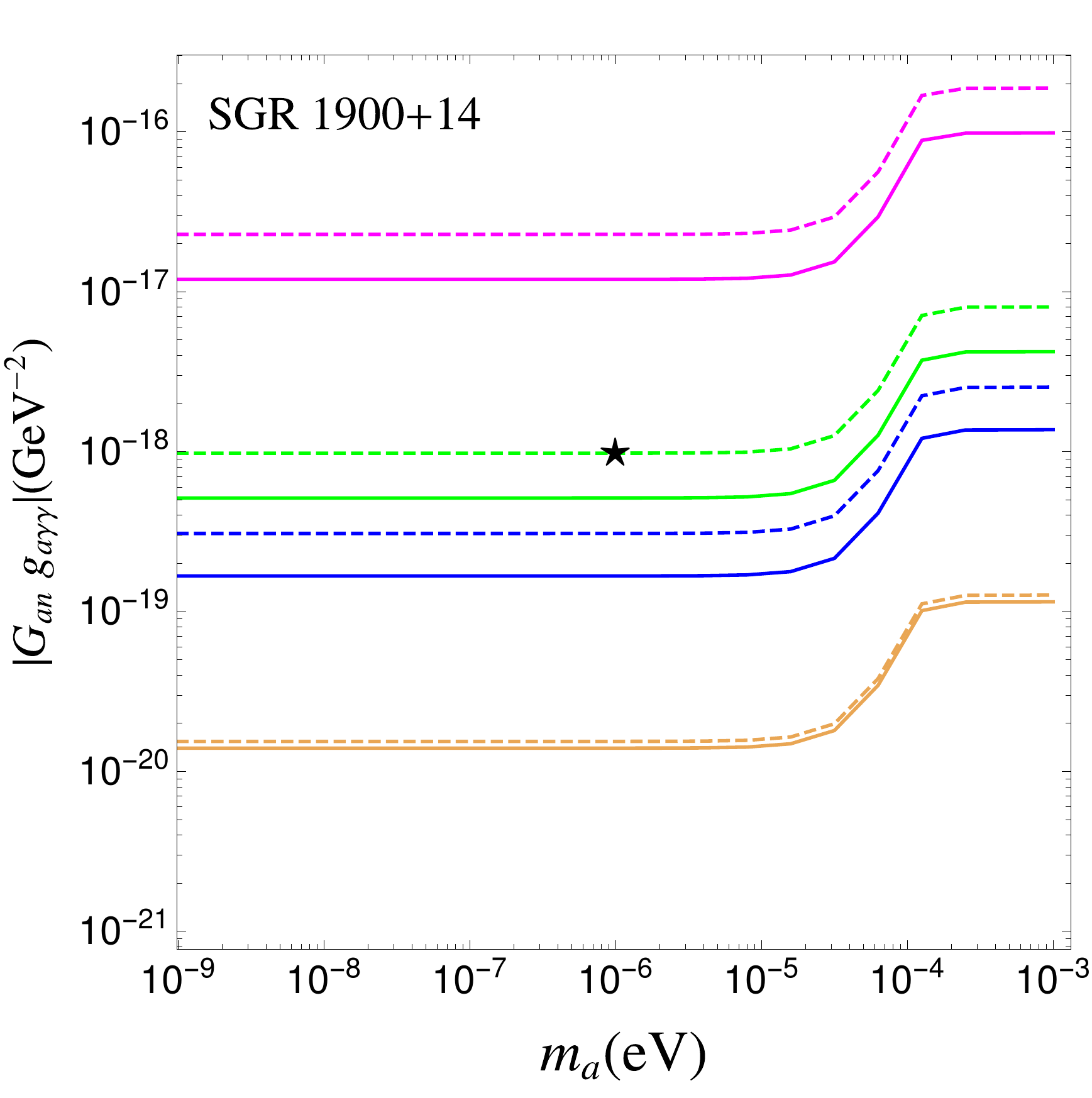}
    
    \caption{\label{SGR190014}\textbf{Left panel}: The theoretical spectrum (black dashed line) and observational data on the $\nu F_{\nu}$ plane for \sgraa\ . The observational data is obtained from \suzaku\ (red) \cite{Enoto:2017dox}, and \xmm\ (Observation ID 0790610101, October 20, 2016) plus \nustar\ (Observation ID 30201013002, October 20, 2016)(blue)\cite{Tamba:2019als}. \commoncaption{9.8\times 10^{-19}}
    }
\end{figure}

The hard X-ray spectrum of the magnetar \sgraa\ has been studied by several groups. The authors of \cite{Ducci:2015pfa} utilized \integral\ data from 2003-2013, where \sgraa\ was within 12\degree\, from the center of the field of view of IBIS/ISGRI. The spectrum between 22-150 keV was fit by a power-law with photon index $\Gamma = 1.9 \pm 0.3$ and the flux was $F_{20-100}$\,$=$\,$1.11 \pm 0.17$\,$\times$\,$10^{-11}$ \flux. The 20-100 keV luminosity was found to be smaller than that found by \textit{BeppoSAX} before the giant flare of 1998 by a factor of $\sim 5$. 

The authors of \cite{Tamba:2019als} analyzed data from \xmm\ (Observation ID 0790610101, October 20, 2016) and \nustar\ (Observation ID 30201013002, October 20, 2016) of \sgraa\ . The \nustar\ data was processed using \texttt{nupipeline} and \texttt{nuproducts} in HEASoft 6.23. Only data from the FPMA detector of \nustar\ was retained, while that corresponding to FPMB was rejected due to contamination from stray light. After background subtraction, the spectrum was binned to have at least 50 counts per bin. Data from \xmm\ was utilized to study the spectrum 1-10 keV range. A joint fitting of \xmm\ and FPMA data from \nustar\  for the 1-70 keV range with a blackbody plus power-law model yielded $\Gamma = 1.21 \pm 0.06$, and absorbed 1-70 keV flux of  $F_{1-70}$\,$=$\,$1.21 $\,$\times$\,$10^{-11}$ \flux\ (see Fig.~4 of \cite{Tamba:2019als}). For our work, we will use data corresponding to the FPMA detector in the hard X-ray range, extracted from Fig.~4 of \cite{Tamba:2019als} and displayed in terms of $\nu F_{\nu}$.

We also consider the analysis of \cite{Enoto:2017dox}  using data from \suzaku\ (Observation ID 404077010, April 26, 2009). For details of the data extraction and processing, as well as the fit, we refer to the original paper. The fit yields $\Gamma = 0.78$ and $F_{15-60} = 16.5  \times \,10^{-12}$ \flux\ (see Tab.~5 of \cite{Enoto:2017dox}). For the spin period, period derivative, and magnetic field we use the values obtained from \suzaku\ Observation ID 401022010, April 2006. The values are 5.2 s, $9.2 \times 10^{-11}$ s s$^{-1}$, and $7.0 \times 10^{14}$ G, respectively (see Tab.~6 of \cite{Enoto:2017dox}), which are the values we take.

The data and the $95\%$ CL upper limits on $G_{an} \times g_{a\gamma\gamma}$ for this magnetar are shown in the left and right panels of Fig.~\ref{SGR190014}, respectively.

\subsection{\texorpdfstring{1E$\,$1841$-$045}{1E 1841-045}}

\begin{figure}
    \centering
    \includegraphics[width=0.48\textwidth]{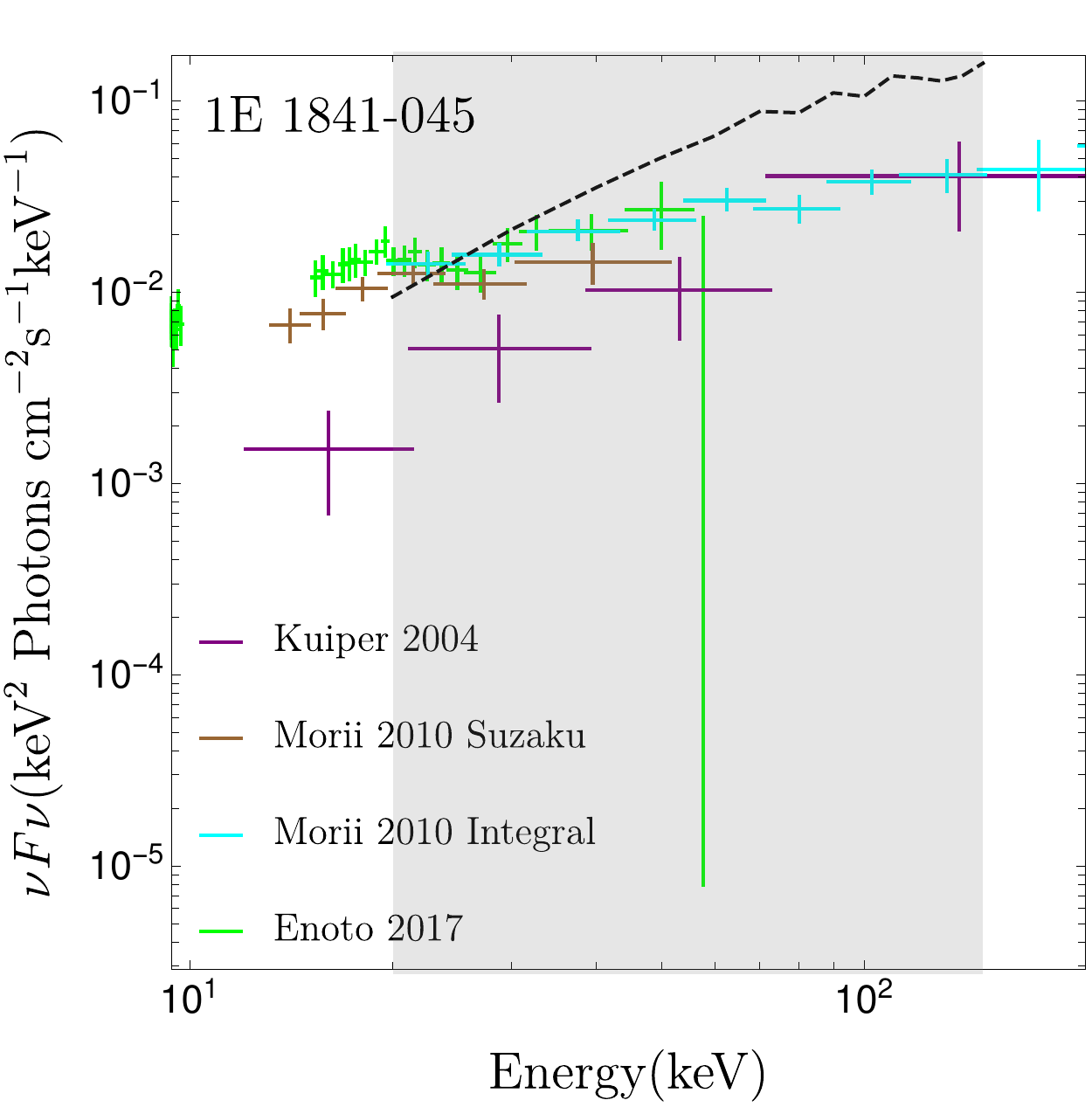}
    \quad
     \includegraphics[width=0.485\textwidth]{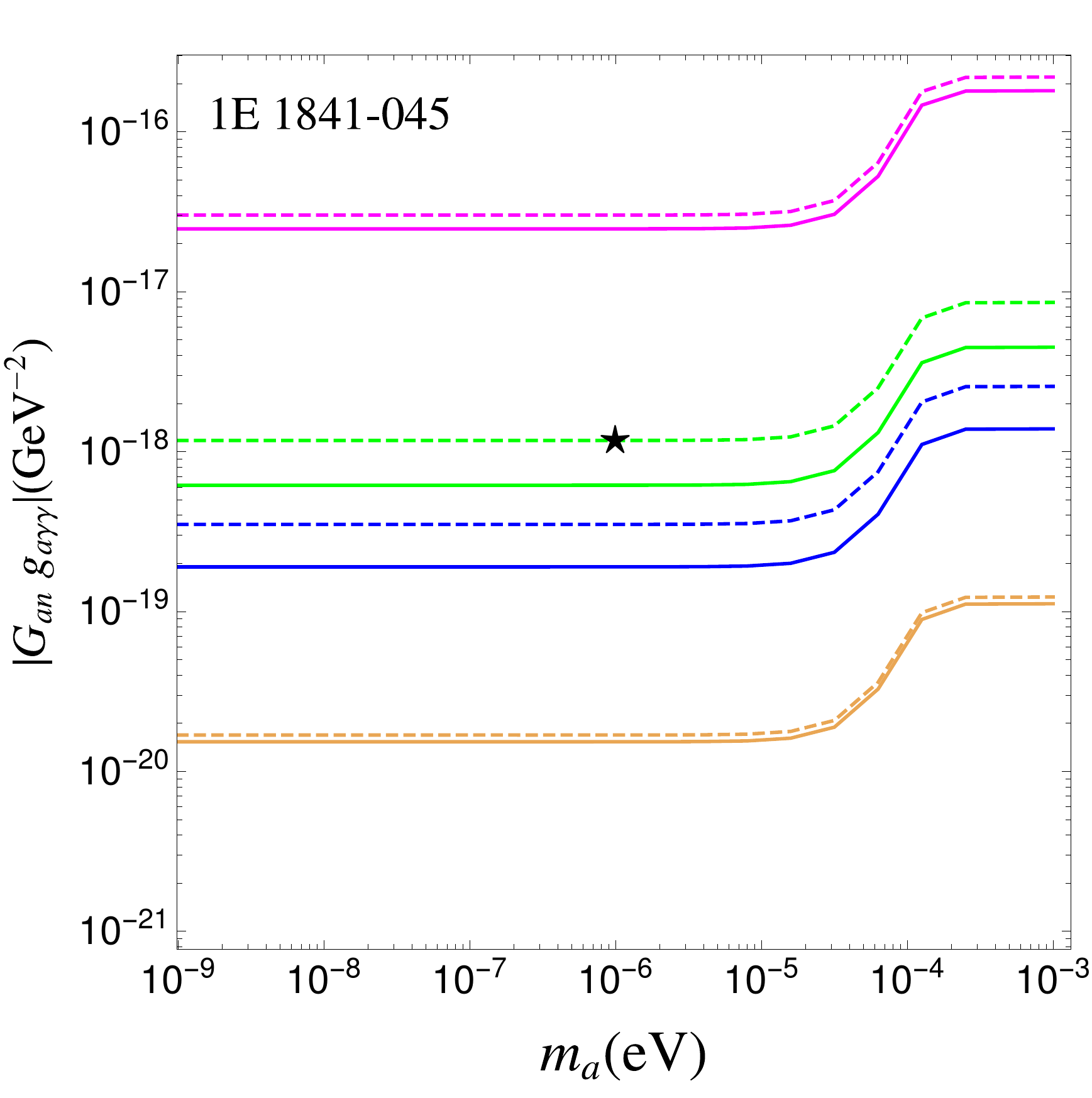}
      \caption{\label{fig:1E1841-045}\textbf{Left panel}: The theoretical spectrum (black dashed line) and observational data on the $\nu F_{\nu}$ plane for \e\ . The observational data is obtained from \rxte\ - HEXTE (purple) \cite{Kuiper:2004ya}, \suzaku\ (brown) \cite{Morii:2010vi}, \suzaku\ (green) \cite{Enoto:2017dox}, and \integral\ (cyan)  \cite{Morii:2010vi}. \commoncaption{11.7\times 10^{-19}}
    }
\end{figure}

One of the earliest studies of hard X-ray emission from the  magnetar \e\ was performed by the authors of \cite{Kuiper:2004ya}, who analyzed data from the period 1999-2003 from the PCA (2-60 keV) and the High Energy X-ray Timing Experiment HEXTE (15-250 keV) instruments aboard \rxte. We will be utilizing the results from \rxte\ - HEXTE in our analysis. The relevant Observation IDs, listed in Tab.~1 of \cite{Kuiper:2004ya}, are 40083, 50082, 60069, and 70094. The total number of pulsed counts in the differential HEXTE energy bands was determined by fitting a truncated Fourier series to the measured pulse phase distributions; for details of this fitting, we refer to the original paper. A power-law model fitted to the hard X-ray emission up to $\sim 150$ keV yielded a photon index of $\Gamma = 0.94 \pm 0.16$. 

\e\ was also  observed by \suzaku\ on April 19–22, 2006, and this data was analyzed by the authors of \cite{Morii:2010vi}. From HXD-PIN data, the phase-averaged flux in the range 1-50 keV was fit by a blackbody plus double power-law model, obtaining $\Gamma = 1.62$ and $F_{1-50}$\,$=$\,$43.7 $\,$\times$\,$10^{-12}$ \flux\ (see Tab.~1 of \cite{Morii:2010vi}). 

The authors of  \cite{Enoto:2017dox}  used data from  \nustar\ (Observation ID 30001025, September 5-23, 2013) and \suzaku\ (Observation ID 401100010, April 19, 2006). We refer to the original paper for details of data extraction, background subtraction and modeling of the spectrum. We will use \suzaku\ Observation ID 401100010 for our work. The spectrum was fit to obtain $\Gamma = 0.87$ and $F_{15-60} = 48.9  \times \,10^{-12}$ \flux\ (see Tab.~5 of \cite{Enoto:2017dox}). For the spin period, period derivative, and magnetic field we use the values obtained from \suzaku\ Observation ID 30001025. The values are 11.789 s, $4.1 \times 10^{-11}$ s s$^{-1}$, and $7.0 \times 10^{14}$ G, respectively (see Tab.~6 of \cite{Enoto:2017dox}). 

The data and the derived $95\%$ CL upper limits on $G_{an} \times g_{a\gamma\gamma}$ are shown
in the left and right panels of Fig.~\ref{fig:1E1841-045}, respectively.

\subsection{\texorpdfstring{1E$\,$ 2259+586}{1E 2259+586}}

\begin{figure}
    \centering
    \includegraphics[width=0.48\textwidth]{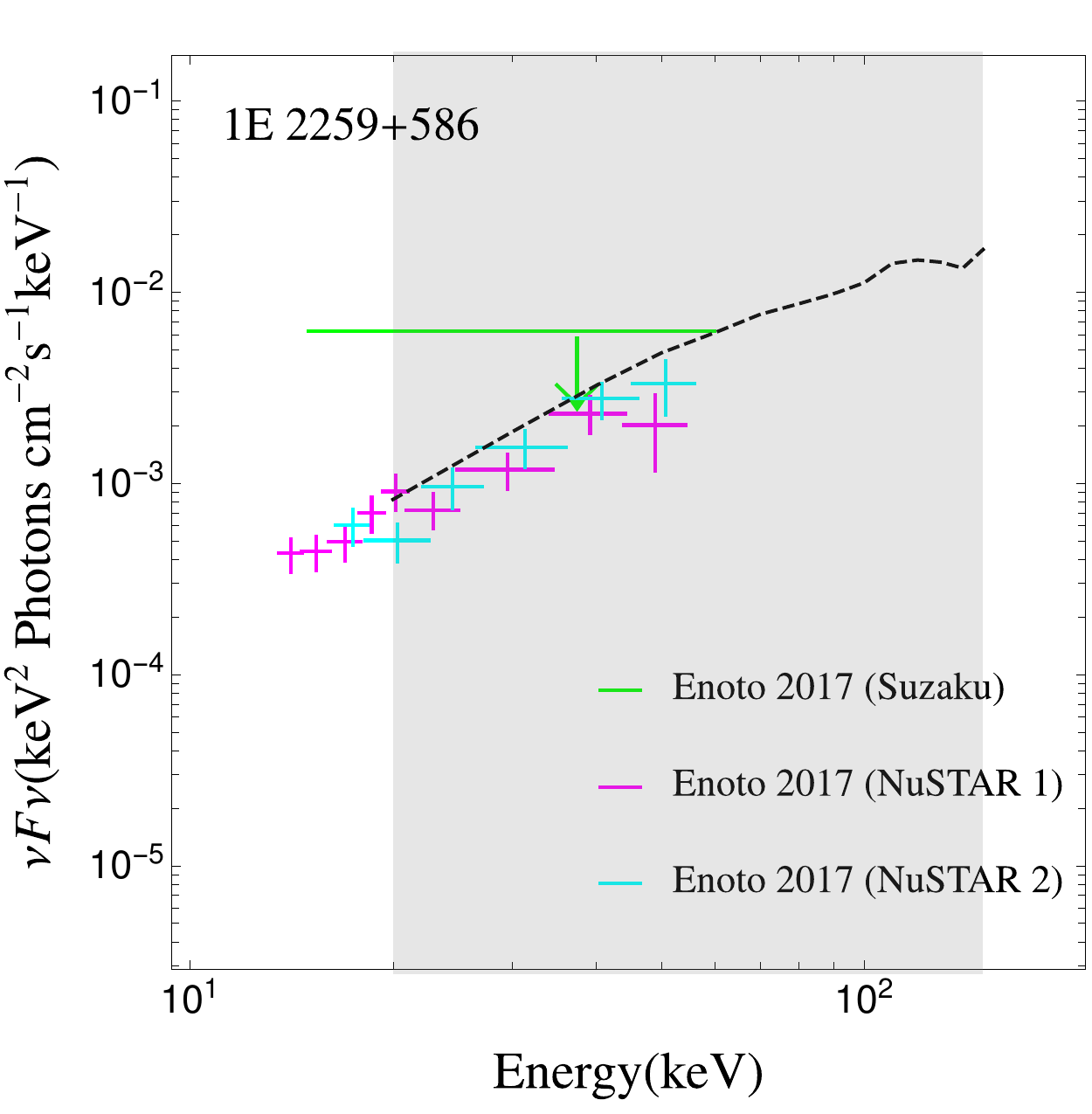}\quad
     \includegraphics[width=0.485\textwidth]{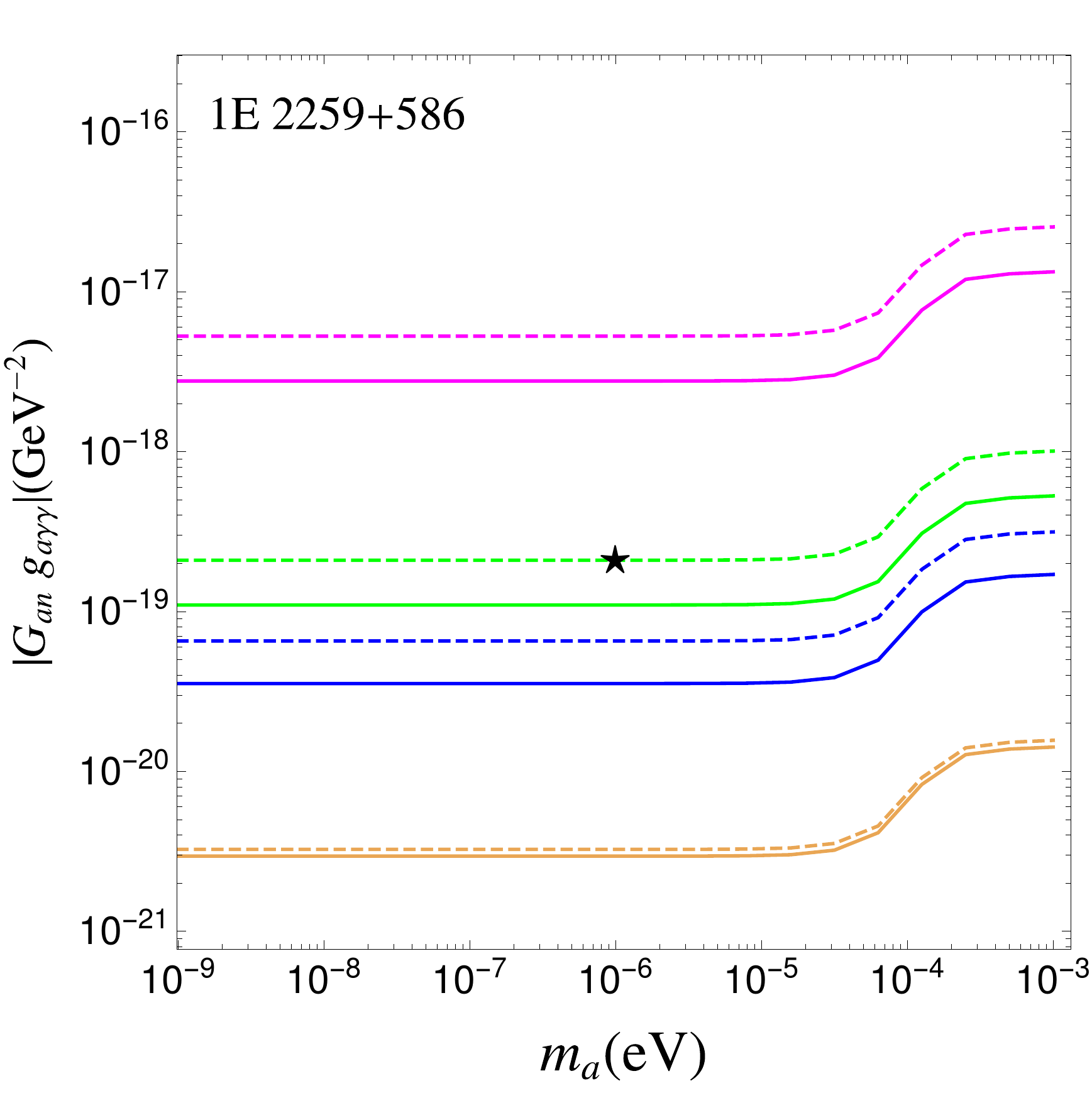}
       \caption{\label{fig:1E2259+586}\textbf{Left panel}: The theoretical spectrum (black dashed line) and observational data on the $\nu F_{\nu}$ plane for \enew\ . The observational data is obtained from \suzaku\ (green) \cite{Enoto:2017dox} and \nustar\ Observation ID 30001026002 and 30001026007 (magenta and cyan) \cite{Enoto:2017dox}. \commoncaption{2.1\times 10^{-19}}
    }
\end{figure}

The hard X-ray spectrum of \enew\ has been studied by the authors of \cite{Enoto:2017dox} using data from  \nustar\ (Observation IDs 30001026002 and 30001026007 from April 24-27 and May 16-18, 2013, respectively) and \suzaku\ (Observation ID 404076010, May 2009). The hard X-ray component has been observed by \nustar, but not by \suzaku; we will use \nustar\ data.  The  soft and hard X-ray spectrum were fit using a quasi-thermal Comptonization-like power-law tail and a hard power-law model. The photon index was obtained as $\Gamma = 0.58$ and $\Gamma = 0.66$ (Observation IDs 30001026002 and 30001026007, respectively, in Tab.~5 of \cite{Enoto:2017dox}). The corresponding fluxes were obtained as $F_{15-60} = 3.1  \times \,10^{-12}$ \flux\ and $F_{15-60} = 3.0 \times \,10^{-12}$ \flux\ , respectively. 

The authors of \cite{Weng:2015fnf} also considered the \nustar\ observations studied by \cite{Enoto:2017dox}, but in addition took into account two further sets: Observation IDs 30001026003 and 30001026005 (see Tab.~1 of \cite{Weng:2015fnf}). The soft and hard X-ray spectrum was fit with a  three-dimensional magnetar emission model and a power-law. Above 15 keV, a photon index of $\Gamma = 0.25$ was obtained, while the flux of the unabsorbed power-law component was obtained as $F_{1-79} = 0.58 \times \,10^{-11}$ \flux\ (see Tab.~2 of \cite{Weng:2015fnf}).

In our work, we display the \nustar\ data corresponding to the observations studied by \cite{Enoto:2017dox} (see the left panel of Fig.~\ref{fig:1E2259+586}). Incorporating the two extra observations from \cite{Weng:2015fnf} would not significantly change our results. For the spin period, period derivative, and magnetic field we use the values obtained from \nustar\ Observation ID 30001026002. The values are 6.979 s, $0.048 \times 10^{-11}$ s s$^{-1}$, and $0.59 \times 10^{14}$ G, respectively (see Tab.~6 of \cite{Enoto:2017dox}). 
The $95\%$ CL upper limits on $G_{an} \times g_{a\gamma\gamma}$ are shown in the right panel of
Fig.~\ref{fig:1E2259+586}.
In this case the result from the experimental data set yields a stronger constraint than the single upper limit. This is obvious from the left panel, as
the single upper limit there is well above the
other data points.

\subsection{\texorpdfstring{1RXS$\,$J1708$-$4009}{1RXS J1708-4009}}

\begin{figure}
    \centering
    \includegraphics[width=0.48\textwidth]{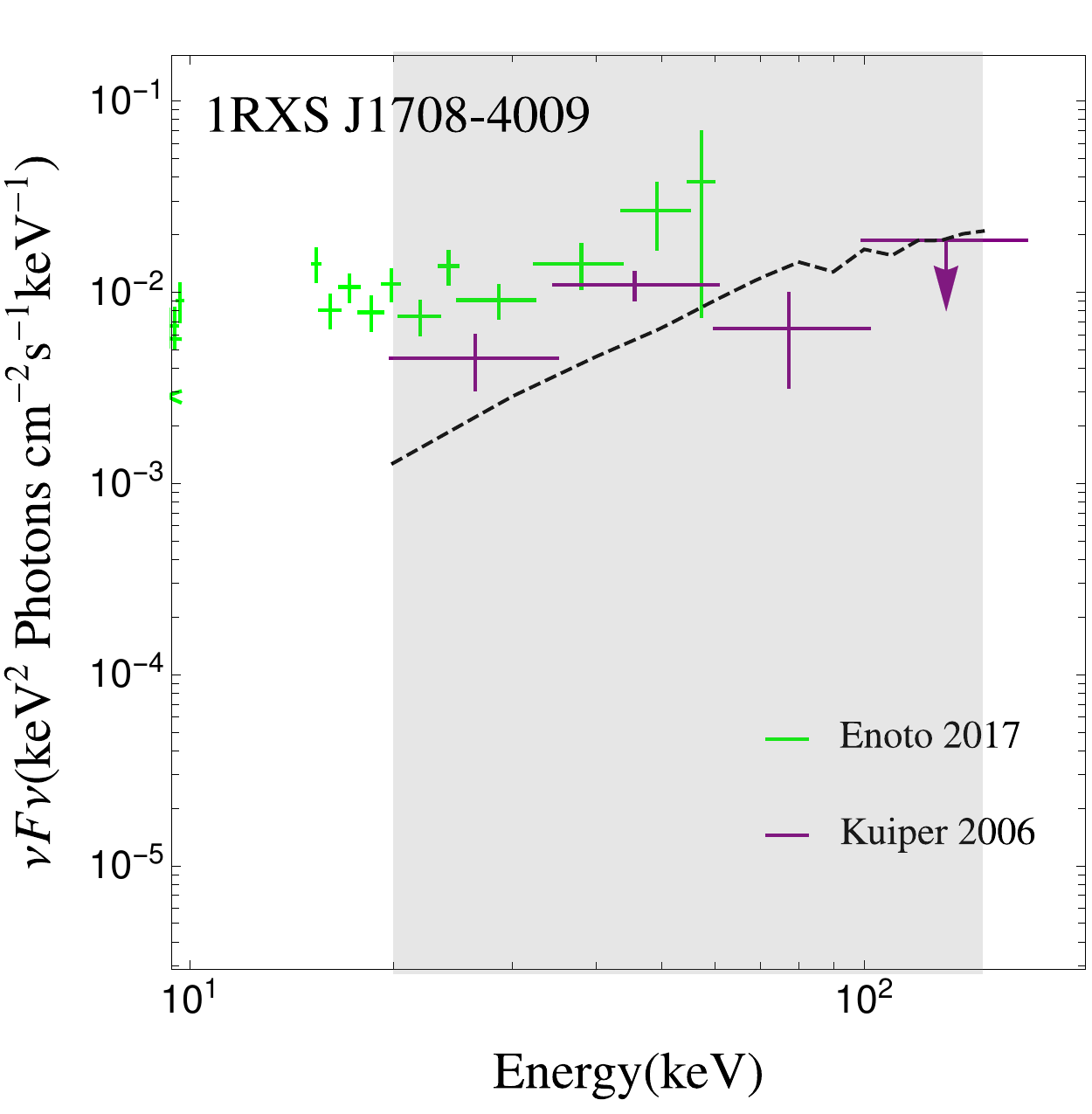}\quad
     \includegraphics[width=0.485\textwidth]{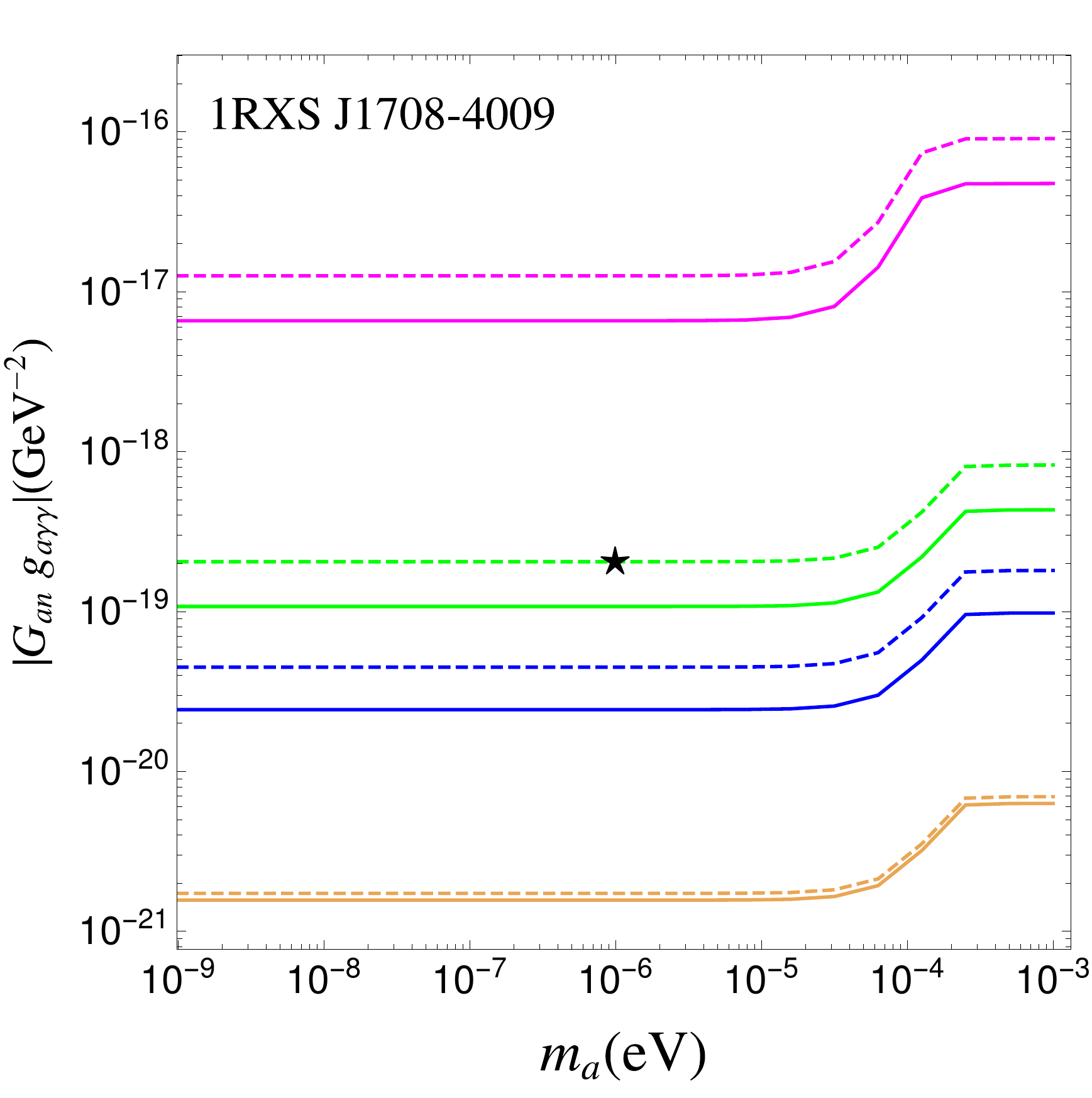}
        \caption{\label{fig:1RXSJ1708-4009}\textbf{Left panel}: The theoretical spectrum (black dashed line) and observational data on the $\nu F_{\nu}$ plane for \rxs\ . The observational data is obtained from \suzaku\ (green) \cite{Enoto:2017dox} and  \integral\ IBIS ISGRI (purple) \cite{Kuiper:2006is}. \commoncaption{2.0\times 10^{-19}}
    }
\end{figure}

The hard X-ray spectrum of the magnetar \rxs\ has been studied by the authors of \cite{Enoto:2017dox} using data from  \suzaku\ (Observation ID 404080010, August 23, 2009 and Observation ID 405076010, September 27, 2010).  We refer to the original paper for details about the data extraction, processing, and spectral fitting. The photon index was obtained as $\Gamma = 0.67$ and $\Gamma = 1.14$ (Observation IDs 404080010 and 40507601, respectively in Tab.~5 of \cite{Enoto:2017dox}). The corresponding flux was obtained as $F_{15-60} = 24.4  \times \,10^{-12}$ \flux. For the spectral data, we will utilize Observation ID 405076010 (see Fig.~8 of \cite{Enoto:2017dox}). For the spin period, period derivative, and magnetic field we use the values obtained from \suzaku\ Observation ID 404080010. The values are 11.005 s, $1.9 \times 10^{-11}$ s s$^{-1}$, and $4.7 \times 10^{14}$ G, respectively (see Tab.~6 of \cite{Enoto:2017dox}).

We also utilize the work by the authors of \cite{Kuiper:2006is}, who used \rxte\ PCA (2-60 keV), HEXTE (15-250 keV), and \integral\ IBIS ISGRI (20-300 keV) to study the spectrum of \rxs. We will especially focus on the time-averaged total (pulsed and unpulsed components) emission from \integral\ IBIS ISGRI analyzed by the authors and presented in Fig.~4 of \cite{Kuiper:2006is}. The data spans from January 29 - August 29, 2003 (Revs. 36-106) and was analyzed using \integral\ off-line scientific analysis version 4.1. The total on-axis exposure after screening was 974 ks. Mosaic images were made in the energy bands 20-35, 35-60, 60-100, 100-175 and 175-300 keV, and \rxs\ was detected significantly in the  20-35 and 35-60 keV bands. The results were displayed in Fig.~4 (magenta dots) of  \cite{Kuiper:2006is}; for more details, we refer to the original paper. 

The data and the $95\%$ CL upper limits on $G_{an} \times g_{a\gamma\gamma}$ are shown in the left and right panels of
Fig.~\ref{fig:1RXSJ1708-4009}, respectively. 
Here two types of experimental data, both upper limits and measured fluxes, play a big role.
For the lowest core temperature $10^8$ K, the constraint on the couplings derived from the 
measured fluxes is stronger than that from the upper limits by about an order of magnitude. 
The reason is for this temperature the peak positions of the spectra for both cases with and without considering superfluidity lie at around $30$ keV, which have much reduced spectral amplitudes at the experimental upper limit. For the other higher temperatures,
the peaks are at a position larger than $150$ keV, and show rising spectral shape 
in the energy range $20-150$ keV, which thus leads to a better result when using 
the single experimental upper limit.

\begin{figure}
    \centering
    \includegraphics[width=0.6\textwidth]{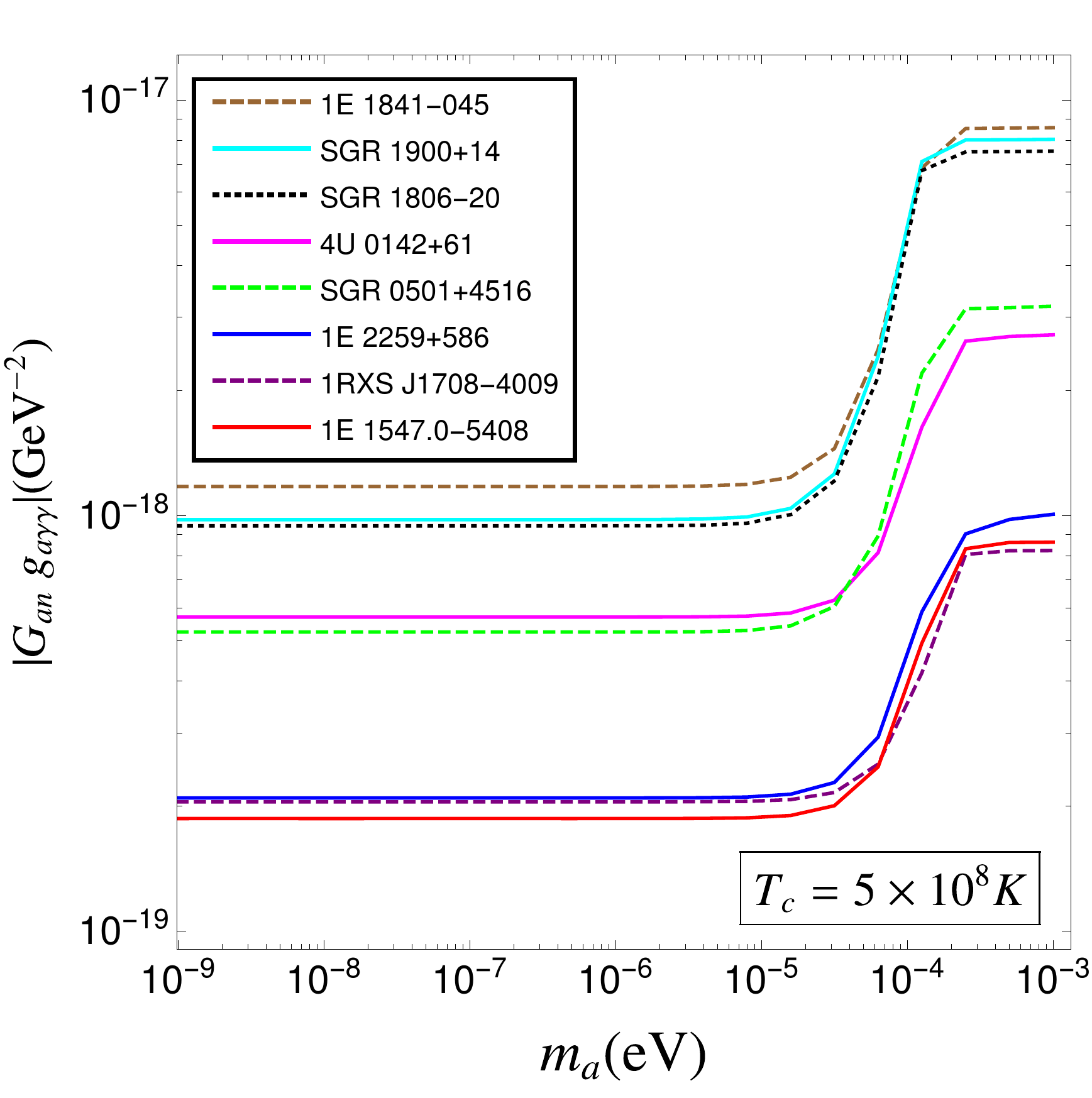}
    \caption{
The $95\%$ CL upper limits from all 8 magnetars assuming a core temperature of $5 \times 10^8$ K and with superfluidity.
    }
    \label{fig:combined}
\end{figure}

\begin{table}
    \centering
    \begin{tabular}{ c|c|c }
        \hline
        \hline
        Name & \multicolumn{2}{c}{$95\%$CL 
        Upper Limits on $G_{an} \times g_{a\gamma\gamma}$ ($10^{-19}$ GeV$^{-2}$)} \\
         &\quad \quad  ($T_c = 5 \times 10^8$ K) \quad\quad  & ($T_c = 10^9$ K) \\
        \hline
        \sgr & 9.4 & 3.0\\
        \aa & 1.9 & 0.44 \\
        \sgrnew & 5.2 & 1.6 \\
        \fu & 5.7 & 1.7 \\
        \sgraa & 9.8 & 3.1 \\
        \e & 11.7 & 3.5 \\
        \enew & 2.1 & 0.65 \\
        \rxs & 2.0 & 0.45 \\
        \hline
        \hline
    \end{tabular}
    \caption{The $95\%$ CL upper limits from all 8 magnetars with superfluidity for two core temperature values at $m_a=10^{-6}$eV.
    \label{tab:final_table}
    }
\end{table}

\section{Conclusions}

\begin{figure}
    \centering
    \includegraphics[width=0.6\textwidth]{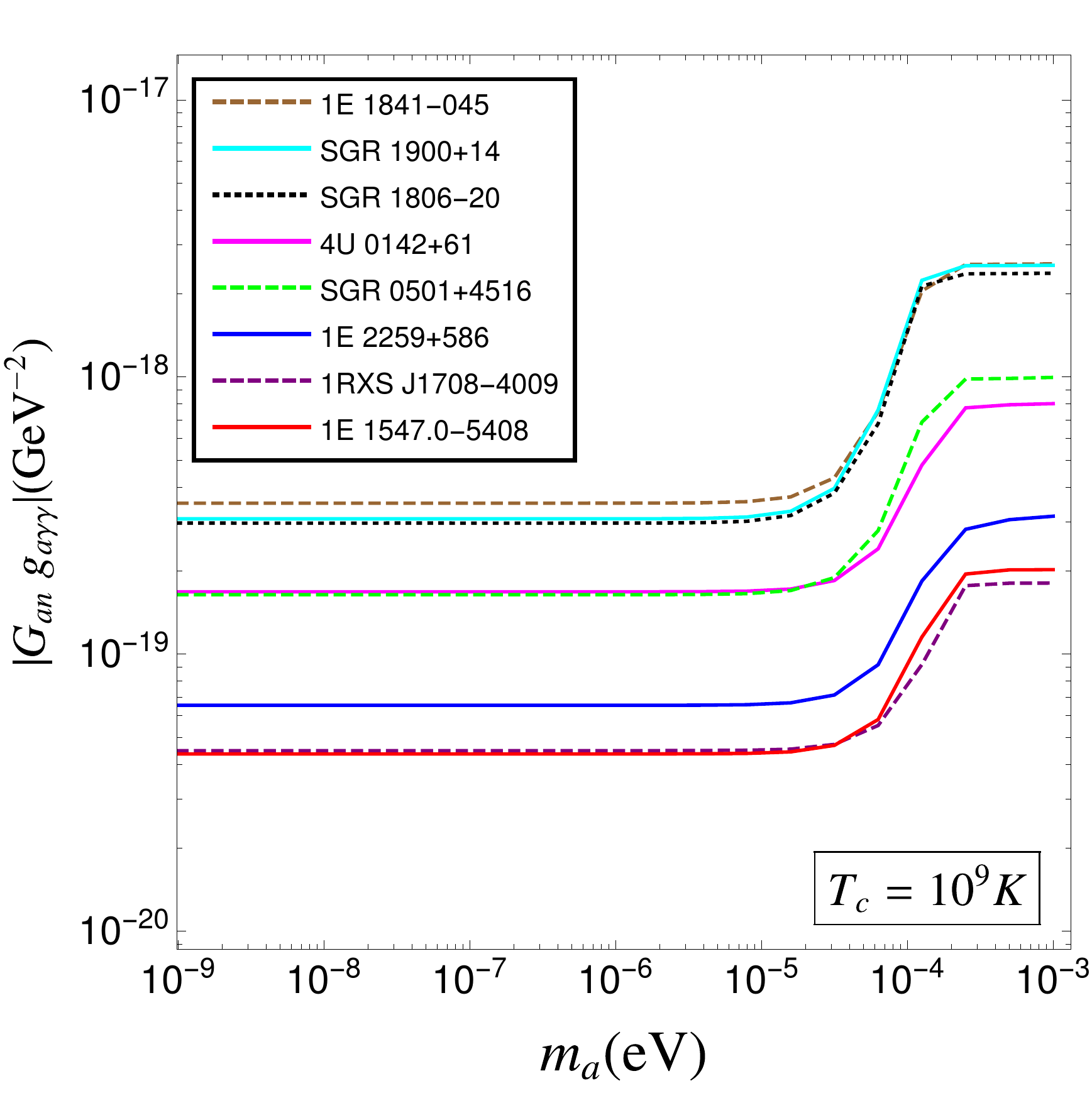}
    \caption{
The $95\%$ CL upper limits from all 8 magnetars assuming a core temperature of $10^9$ K and with superfluidity.
    }
    \label{fig:combined2}
\end{figure}

The final results of our analysis are presented in Fig.~\ref{fig:combined} for a core temperature of $T = 5 \times 10^8$ K and Fig.~\ref{fig:combined2} for a core temperature of $T = 10^9$ K. The constraints on the ALP parameter space are also listed in Tab.~\ref{tab:final_table} for $m_a=10^{-6}$ eV.
For both temperatures, the best constraint comes from 
 \aa \ in the majority of the mass regime and \rxs\ is slightly
 better for higher mass.

The intersection between the study of magnetars  and topics in fundamental physics has traditionally been in the arena of strong-field QED \cite{Baring:2020zke, Heyl:1997hr, Lai:2006af}. The near-critical magnetic field near magnetars holds out the promise of observing exotic Standard Model phenomena: magnetic photon splitting and single-photon pair creation, as well as  birefringence of the magnetized quantum vacuum. 

To these goals, we add the possibility of investigating one of the most sought-after particles beyond the Standard Model: the axion. As we have demonstrated, existing data from magnetars are capable of putting robust limits on ALPs. The discovery of more magnetars, and the measurement of hard X-ray emission from them, will lead to stronger constraints on ALPs. A further important future direction is to carefully incorporate astrophysical modelling of the hard X-ray emission following the advances made in \cite{Wadiasingh:2017rcq} as a ``background" to the ALP-induced spectrum, and probe the morphology of the total spectrum.

The most  crucial difference between the  emission from axions and emission predicted by astrophysical models is the stark difference in polarization, with ALPs producing a clean O-mode polarization, and astrophysics models producing an X-mode polarization that is rendered clean by strong QED effects, as emphasized in \cite{Heyl:2018kah}. The future era of hard X-ray and soft gamma-ray polarimetry will offer a window to these differences \cite{Krawczynski:2019ofl}.

\section{Acknowledgments}

JFF is supported by NSERC and FRQNT. HG and KS are supported by the U.~S. Department of Energy grant DE-SC0009956.  The work of SPH is supported by the U.~S. Department of Energy grants DE-FG02-05ER41375 and DE-FG02-00ER41132 as well as the National
Science Foundation grant No.~PHY-1430152 (JINA Center for the Evolution of the Elements).   We would like to thank Matthew Baring, Teruaki Enoto, Alex Haber, Sanjay Reddy, Armen Sedrakian, Tsubasa Tamba, and George Younes for helpful discussions.

\appendix

\section{Magnetar core temperature} \label{a1}
\begin{figure}
    \centering
    \includegraphics[width=0.6\textwidth]{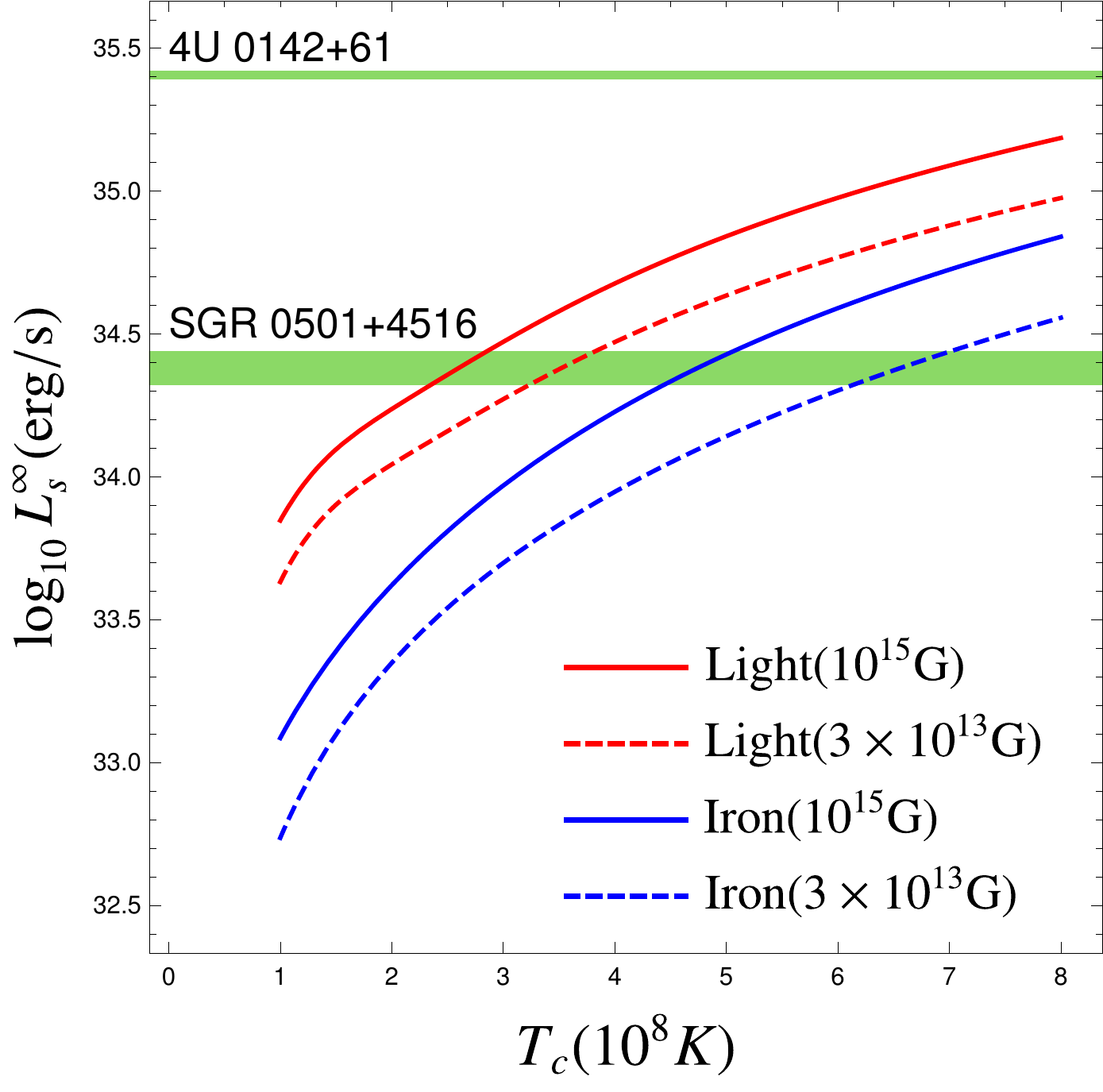}
    \caption{Luminosity versus core temperature for four combinations of the two choices of surface magnetic field: $B=3 \times 10^{13} G$ and $B=10^{15}G$
and two accreted magnetized envelopes: the iron envelope and the accreted envelope composed of a light element.  The relation between the core temperature $T_c$ and surface temperature $T_s$ is calculated from the formula provided in Ref.~\cite{Potekhin:2003aj}, and is consistent with Fig.~1 in Ref.~\cite{Beloborodov:2016mmx}.  Also overlaid on this plot are the observed luminosities for two McGill-catalog magnetars: 4U 0142+61 in the range $(2-10)$keV plus $(20-150)$keV, 
and SGR 0501+4516 in the range $(2-10)$keV plus $(20-100)$keV.}
    \label{fig:coretemp}
\end{figure}
The spectrum of X-rays emitted from the magnetar surface is measured and is converted into a blackbody temperature of the surface of the magnetar \cite{Olausen:2013bpa}.  Theoretical modeling is required to deduce the core temperature from the surface temperature.  The relationship between the core temperature and the surface temperature -- the temperature at the top of the heat-blanketing envelope surrounding the neutron star -- has been long studied \cite{Potekhin:2006iw,Potekhin:2003aj,Potekhin:1997mn,1983ApJ...272..286G} (see also the reviews \cite{Potekhin:2015qsa,Yakovlev:2004iq}) and is understood to be a function of the envelope composition, the magnetic field magnitude and direction, and the mass of the neutron star.  However, the surface temperature of magnetars is unexpectedly high \cite{Vigano:2013lea,2009MNRAS.395.2257K}, seemingly requiring the magnetar core to have a temperature in excess of $10^9 \text{ K}$.  Such high core temperatures are not sustainable for the lifetime of a magnetar due to the large neutrino emission from such a hot core \cite{2012MNRAS.422.2632H,Potekhin:2006iw}.  This has led to the proposal of mechanisms whereby the internal magnetic field leads to the high surface temperature \cite{Beloborodov:2016mmx}.

In Fig.~\ref{fig:coretemp}, we show the relationship between surface luminosity (which can easily be converted to surface temperature) and core temperature.  Four different models of the neutron star envelope have been used, corresponding to different elemental composition of the envelope and different values of the magnetic field.  The surface luminosity - core temperature relationship is from Ref.~\cite{Potekhin:2003aj}, and is consistent with Fig.~1 in Ref.~\cite{Beloborodov:2016mmx}.  Overlaid on Fig.~\ref{fig:coretemp} are the observed luminosities for two representative magnetars: 4U 0142+61 and SGR 0501+4516.  We can see that  SGR 0501+4516 has a surface luninosity that is consistent with the models presented in Ref.~\cite{Potekhin:2003aj}, although depending on the assumptions of properties of the envelope, the core temperature could vary by a factor of a few. Indeed, the core temperature of SGR 0501+4516 could range from $\sim 3 \times 10^8$ K to $\sim 7 \times 10^8$ K, depending on the model.  Other magnetars, for example, 4U 0142+61, have surface luminosities that require core temperatures above what is expected for a magnetar that has existed for kiloyears. The core temperature in that case could plausibly be $10^9$ K or higher.

The main point of this appendix and Fig.~\ref{fig:coretemp} is to advance a plausible range of magnetar temperatures to investigate, without necessarily choosing a specific temperature. From Fig.~\ref{fig:coretemp}, it is apparent that a judicious choice of ranges is $10^8$ K, $5 \times 10^8$ K, $10^9$ K, and $5 \times 10^9$ K. We therefore present our analysis for this range of temperatures. 

%%%%%%%%%%%%%%%%%%%%%%%%%%%%%%%%%%5

\section{Axion emissivity in degenerate nuclear matter}
\label{sec:np_emissivity}
The axion emissivity in strongly degenerate nuclear matter due to $n+n\rightarrow n+n+a$ and $p+p\rightarrow p+p+a$ was originally calculated by \cite{Iwamoto:1984ir}, and improved on by \cite{Iwamoto:1992jp,Stoica:2009zh}.  Ref.~\cite{Harris:2020qim} has a detailed discussion of the emissivity calculation in both degenerate and non-degenerate nuclear matter.  We provide here the derivation of the emissivity due to $n+p\rightarrow n+p+a$, improving upon the results in the literature (see footnote \ref{footnote:raffelt}).  The axion emissivity from $n+p\rightarrow n+p+a$ is given by
\begin{align}
    Q^0_{np} &= \int \frac{\mathop{d^3p_1}}{(2\pi)^3}\frac{\mathop{d^3p_2}}{(2\pi)^3}\frac{\mathop{d^3p_3}}{(2\pi)^3}\frac{\mathop{d^3p_4}}{(2\pi)^3}\frac{\mathop{d^3\omega}}{(2\pi)^3} \frac{S\sum_{\text{spins}}\vert\mathcal{M}\vert^2}{2^5E_1^*E_2^*E_3^*E_4^*\omega}\omega f_1f_2(1-f_3)(1-f_4)\nonumber\\
    &\times (2\pi)^4\delta(E_1+E_2-E_3-E_4-\omega)\delta^3(p_1+p_2-p_3-p_4-p_a),\label{eq:Q_np_appendix}
\end{align}
using the matrix element displayed in Eq.~\ref{eq:npmatrixelement}.
We consider the integral over axion energy in spherical coordinates and do the angular part, giving a factor of $4\pi$.  Then, following \cite{Harris:2020qim,1979ApJ...232..541F}, we multiply Eq.~\ref{eq:Q_np_appendix} by one in the form
\begin{align}
    1 &= \int_0^{\infty}\mathop{dp_1}\mathop{dp_2}\mathop{dp_3}\mathop{dp_4}\delta(p_1-p_{Fn})\delta(p_2-p_{Fp})\delta(p_3-p_{Fn})\delta(p_4-p_{Fp})\\
    &= \frac{1}{p_{Fn}^2p_{Fp}^2}\int\mathop{dE_1}\mathop{dE_2}\mathop{dE_3}\mathop{dE_4}E_1^*E_2^*E_3^*E_4^*\delta(p_1-p_{Fn})\delta(p_2-p_{Fp})\delta(p_3-p_{Fn})\delta(p_4-p_{Fp}),
\end{align}
which enables us to separate the phase space integral into two parts\footnote{This factorization, known as ``phase space decomposition'' \cite{Shapiro:1983du} is an approximation, but is known to converge to the exact result -- obtained by doing the integration over the full phase space -- as the temperature drops below 10-20 MeV \cite{Harris:2020qim}, though the precise temperature where the Fermi surface approximation becomes valid depends on the reaction \cite{Alford:2018lhf}.}
\begin{equation}
    Q^0_{np} = \frac{(m_n/m_{\pi})^4f^4G_{an}^2}{48\pi^{10}p_{Fn}^2p_{Fp}^2}AI.
\end{equation}
Noting that for nucleon bremsstrahlung processes $E_1+E_2-E_3-E_4=E_1^*+E_2^*-E_3^*-E_4^*$ (that is, the nuclear mean field terms cancel), the energy integral is
\begin{equation}
I = \int \mathop{d\omega}\mathop{dE_1}\mathop{dE_2}\mathop{dE_3}\mathop{dE_4} \omega^2 \delta(E_1^*+E_2^*-E_3^*-E_4^*-\omega)f_1f_2(1-f_3)(1-f_4).
\end{equation}
This integral is evaluated by changing variables to $x_i = (E_i-\mu_i)/T= (E_i^*-\mu_i^*)/T$ and $x=\omega/T$, yielding
\begin{equation}
    I=\frac{T^6}{6}\int_0^{\infty}\mathop{dx}\frac{x^3(x^2+4\pi^2)}{e^x-1} = \frac{62}{945}\pi^6T^6.
\end{equation}
The angular integral is 
\begin{align}
A &= \int \mathop{d^3p_1}\mathop{d^3p_2}\mathop{d^3p_3}\mathop{d^3p_4}\delta^3(\mathbf{p_1}+\mathbf{p_2}-\mathbf{p_3}-\mathbf{p_4})\delta(p_1-p_{Fn})\delta(p_2-p_{Fp})\delta(p_3-p_{fn})\nonumber\\
&\times\delta(p_4-p_{Fp})\left[\frac{\mathbf{k}^4}{(\mathbf{k}^2+m_{\pi}^2)^2}+\frac{4\mathbf{l}^4}{(\mathbf{l}^2+m_{\pi}^2)^2}-2\frac{\mathbf{k}^2\mathbf{l}^2-3(\mathbf{k}\cdot \mathbf{l})^2}{(\mathbf{k}^2+m_{\pi}^2)(\mathbf{l}^2+m_{\pi}^2)}\right],
\end{align}
where we have neglected the axion momentum in the momentum-conserving delta function.
By multiplying the integral by $\int \mathop{d^3k}\mathop{d^3l}\delta^3(\mathbf{k}-\mathbf{p_2}+\mathbf{p_4})\delta^3(\mathbf{l}-\mathbf{p_2}+\mathbf{p_3})$,
we convert the integral to one over $\mathbf{p_1}$, $\mathbf{p_2}$, $\mathbf{k}$, and $\mathbf{l}$.  Doing the integral over the 3-momentum conserving delta function and relabelling $\mathbf{p_1}$ as $\mathbf{p}$, we find
\begin{align}
    A &= \int\mathop{d^3p}\mathop{d^3k}\mathop{d^3l}\delta(\vert \mathbf{p}\vert-p_{Fn})\delta(\vert \mathbf{p}+\mathbf{k}+\mathbf{l}\vert-p_{Fp})\delta(\vert \mathbf{p}+\mathbf{k} \vert - p_{Fn})\delta (\vert \mathbf{p}+\mathbf{l} \vert - p_{Fp})\nonumber\\
    &\times \left[\frac{\mathbf{k}^4}{(\mathbf{k}^2+m_{\pi}^2)^2}+\frac{4\mathbf{l}^4}{(\mathbf{l}^2+m_{\pi}^2)^2}-2\frac{\mathbf{k}^2\mathbf{l}^2-3(\mathbf{k}\cdot \mathbf{l})^2}{(\mathbf{k}^2+m_{\pi}^2)(\mathbf{l}^2+m_{\pi}^2)}\right].\label{eq:A_pkl}
\end{align}
We convert to spherical coordinates and choose $\mathbf{p}$ to lie along the $z$ axis and $\mathbf{k}$ to lie in the $x-z$ plane, thus $\mathbf{p}=p(0,0,1), \mathbf{k}=k(\sqrt{1-r^2},0,r), \text{ and } \mathbf{l}=l(\sqrt{1-s^2}\cos{\phi},\sqrt{1-s^2}\sin{\phi},s)$.  By virtue of the coordinates chosen, three integrals in Eq.~\ref{eq:A_pkl} become trivial, giving a factor of $4\pi\times 2\pi$.  Then the integrals can be done in the order $r,s,\phi$, which leaves integrals over $k$ and $l$.  We recommend making the integral nondimensional.  While quite complicated, it can be done analytically.  We find
\begin{equation}
    A = 32\pi^3 p_{Fn}^2p_{Fp}^3 G(c,d),
\end{equation}
where
\begin{align}
    &G(c,d) = 5+\frac{1}{2}\frac{d^2}{1+d^2}+\frac{d}{4}\left(\pi - \arctan{\left(\frac{2d}{1-d^2}\right)}\right)-4d\arctan{\left( \frac{4c^2d}{4c^2d^2+d^2-c^2}\right)}\nonumber\\
    &-\frac{(c-d)^3}{c(4c^2d^2+(c-d)^2)}-\frac{(c+d)^3}{c(4c^2d^2+(c+d)^2)}\\
    &-\frac{4c^2d^3}{H(c,d)}\left( \pi + \arctan{\left(\frac{2d(d^2(1+4c^2)-c^2)H(c,d)}{(1+4c^2)^2d^6-(1+10c^2+40c^4)d^4+c^2(2-7c^2)d^2-c^4}\right)}  \right),\nonumber
\end{align}
with $c=m_{\pi}/(2p_{Fn})$, $d=m_{\pi}/(2p_{Fp})$, and
\begin{equation}
    H(c,d) = \sqrt{(1+8c^2+32c^4)d^4+2c^2(4c^2-1)d^2+c^4}.
\end{equation}
In this derivation, we have assumed that $p_{Fn}>p_{Fp}$ (corresponding to $c<d$).  We have also assumed  $c<1$ and $d<1$, conditions which are met in the neutron star core due to its high density.  Thus, the total emissivity from $n+p\rightarrow n+p+a$ is given by Eq.~\ref{eq:Q0_np} and the differential emissivity is given by Eq.~\ref{eq:axion_np_emissivity}.  

A factor of $C_{\pi}$ can be included in all expressions for the emissivity in this section to account for improvements to the one-pion exchange approximation, discussed in Appendix \ref{a3}.
%%%%%%%%%%%%%%%%%%%%%%%%%%%%%%%%%%%%%%%%%%%%%%%%%%
\section{Beyond one-pion exchange} \label{a3}
Friman \& Maxwell \cite{1979ApJ...232..541F} treated the strong interaction that occurs in a nucleon bremsstrahlung process as the exchange of a single pion.  This approximation of the nuclear interaction was used by Brinkmann \& Turner \cite{PhysRevD.38.2338} in their calculation of the axion emission rate from $N+N'\rightarrow N+N'+a$.  It is well known that the one-pion exchange approximation overestimates the rate of the bremsstrahlung process \cite{Hanhart:2000ae,Rrapaj:2015wgs}.  For example, Hanhart \textit{et al.}\ \cite{Hanhart:2000ae} calculated the rate of $n+n\rightarrow n+n+a$ in the soft radiation approximation, where the bremsstrahlung rate is directly related to the on-shell nucleon-nucleon scattering amplitude.  The calculation is model-independent, but does not include many-body effects.  This treatment of the nuclear interaction results in approximately a factor of 4 decrease in the matrix element squared (the modification is weakly momentum-dependent), and thus ends up as a roughly uniform (independent of density or temperature) modification of the emissivity by a factor $C_{\pi}=1/4$ \cite{Beznogov:2018fda}.

In the limit where the emitted axion energy tends to zero, the intermediate nucleon propagator becomes close to on-shell.  However, the emission rate does not diverge in this limit because the divergence of the nucleon propagator is regulated by the finite nucleon decay width.  This is known as the Landau-Pomeranchuk-Migdal (LPM) effect \cite{Landau:1953gr,PhysRev.103.1811}.  However, the nucleon decay width is quite small at magnetar temperatures, and the LPM effect is only significant when the energy of the emitted axion is less than the nucleon decay width, which only occurs once the temperature rises above 5-10 MeV \cite{vanDalen:2003zw}.
%%%%%%%%%%%%%%%%%%%%%%%%%%%%%%%%%%%%%%%
\section{Axion emissivity with superfluid protons} \label{a4}
In the calculation of the axion emissivity due to nucleon bremsstrahlung where one nucleon species (the proton) is superfluid, we are still able to use the Fermi surface approximation and so the calculation strongly resembles the emissivity calculation in ungapped nuclear matter (Appendix \ref{sec:np_emissivity} and Ref.~\cite{Harris:2020qim}).  In the Fermi surface approximation, we need only consider energies of nucleons near the Fermi surface, so the superfluid energy dispersion relations Eq.~\ref{eq:superfluid_energy} near the Fermi surface are sufficient to use for the protons in the bremsstrahlung process.  The neutron dispersion relations remain those in ungapped nuclear matter Eq.~\ref{eq:E_ungapped}.  The phase space integral in the emissivity calculation is broken up into an angular integral and an energy integral.  The angular part is unchanged by the gap in the proton energy spectrum, because in the presence of nucleon superfluidity, $\sum_{\text{spins}}\vert\mathcal{M}\vert^2 $ is unchanged \cite{Yakovlev:2000jp}.  However, the energy integral is altered and no longer can be fully simplified analytically.

The axion emissivity from $p+p\rightarrow p+p+a$ is $Q^S_{pp} = Q^0_{pp}R_{pp}(T,n_B)$ (Eq.~\ref{eq:QSppwithR}), with
\begin{align}
    R_{pp}(T,n_B) &= \frac{945}{62\pi^6}\int_0^{\infty}\mathop{dx} x^2 \int_{-\infty}^{\infty}\mathop{dx_1}\mathop{dx_2}\mathop{dx_3}\mathop{dx_4}\delta(z_1+z_2-z_3-z_4-x)\label{eq:ppsuperfluidintegral}\\
    &\times f(z_1)f(z_2)(1-f(z_3))(1-f(z_4)),\nonumber
\end{align}
where 
\begin{equation}
    z_i = \sign(x_i)\sqrt{x_i^2+(\Delta(T,n_B)/T)^2}.
    \label{eq:z}
\end{equation}
The integral \ref{eq:ppsuperfluidintegral}, after analytically integrating over $x$ to remove the delta function, must be done numerically.

In the process $n+p\rightarrow n+p+a$, only two of the four nucleons (the two protons) are gapped.  The emissivity can be written $Q^S_{np} = Q^0_{np}R_{np}(T,n_B)$ (Eq.~\ref{eq:QSnpwithR}), where
\begin{align}
    R_{np}(T,n_B) &= \frac{945}{62\pi^6}\int_0^{\infty}\mathop{dx} x^2 \int_{-\infty}^{\infty}\mathop{dx_1}\mathop{dx_2}\mathop{dx_3}\mathop{dx_4}\delta(x_1+z_2-x_3-z_4-x)\label{eq:npsuperfluidintegral}\\
    &\times f(x_1)f(z_2)(1-f(x_3))(1-f(z_4)).\nonumber
\end{align}
After analytically integrating over $x_1$ and $x_3$, the integral \ref{eq:npsuperfluidintegral} must be done numerically.

The functions $R_{nn}$ and $R_{pp}$ depend on density and temperature because of the presence of the superfluid gap in the integral.  When the gap $\Delta=0$, the variable $z_i = x_i$ and $R_{pp}(T,n_B)=R_{np}(T,n_B)=1$.
%%%%%%%%%%%%%%%%%%%%%%%%%%%%%%%%%%%%%%%
\section{Axion-Photon Conversion} \label{appconv}
\begin{figure}[t!]
\centering
%\includegraphics[width=0.42\textwidth]{Figures/chi_x_HardX.pdf}
%\quad
%\includegraphics[width=0.39\textwidth]{Figures/sin_cos_x_HardX.pdf}
%\\
\includegraphics[width=0.48\textwidth]{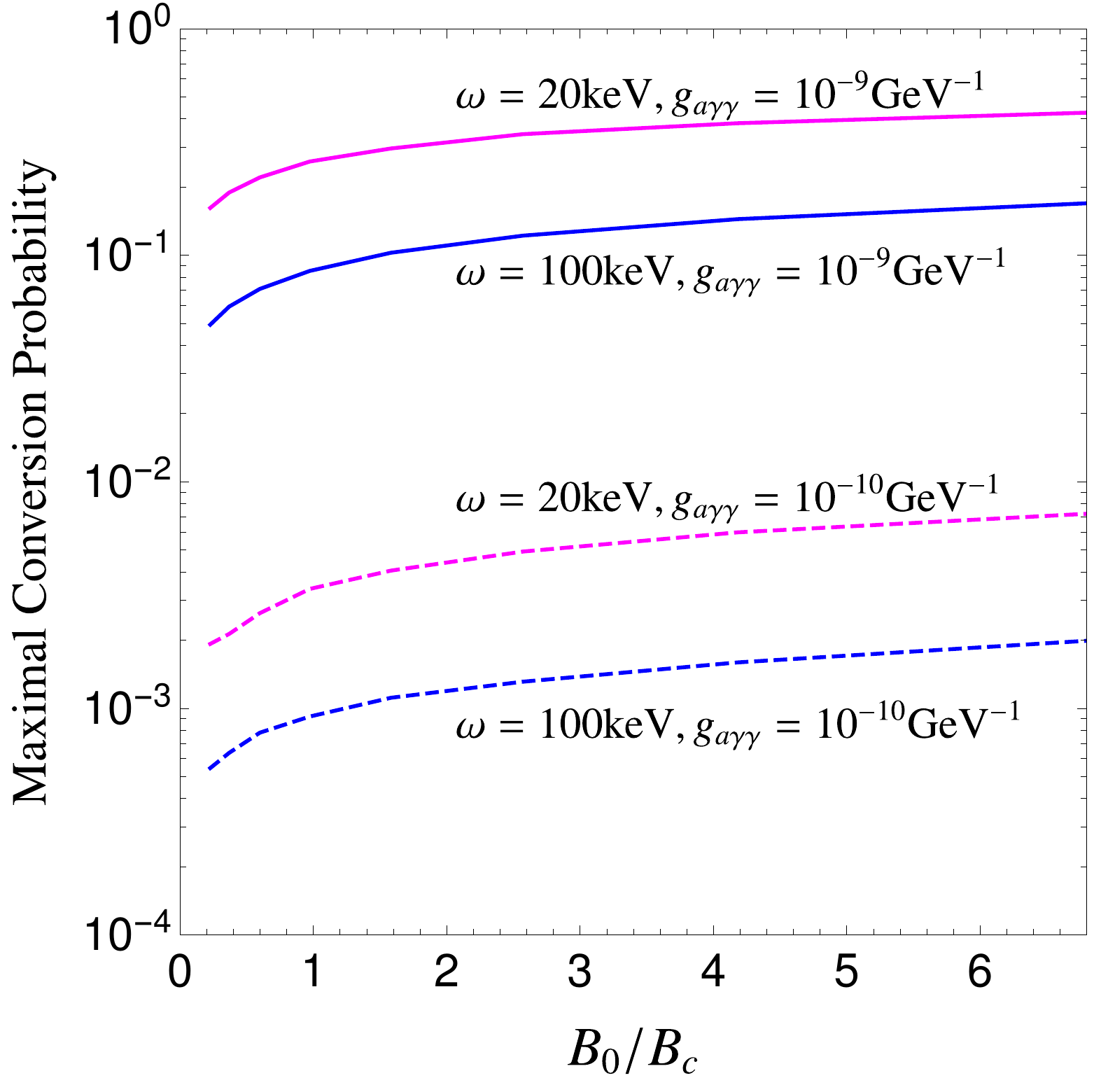}
\quad
\includegraphics[width=0.47\textwidth]{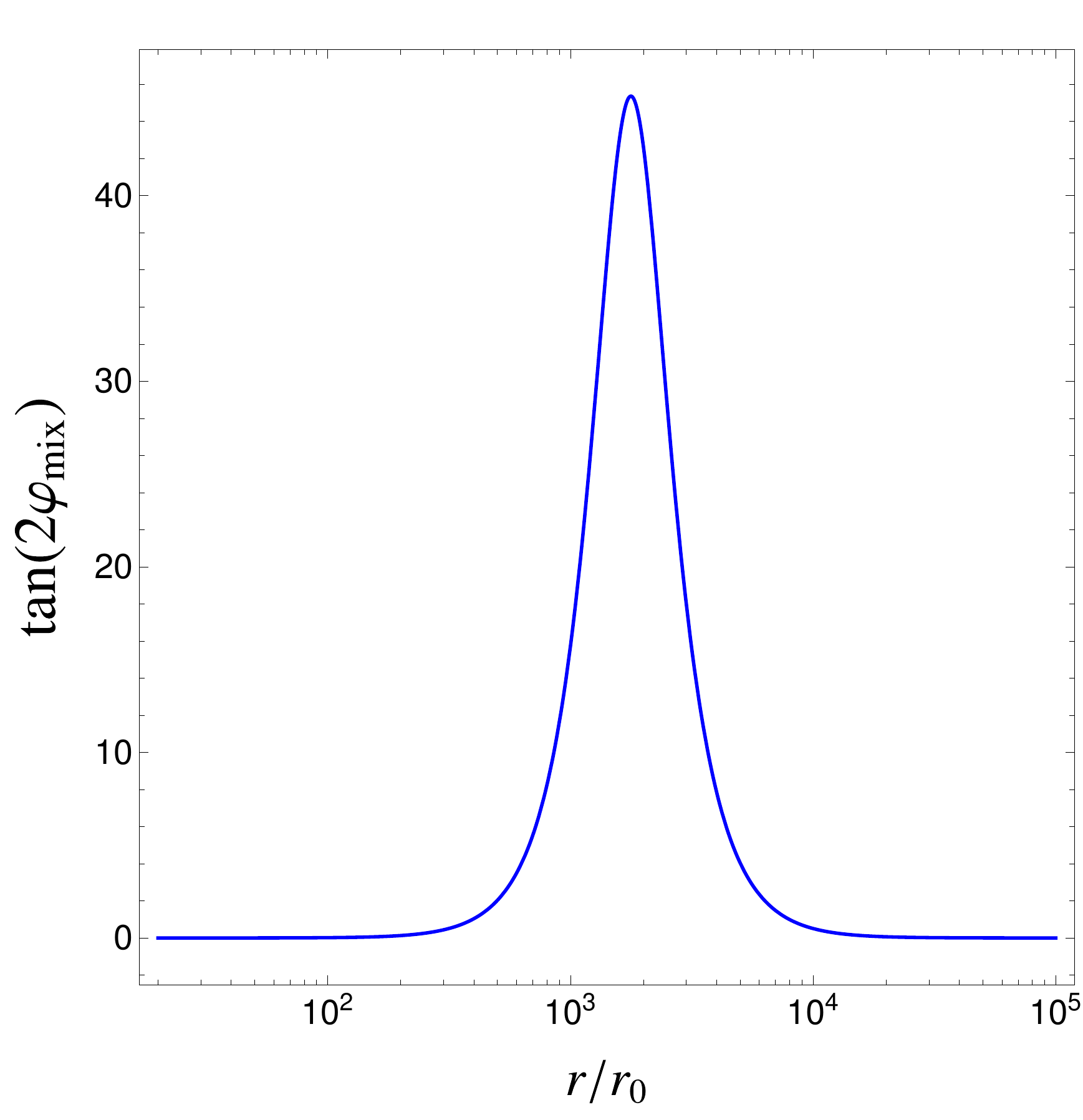}
\\
\includegraphics[width=0.48\textwidth]{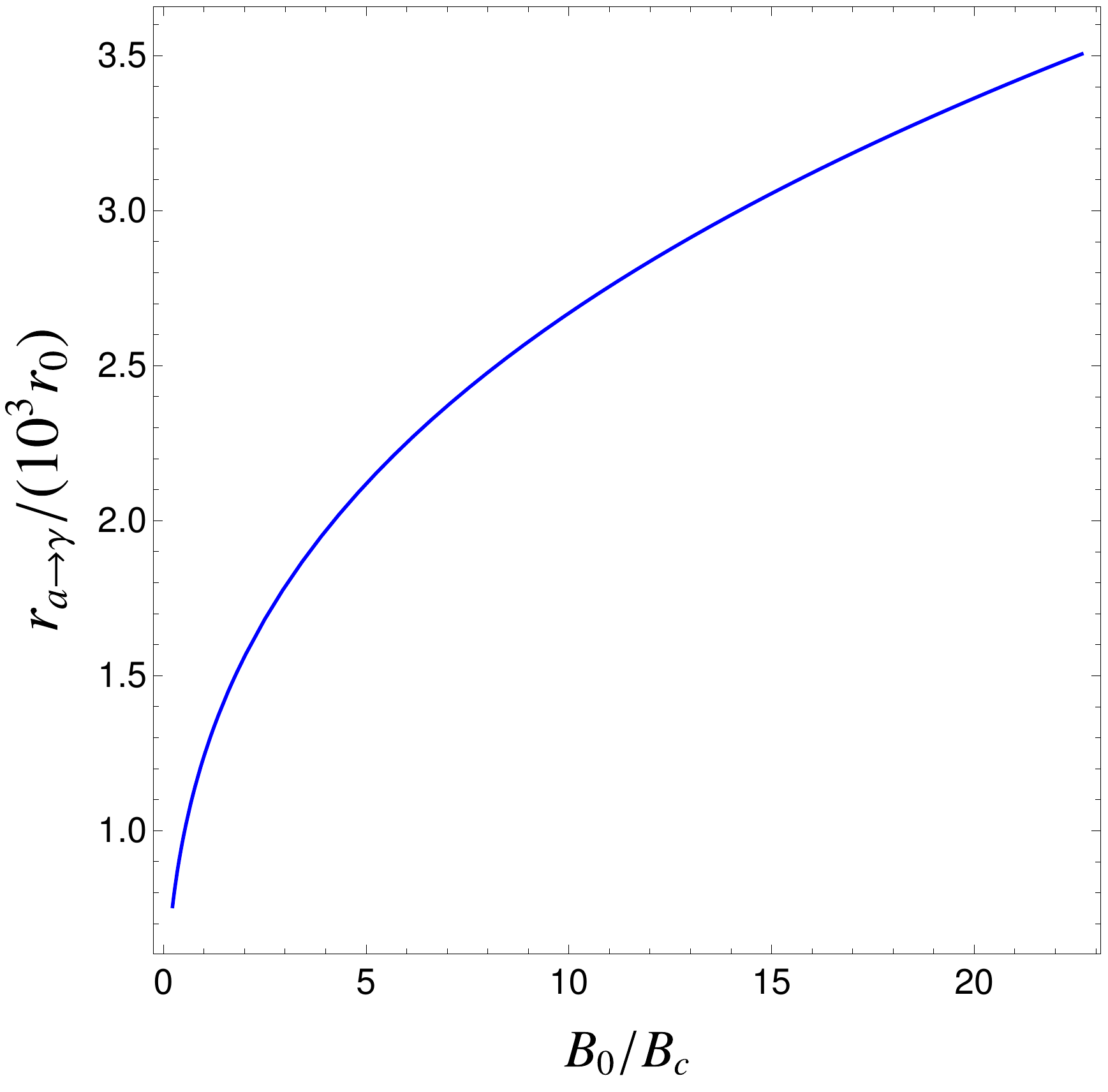}
\quad
\includegraphics[width=0.47\textwidth]{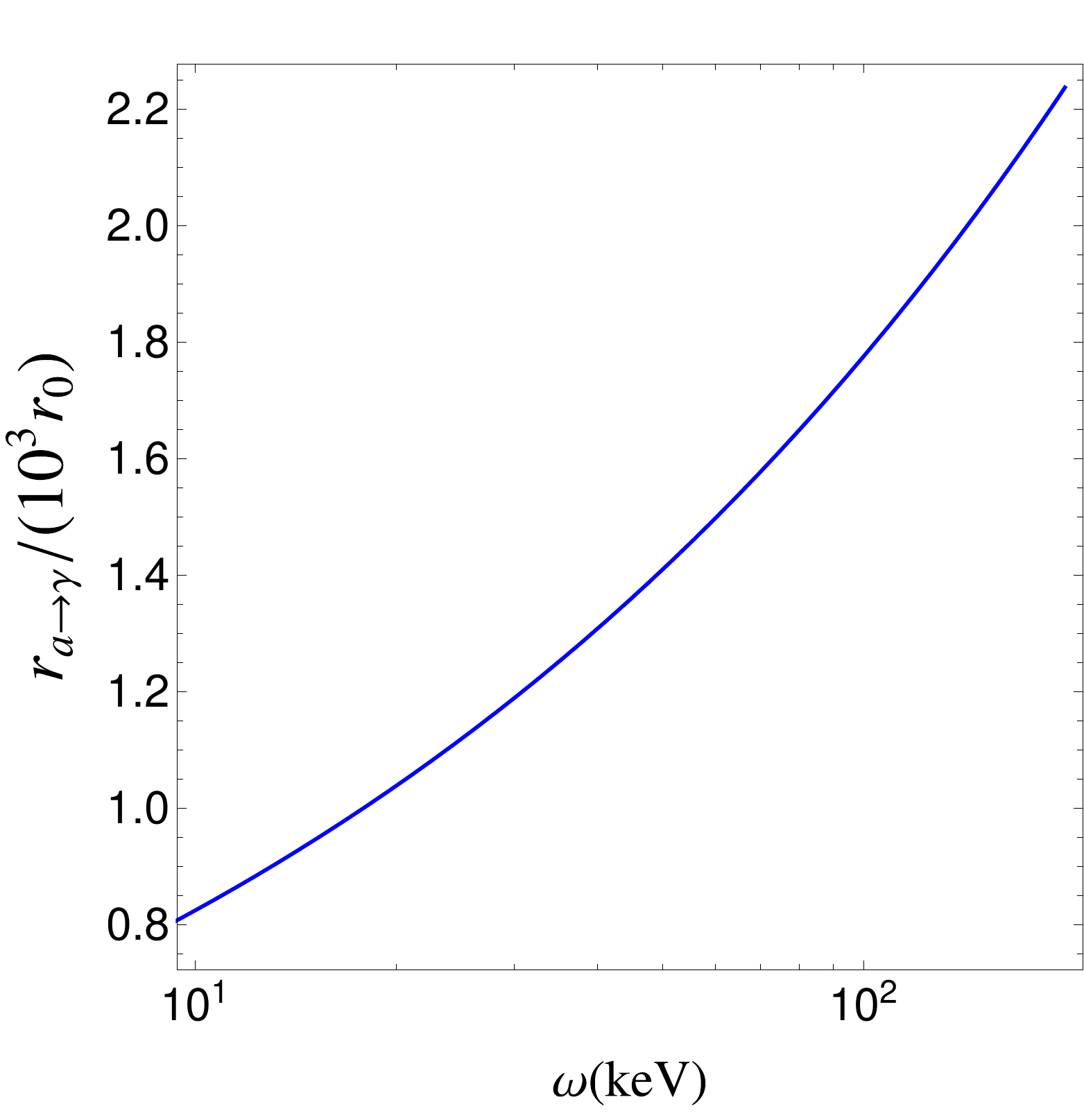}
\caption{
We show the maximal conversion probability (top left) as a function of $B_0/B_c$ with $B_0$ the surface magnetic field and $B_c$ the critical QED field strength for different combinations of the axion energy $\omega$ and axion-photon coupling $g_{a\gamma\gamma}$.
In top right, we show the mixing angle as a function of the dimensionless distance $r/r_0$ from the magnetar surface with $\omega=100\,\text{keV}$, $g_{a \gamma\gamma}=10^{-9}\,\text{GeV}^{-1}$ and 
$B_0=1.3\times 10^{14}$ G corresponding to Magnetar 4U 0142+61.  The bottom panel show the radius of conversion as a function of $B_0/B_c$ for $\omega=100\,\text{keV}$ (left) and as a function of the axion energy $\omega$ for fixed $B_0=1.3\times 10^{14}$ G (right).  In all plots we fix $\theta=\pi/2$, $m_a=10^{-6}$eV, $r_0=10\,\text{km}$ and we use a full numerical solution to the evolution equations when appropriate.
}
\label{fig:FigEvol}
\end{figure}
In this Appendix, we provide details of the ALP-photon conversion.

Firstly, we note that the Euler-Heisenberg approximation for refractive indices Eq.~\ref{EqnEH} are corrected when the magnetic field $B$ is large compared to $B_c$, which technically is the case close to the magnetar surface where $B\gtrsim B_c$.  However, the ALP-to-photon conversion occurs a few thousand radii $r_0$ away from the magnetar where $B\ll B_c$ following Eq.~\ref{dipolarB}, therefore the approximations \ref{EqnEH} are sufficient.

Since the decay of ALPs can be neglected, the ALP-to-photon conversion probability $P_{a\to\gamma}$ can be computed following the methods developed in \cite{Fortin:2018aom,Fortin:2018ehg}. A detailed numerical analysis of ALP conversion in the soft X-ray thermal emission band was performed in \cite{Lai:2006af,Perna:2012wn}.  As expected, the different results agree in the appropriate limits.

From the propagation equations \ref{EqnDiffMat}, one first notices that the perpendicular electric field decouples.  Hence conversion occurs only between ALPs and parallel photons.  By analogy with quantum-mechanical conservation of probability $\frac{d}{dx}[|a(x)|^2+|E_\parallel(x)|^2]=0$, it is convenient to parametrize the two remaining complex fields as
\be\label{EqnSoln}
a(x)=\cos[\chi(x)]e^{-i\phi_a(x)},\qquad\qquad E_\parallel(x)=i\sin[\chi(x)]e^{-i\phi_E(x)},
\ee
where $\chi(x)$, $\phi_a(x)$ and $\phi_E(x)$ are real functions.  With these new functions, the evolution equations Eq.~\ref{EqnDiffMat} simplify to
\be\label{EqnEvol}
\begin{aligned}
\frac{d\chi(x)}{dx}&=-\Delta_M(x)r_0\cos[\Delta\phi(x)],\\
\frac{d\Delta\phi(x)}{dx}&=[\Delta_a-\Delta_\parallel(x)]r_0+2\Delta_M(x)r_0\cot[2\chi(x)]\sin[\Delta\phi(x)],
\end{aligned}
\ee
where we defined the relative phase as $\Delta\phi(x)=\phi_a(x)-\phi_E(x)$.  There obviously is an analogous equation for the total phase $\phi_a(x)+\phi_E(x)$, but by focusing on the conversion probability it too decouples from the system.  It can thus be forgotten altogether.

The function $\chi(x)$ parametrizes the overall mixture of the state at the dimensionless distance $x$, with $\chi=0$ corresponding to a pure ALP state and $\chi=\pi/2$ corresponding to a pure parallel photon state.  Hence, with the boundary condition $\chi(1)=0$, \textit{i.e.}\ with a pure ALP initial state at the surface of the magnetar, the ALP-photon conversion probability is simply given by
\be\label{EqnProb}
P_{a\to\gamma}(x)=\sin^2[\chi(x)].
\ee
For the evolution equations \ref{EqnEvol} to make sense, we note that $\Delta\phi(1)$ must satisfy $\Delta\phi(1)=m\pi$ with $m\in\mathbb{Z}$.  It is now straightforward to compute the conversion probability Eq.\ref{EqnProb} numerically from the evolution equations \ref{EqnEvol}.

To reach a more intuitive understanding of the physics at play, one can first obtain a semi-analytic approximation to the conversion probability in the regime of small mixing, \textit{i.e.}\ for small $g_{a\gamma\gamma}$.  Using again a quantum-mechanical analogy, this time with time-dependent perturbation theory (see \cite{Raffelt:1987im}), the conversion probability Eq.~\ref{EqnProb} can be approximated by
\bea \label{EqnPapprox}
P_{a\to\gamma}(x)&=&\left|\int_1^xdx'\,\Delta_M(x')r_0\,\text{exp}\left\{i\int_1^{x'}dx''\,[\Delta_a-\Delta_\parallel(x'')]r_0\right\}\right|^2\nonumber\\
&=&(\Delta_{M0}r_0)^2\left|\int_1^xdx'\,\frac{1}{x'^3}\,\text{exp}\left[i\Delta_ar_0\left(x'-\frac{x_{a\to\gamma}^6}{5x'^5}\right)\right]\right|^2,
\eea
in a dipolar magnetic field Eq.~\ref{dipolarB}.  As mentioned above, Eq.~\ref{EqnPapprox} is accurate when $g_{a\gamma\gamma}$ is small enough, which is in our regime of interest.  We note the appearance of the dimensionless conversion radius $x_{a\to\gamma}$ in Eq.~\ref{EqnPapprox}, for which the conversion probability peaks, corresponding to
\be \label{EqnConvRadius}
x_{a\to\gamma}=\frac{r_{a\to\gamma}}{r_0}=\left(\frac{7\alpha}{45\pi}\right)^{1/6}\left(\frac{\omega}{m_a}\frac{B_0}{B_c}|\sin\theta|\right)^{1/3}.
\ee
Since it relies on the Euler-Heisenberg approximation Eq.~\ref{EqnEH}, the conversion radius $r_{a\to\gamma}$ Eq.~\ref{EqnConvRadius} is valid when it is much larger than the radius of the magnetar.  By self-consistency, the conversion radius must thus be large for the Euler-Heisenberg approximation to make sense.  It is easy to check that it is indeed the case for a typical magnetar, with the conversion radius several thousand times the magnetar radius.  This observation justifies the use of a dipolar magnetic field and the Euler-Heisenberg approximation.

Another intuitive way to understand the conversion radius is to analyse the mixing angle between the ALP field and the perpendicular photon field.  The mixing angle $\varphi_\text{mix}$ is given by
\begin{equation} \label{mixingangle}
\tan(2\varphi_\text{mix})=\frac{2\Delta_M(x)}{\Delta_\parallel(x)-\Delta_a}.
\end{equation}
For a typical magnetar as of relevance to this work, the mixing angle Eq.~\ref{mixingangle} at the magnetar surface is negligible.  But the position dependence of the different quantities lead to an increase of the mixing angle (and thus of the conversion probability) away from the magnetar.  Indeed, the photon refractive index appearing in the denominator of Eq.~\ref{mixingangle} leads to $\Delta_\parallel(x)\sim1/r^6$, whereas the ALP-photon mixing term appearing in the numerator of Eq.~\ref{mixingangle} goes as $\Delta_M(x)\sim1/r^3$.  Thus the denominator decreases faster than the numerator, leading to a maximum in the mixing angle corresponding to the conversion radius Eq.~\ref{EqnConvRadius}.  Therefore the mixing angle and the probability of conversion peak at the conversion radius $r_{a\to\gamma}\gg r_0$, before decreasing again.  From this we conclude that most of the conversion occurs at the conversion radius.

Returning to the probability of conversion, a simple re-scaling shows that the conversion probability at infinity $P_{a\to\gamma}\equiv P_{a\to\gamma}(\infty)$ can be expressed into the following form,
\begin{equation}\label{EqnPinfapprox}
P_{a\to\gamma}=\left(\frac{\Delta_{M0}r_0^3}{r_{a\to\gamma}^2}\right)^2\left|\int_\frac{r_0}{r_{a\to\gamma}}^\infty dt\,\frac{1}{t^3}\,\text{exp}\left[i\Delta_ar_{a\to\gamma}\left(t-\frac{1}{5t^5}\right)\right]\right|^2.
\end{equation}
For large conversion radius where $r_{a\to\gamma}\gg r_0$, the lower bound on the integral Eq.~\ref{EqnPinfapprox} can be set to zero, and the probability of conversion takes an even simpler form in both the small and large $|\Delta_ar_{a\to\gamma}|$ regimes,
\bea \label{EqnPinfapproxanal}
P_{a\to\gamma}=\left(\frac{\Delta_{M0}r_0^3}{r_{a\to\gamma}^2}\right)^2\times\begin{cases}\frac{\pi}{3|\Delta_ar_{a\to\gamma}|}e^{\frac{6\Delta_ar_{a\to\gamma}}{5}}&|\Delta_ar_{a\to\gamma}|\gtrsim0.45\\
\frac{\Gamma\left(\frac{2}{5}\right)^2}{5^\frac{6}{5}|\Delta_ar_{a\to\gamma}|^\frac{4}{5}}&|\Delta_ar_{a\to\gamma}|\lesssim0.45\end{cases}.
\eea

In Fig.~\ref{fig:FigEvol}, we display the maximal conversion probability as a function of $B_0/B_c$ for different combinations of the axion energy $\omega$ and axion-photon coupling $g_{a\gamma\gamma}$ (top left), the mixing angle as a function of the dimensionless distance $r/r_0$ for $\omega=100\,\text{keV}$, $g_{a\gamma\gamma}=10^{-9}\,\text{GeV}^{-1}$ and
$B_0=1.3\times 10^{14}$ G corresponding to Magnetar 4U 0142+61 (top right), as well as the radius of conversion as a function of $B_0/B_c$ for $\omega=100\,\text{keV}$ (bottom left) and as a function of the axion energy $\omega$ for fixed $B_0=1.3\times 10^{14}$ G (bottom right).
%%%%%%%%%%%%%%%%%%%%%%%%
\section{Uncertainty estimates from varying magnetar properties}\label{sec:uncertainties}
In this appendix, we consider, in turn, the effects on the axion production spectrum $\mathop{dN}/{d\omega}$ from uncertainties in the nuclear equation of state, the mass of the magnetar, and the model of proton singlet superfluidity chosen.  Then we combine these uncertainties to get a range of expected variation in $\mathop{dN}/\mathop{d\omega}$ and turn that into an uncertainty band on the $ G_{an}  \times g_{a\gamma\gamma}$ versus $m_a$ plane.  
%%%%%%%%%%%%%%%%%%%%%%%%%%
\subsection{Varying EoS}
\begin{figure}
    \centering
    \includegraphics[width=0.6\textwidth]{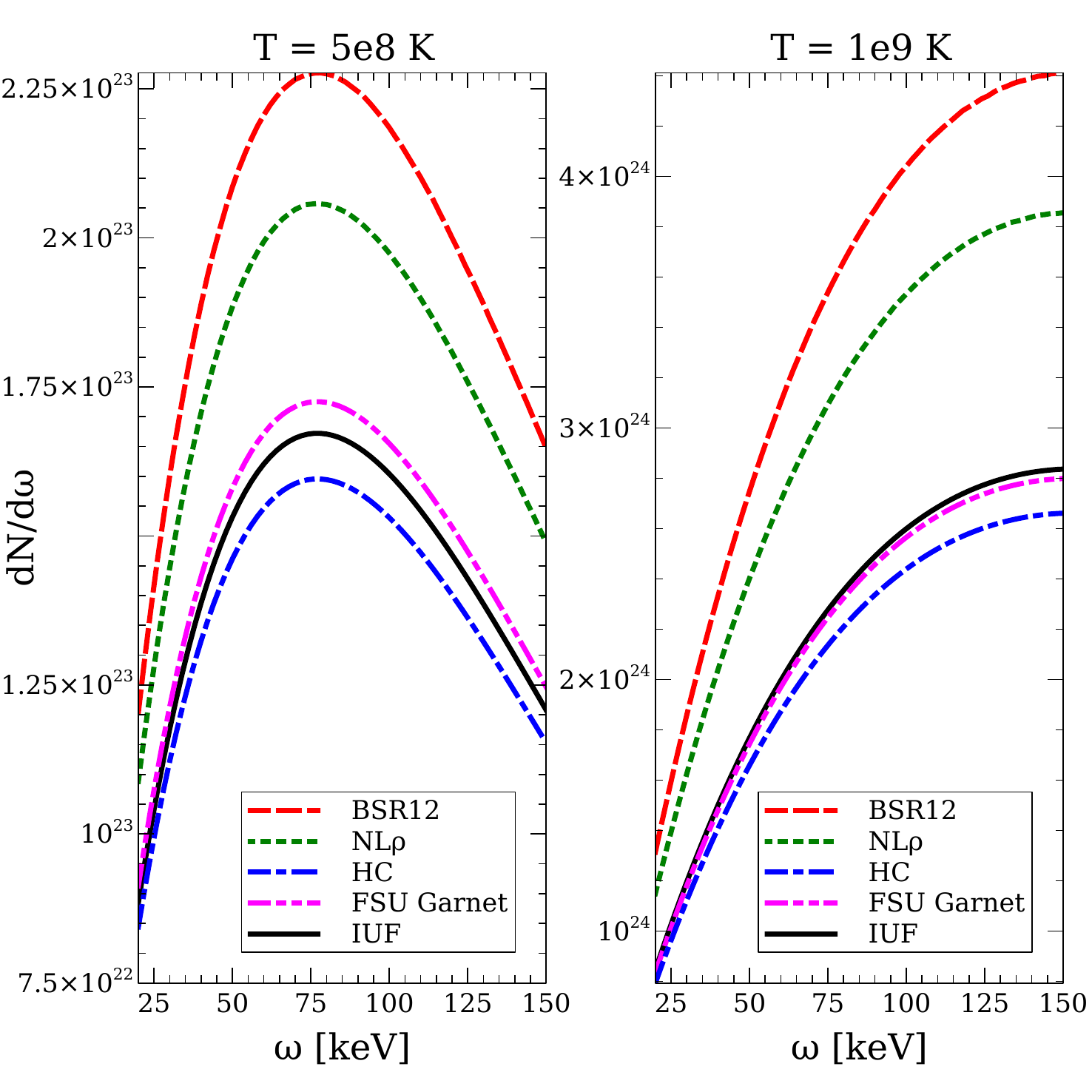}
    \caption{Axion spectrum emitted from a $1.4 M_{\odot}$ neutron star with proton $^1S_0$ CCDK proton superfluidity, for nuclear matter with five different equations of state.}
    \label{fig:dNdw_vary_EOS}
\end{figure}
To understand the influence of the nuclear equation of state on the axion production rate, we consider the same $1.4M_{\odot}$ neutron star with CCDK $^1S_0$ proton superfluidity that we used in the main text, but we choose a variety of nuclear EoSs to see the uncertainty that this brings to the axion spectrum $\mathop{dN}/\mathop{d\omega}$.  We compare the IUF EoS \cite{Fattoyev:2010mx} that we used in the main text with the four additional RMF-based EoSs BSR12 \cite{Dhiman:2007ck}, NL$\rho$ \cite{Liu:2001iz}, HC \cite{Bunta:2003fm}, FSU Garnet \cite{Utama:2016tcl}, each of which is roughly consistent with a variety of nuclear and astrophysical constraints as discussed in \cite{Nandi:2018ami}.  In Fig.~\ref{fig:dNdw_vary_EOS} we plot the axion spectrum at two different core temperatures, for several different equations of state.  The differences in the axion emission spectra stems from the small differences in the proton fraction predicted by the EoSs, as, depending on the temperature, axion emission from processes involving protons are Boltzmann suppressed by the superfluid gap over the temperature.  While different EoSs predict different axion emittion rates (though the differences are less than a factor of two), the spectral shape is almost unchanged.
%%%%%%%%%%%%%%%%%%%%%%%%%%%%%%%
\subsection{Varying neutron star mass}
\begin{figure}
    \centering
    \includegraphics[width=0.6\textwidth]{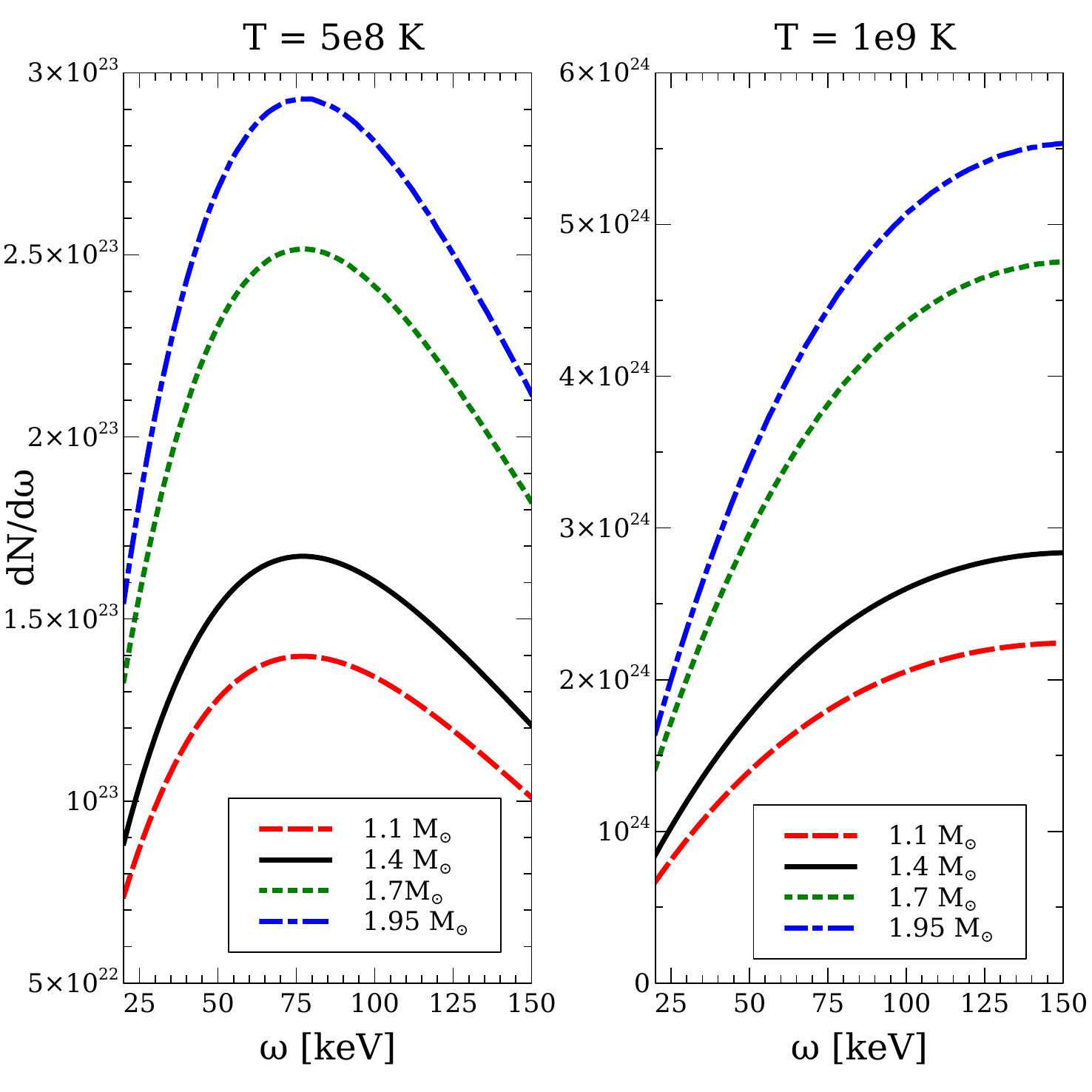}
    \caption{Axion spectrum emitted from a neutron star with matter modeled by the IUF EoS and with $^1S_0$ CCDK proton superfluidity, for different choices of the neutron star mass.}
    \label{fig:dNdw_vary_mass}
\end{figure}
As it is likely that the magnetars we analyzed in this work are not all $1.4M_{\odot}$ stars, it is important to examine how the axion constraints that we derive change as a function of the magnetar mass.  To study the effect of the mass of the neutron star on the axion production rate, we fix the nuclear EoS and the CCDK model for proton $^1S_0$ superfluidity, and vary just the magnetar mass by changing the value of the central pressure used in the solution of the Tolman-Oppenheimer-Volkoff (TOV) stellar structure equation.  The axion spectrum due to nucleon bremsstrahlung processes is plotted in Fig.~\ref{fig:dNdw_vary_mass} for two different assumptions of core temperature.  Higher mass neutron stars contain more baryons, and therefore emit an increased number of axions, but with an almost unchanged spectral shape.  The differences from the $1.4M_{\odot}$ magnetar considered in the text are within a factor of 2.  Of course, neutron stars with masses above $1.4M_{\odot}$ reach higher baryon densities in their cores, raising the possibility that they contain exotic phases of matter.  We do not consider this possibility in this work.
%%%%%%%%%%%%%%%%%%%%
\subsection{Varying proton $^1S_0$ superfluidity model}
\begin{figure}
    \centering
    \includegraphics[width=0.45\textwidth]{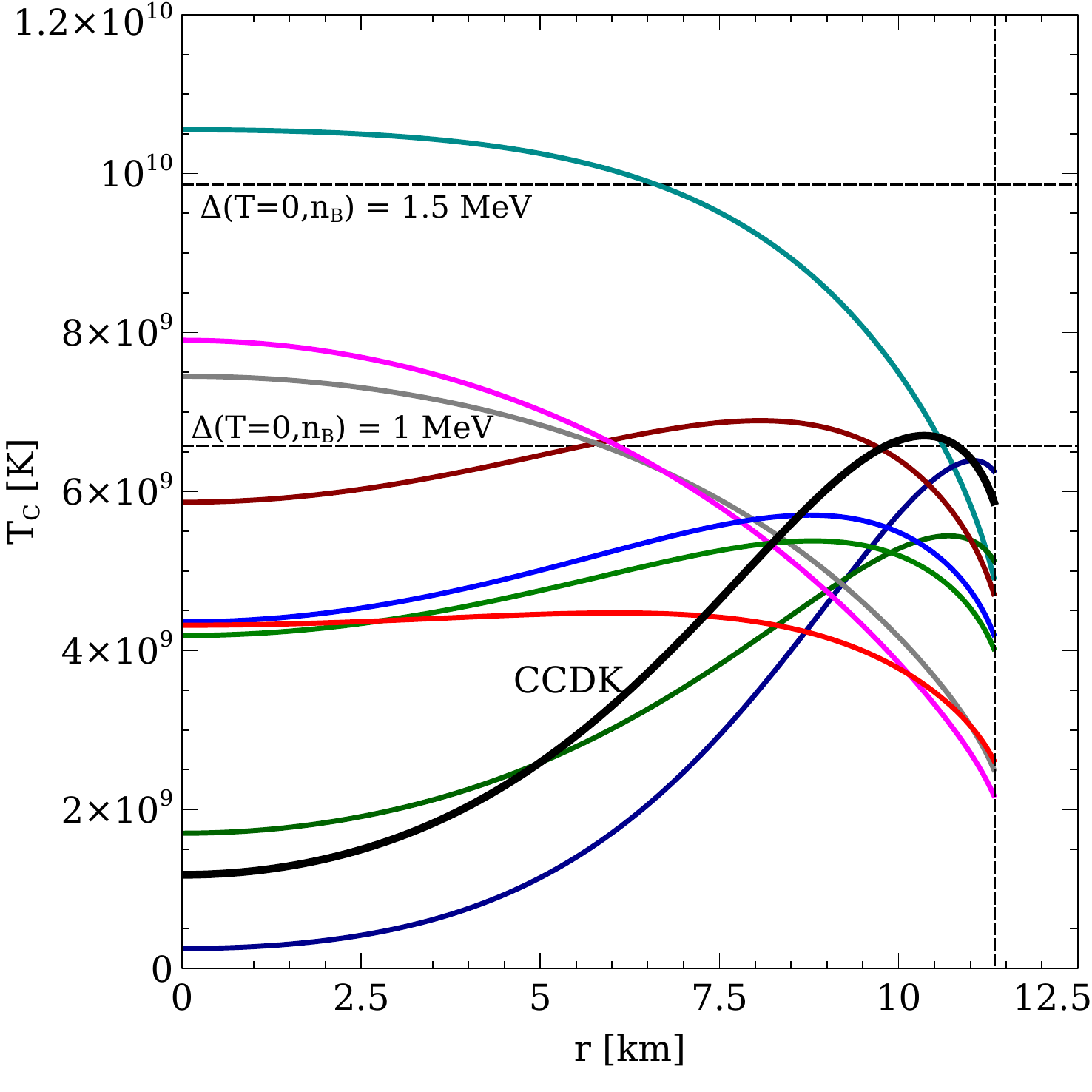}\hspace{1cm}
    \includegraphics[width=0.45\textwidth]{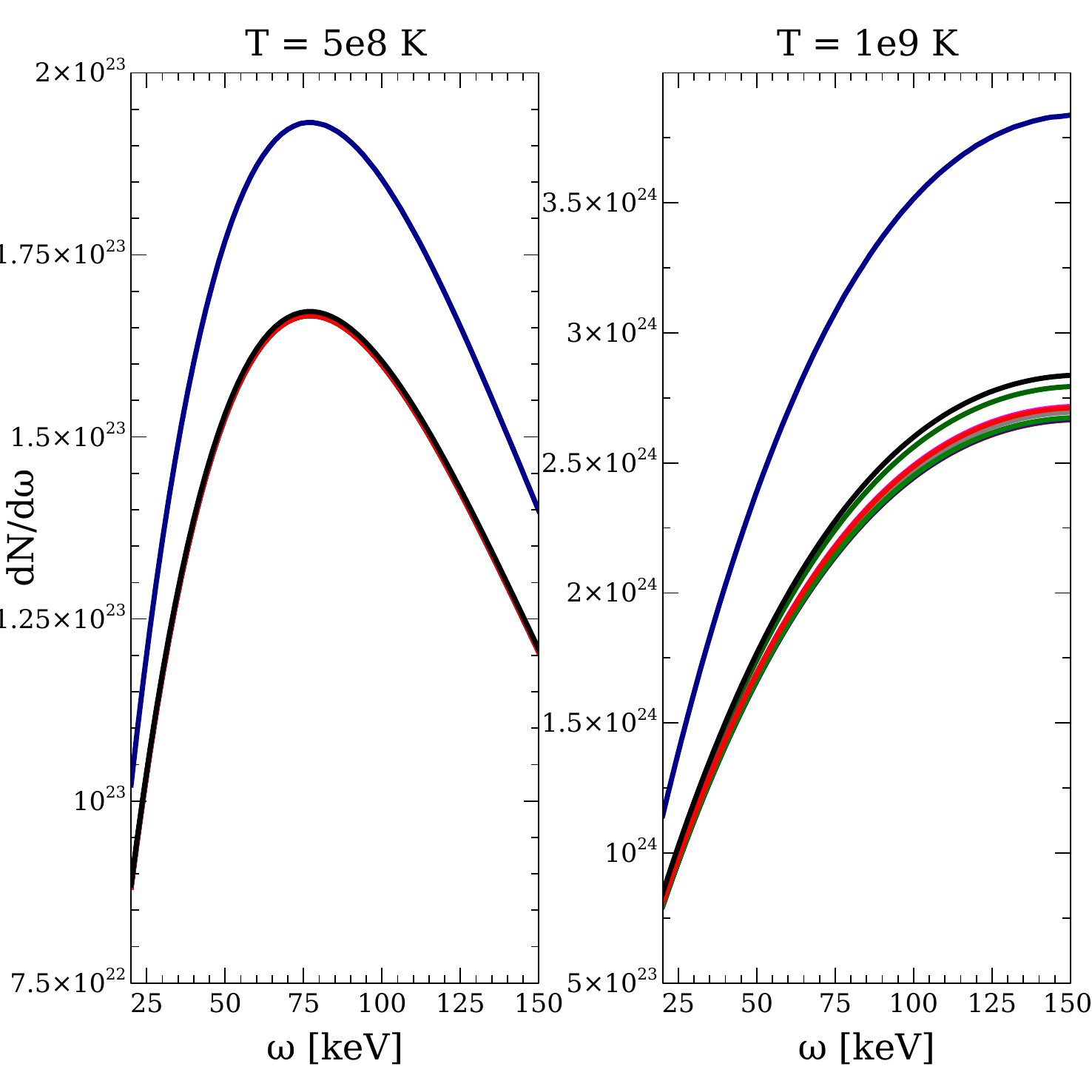}
    \caption{\textbf{Left panel:} Critical temperature profiles of ten models of proton superfluidity, discussed further in the text.  One of those models is CCDK, which is used in the analysis in the main text.  \textbf{Right panel:} Axion spectrum produced by a $1.4M{\odot}$ magnetar with the IUF EoS, with each model of proton superfluidity shown in the left hand panel.  The line colors are consistant between the two plots. }
    \label{fig:vary_superfluidity}
\end{figure}
As the density-dependence of the proton gap is not precisely known, to get an idea of the uncertainty associated with choosing a particular model (CCDK) of proton superfluidity, we study a $1.4M_{\odot}$ magnetar with the IUF EoS and choose a variety of possible $^1S_0$ proton gap profiles throughout the star.  We obtain 9 profiles by taking the gap parametrization
\begin{equation}
\Delta(T=0,n_B) = \Delta_0 \frac{p_{Fp}^2}{p_{Fp}^2+\alpha^2}\frac{(p_{Fp}-\beta)^2}{(p_{Fp}-\beta)^2+\gamma^2}
\end{equation}
given in \cite{Ho:2014pta} and choosing values for the parameters $\Delta_0, \alpha, \beta, \gamma$ between 0 and 1 GeV.  We only keep models where the critical temperature exceeds $4\times 10^9 \text{ K}$ in greater than one half of the volume of the core (following the constraints discussed in Sec.~\ref{superflu} and Ref.~\cite{Beznogov:2018fda}) and also where the gap does not exceed 2 MeV anywhere in the core, as current calculations seem to indicate this is unlikely \cite{Sedrakian:2018ydt}.  The critical temperature profiles of the nine models plus CCDK for a $1.4M_{\odot}$ neutron star are plotted in the left panel of Fig.~\ref{fig:vary_superfluidity} and the resultant axion spectra $\mathop{dN}/\mathop{d\omega}$ are shown in the right panel.  The different models of superfluidity result in spectra that are within 50\% of each other.  This is because only the protons are assumed to be superfluid and so the axion production rate always has a ``baseline'' $n+n\rightarrow n+n+a$ channel for axion production that remanins unchanged.  If we relaxed our assumption that neutron pairing only occurs below temperatures of $10^8 \text{ K}$, the differences in the axion emission spectra would be more dramatic.

%%%%%%%%%%%%%%%%%%%%%%%%%%%%%%%%%%%%%%%
\subsection{Combined errors due to neutron star mass, equation of state, and superfluidity}
\begin{figure}
    \centering
    \includegraphics[width=0.45\textwidth]{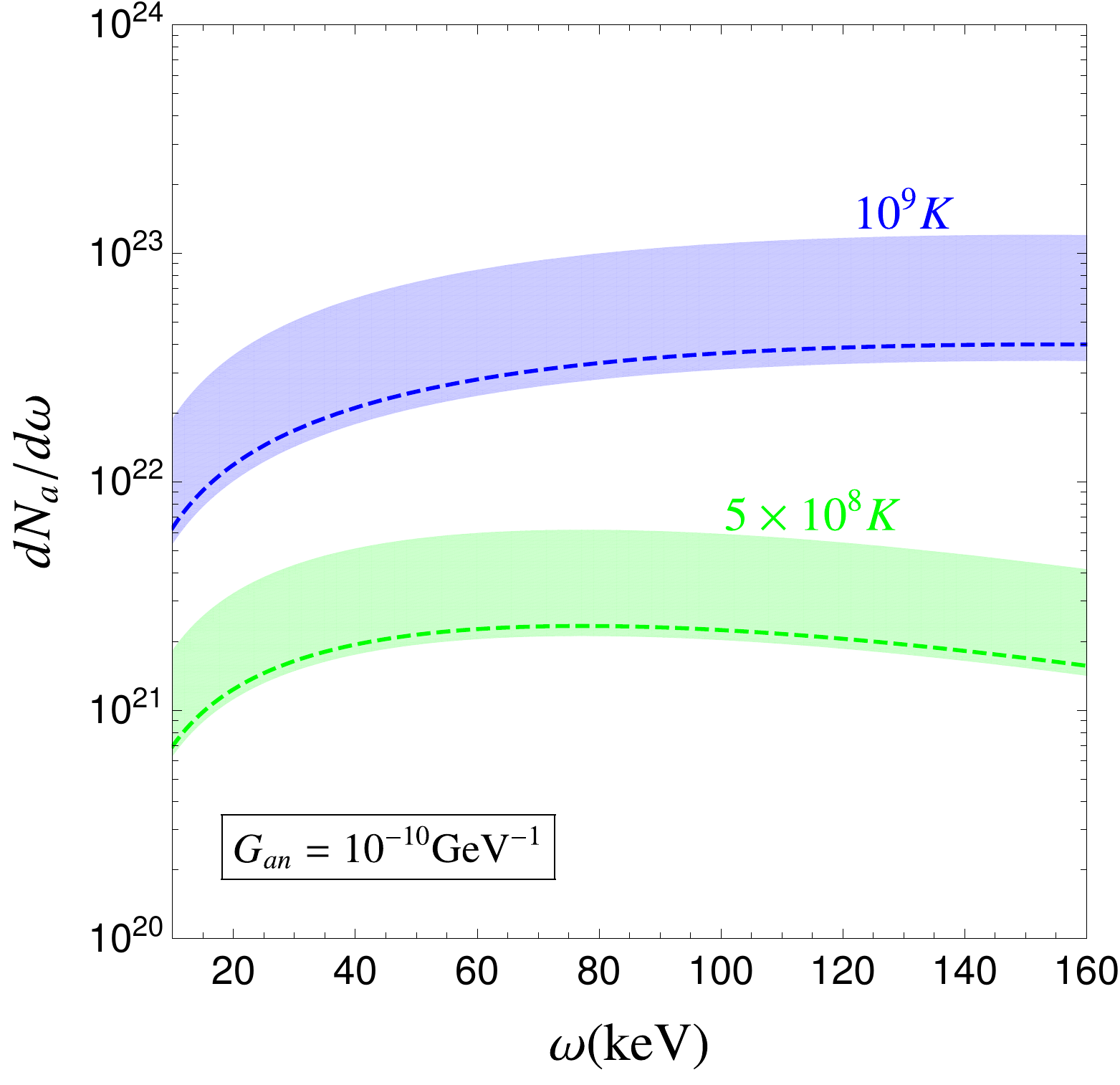}\hspace{1cm}
    \includegraphics[width=0.45\textwidth]{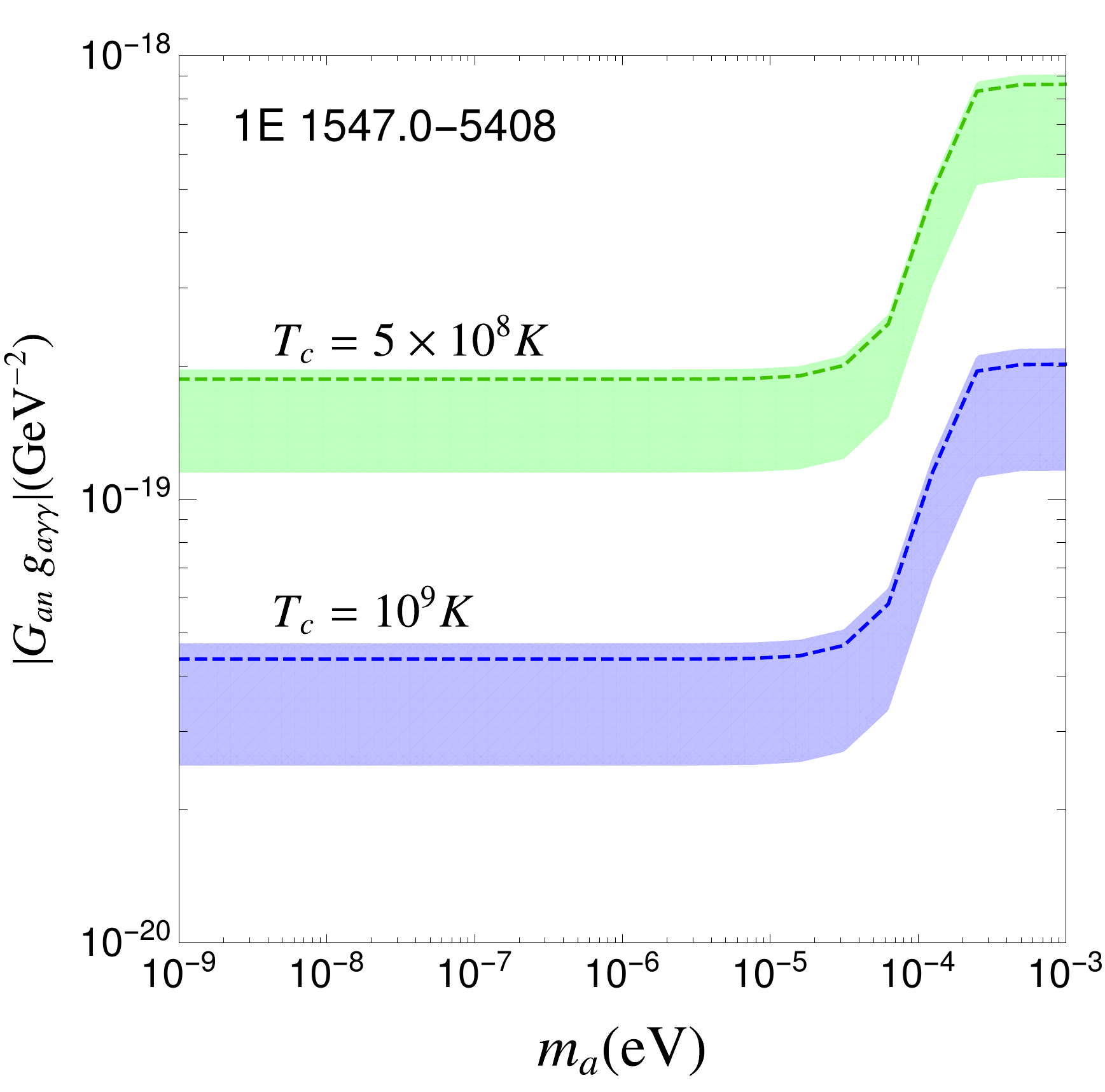}
    \caption{\textbf{Left panel:} The axion spectrum $\mathop{dN}/\mathop{d\omega}$ just as in Fig.~\ref{fig:dNadw}, but for two choices of core temperature and with band representing the uncertainty in $\mathop{dN}/\mathop{d\omega}$ due to uncertainties in the nuclear EoS, the magnetar mass, and the critical temperature of proton $^1S_0$ proton pairing.  The dashed curves correspond to the choices made in our analysis in the main text, i.e., the corresponding two dashed curves in Fig.~\ref{fig:dNadw}.  \textbf{Right panel:}  The 95\% CL upper limits from the magnetar 1E 1547.0-5408 for two different choices of magnetar core temperature.  Around these limits, which were displayed in the main text, we put bands representing the uncertainties in the constraints stemming from the uncertainties in the axion spectrum shown in the left hand panel of this figure.}
    \label{fig:upper_lower_bound_dNdw}
\end{figure}
While in the previous subsections of this appendix we studied the uncertainties in the neutron star mass, the EoS, and the proton superfluidity model independently, we now consider the total uncertainty combining these three factors.  In the left panel of  Fig.~\ref{fig:upper_lower_bound_dNdw} we show, for two choices of core temperature, a band ranging from the highest expected axion emission spectrum (corresponding to a $2.0M_{\odot}$ star with the BSR12 EoS and the smallest proton superfluid gaps) to the lowest expected axion emission spectrum ($1.1M_{\odot}$, HC EoS, and largest proton superfluid gaps).  The spectrum used in the text (IUF, $1.4 M_{\odot}$, CCDK) is shown as a dotted line within the uncertainty band.  This band indicates less than a factor of 3 uncertainty from the axion spectrum used in the text, and the spectrum we used in the main text was on the low side of the uncertainty band.

To see how these uncertainties in the axion emission spectrum influence the constraints we set in the $G_{an}\times g_{a\gamma\gamma}-m_a$ plane, in the right panel of Fig.~\ref{fig:upper_lower_bound_dNdw} (for the magnetar which provides the strongest constraint) we plot the constraint uncertainty bands corresponding to the spectrum uncertainty bands shown in the left panel of Fig.~\ref{fig:upper_lower_bound_dNdw}.  The constraints derived in the main text (see Fig.~\ref{fig:combined} and \ref{fig:combined2}) are shown as dashed lines.  From this plot, it is clear that the dominant uncertainty in our constraints on the axion couplings as a function of axion mass comes from the unknown magnetar core temperature, not the magnetar mass or aspects of nuclear physics.

%%%%%%%%%%%%%%%%%%%%%%%%%%%%%%%%%%%%%%%%%
\bibliographystyle{utphys}
\bibliography{mybib}
\end{document}